\documentclass[twocolumn,aip,reprint,superscriptaddress,english,floatfix]{revtex4-1}
\usepackage{amsmath}
\usepackage{amssymb}
\usepackage{amsfonts}
\usepackage[pdftex]{graphicx}
\usepackage{subfigure} 
\usepackage{array}
\usepackage[english]{babel}
\usepackage{braket}
\usepackage{color}
\usepackage{mathtools}
\usepackage{threeparttable}
\usepackage[colorlinks,linkcolor=blue,anchorcolor=blue,citecolor=blue,urlcolor=blue]{hyperref}
\usepackage{babel}
\makeatother

\usepackage{babel}
\begin{document}

\title{Engineering First-Order Quantum Phase Transitions for Weak Signal Detection}

\author{Li-Ping Yang}
\author{Zubin Jacob}
\email{zjacob@purdue.edu}

\homepage{http://www.electrodynamics.org/}

\selectlanguage{english}%

\affiliation{Birck Nanotechnology Center and Purdue Quantum Science and Engineering Institute,School of Electrical
and Computer Engineering, Purdue University, West Lafayette, IN 47906,
U.S.A.}
\begin{abstract}
The quantum critical detector (QCD), recently introduced for weak signal amplification  [Opt. Express 27, 10482 (2019)], functions by exploiting high sensitivity near the phase transition point of first-order quantum phase transitions. We contrast the behavior of the first-order as well as the second-order quantum phase transitions (QPTs) in the detector. We find that the giant sensitivity to a weak input signal, which can be utilized for quantum amplification, only exists in first-order QPTs. We define two new magnetic order parameters to quantitatively characterize the first-order QPT of the interacting spins in the detector. We also introduce the Husimi $Q$-functions as a powerful tool to show the fundamental change in the ground-state wave function of the detector during the QPTs and especially, the intrinsic dynamical change within the detector during a quantum critical amplification. We explicitly show the high figures of merit of the QCD via the quantum gain and signal-to-quantum noise ratio. Specifically, we predict the existence of a universal first-order QPT in the interacting spin system resulting from two competing ferromagnetic orders. Our results motivate new designs of weak signal detectors by engineering first-order QPTs, which are of fundamental significance in the search for new particles, quantum metrology, and information science.  
\end{abstract}

\maketitle
\section{Introduction}

Detection of weak quantum signals is central to precision metrology\cite{giovannetti2004quantum,Degen2017quantum}, search for new fundamental particles~\cite{ADMX2018}, studying gravitational effects~\cite{LIGO2017}, and quantum information science~\cite{hadfield2009single,eisaman2011invited}. An early example of a weak signal detector is the bubble chamber for charged particle detection in high-energy physics experiments~\cite{Glaser1952some}. Here, superheated liquid vaporizes into gas bubbles, which traces out the path of the charged particle. Another more recent example of a quantum detector is the superconducting nanowire single photon detector (SNSPD)~\cite{gol2001picosecond}. Here, a single photon causes a click event in a critically biased superconductor by inducing a phase change to the normal conducting state. The common theme between these two detection mechanisms is a weak signal inducing a thermodynamic phase transition. This concept recently inspired a new class of detectors that function by a weak signal triggering a quantum phase transition (QPT). To emphasize the need for a critical biasing field and a quantum (as opposed to classical) phase transition, we address this device as the quantum critical detector (QCD)~\cite{yang2018QCD}. 

The word ``critical" in QCDs is used to underscore  the requirement for a critical biasing signal which is not required in widely utilized quantum linear amplifiers. The philosophy of operation is similar to a conventional single photon triggered avalanche process where a weak signal (i.e. single-photon pulse) triggers an avalanche in an optimally biased system leading to a large number of electron hole pairs. The key difference is the read-out mechanism (amplification) in the QCD, which is not an avalanche but is related to the phase transition that causes a macroscopic excitation in a single bosonic output mode or an abrupt change in the long-range order of spins. The concept of a ``click" or single-shot quantum detection event is related to the non-analyticity that occurs near the phase transition point as a parameter in the detector is perturbed. QCDs necessarily require a first-order phase transition to ensure a giant change in macroscopic observables arising from a weak perturbing signal. Our focus is not the critical point of continuous phase transitions which have been recently exploited in quantum critical metrology~\cite{frerot2018quantumcriticalmetrology}.

This approach of exploiting first-order QPTs to detect signals in a single shot is fundamentally different from repeated measurement schemes, parameter estimation or quantum sensing/metrology~\cite{giovannetti2004quantum,giovannetti2006quantum,Degen2017quantum}. The read-out mechanism in these established schemes are related to interferometric processes which can be enhanced through engineering quantum states of light~\cite{Yurke1986su2,Dowling1998correlated} or matter~\cite{Bollinger1996optimal}. In stark contrast, the QCD does not utilize an interferometric read-out but requires a quantum bias that prepares the detector in a pre-determined phase close to the phase transition point~\cite{yang2018QCD}.

We also emphasize that the mechanism of quantum critical amplification is fundamentally different from the well established concept of quantum linear amplifiers~\citep{Caves1980QSL,caves1982quantum}. In this traditional amplification approach,  the weak input signal is directly amplified to generate a large output signal. The information carriers in input and output signals are usually of the same kind (eg: bosonic excitations) and the gain of the amplifier is defined as the output-to-input ratio of signal amplitudes not particle number. The canonical example is the quantum linear amplifier, including the phase-preserving (phase insensitive)~\cite{Haus1962quantum} and parametric (phase-sensitive)  amplifiers~\cite{louisell1961quantum,Mollow1967parametric1,collett1988quantum,roy2016introduction}. In his seminal work~\cite{caves1982quantum}, Caves presented a comprehensive review of the fundamental quantum limit for linear amplification under this scheme. This quantum limit lays a lower bound on the minimum amount of noise added by a high-gain bosonic quadrature amplifier during the amplification~\cite{clerk2004quantum,caves2012quantum}, which has also been generalized to fermionic amplifiers~\cite{Gavish2004generalized}. 

In stark contrast, for the phase transition amplification scheme, the weak input signal functions as a control of an optimally biased phase transition system, which is significantly different from the quantum linear amplifiers. In these critically biased amplifiers, the input and output information carriers can be fundamentally different (eg: input photons and output electrons) and the corresponding gain (i.e., the amplification factor) is defined as the ratio of the outputs with and without the input control signal. The classical critical detectors have been extensively used in practical experiments, such as the SNSPD~\cite{gol2001picosecond}, the single-photon avalanche diode (SPAD)~\cite{bulter2014single}, etc. The key amplification mechanism in the superconducting detectors is based on the thermodynamic (classical) phase transition triggered by the weak input signal. The QCD falls under this critical amplification scheme and it is an open question whether quantum limits can be placed on this class of critically biased amplifiers~\cite{Propp2019nonlinear}.

In this paper, we show universal detection and amplification behavior in a class of models exhibiting quantum phase transitions. We use this to explain the amplification mechanism of a QCD. The detector model we introduce is closely related to the Dicke model~\cite{hepp1973superradiant} and the Lipkin-Meshkov-Glick (LMG) model~\cite{lipkin1965validity,meshkov1965validity,glick1965validity}. We also introduce two new magnetic order parameters (OPs) to rigorously characterize the first-order QPTs in the detector model. By employing a mean-field theory as well as full numerics, we generate the complete phase diagram of the detector. 

We uncover the first-order QPT in the LMG model, which is essential for our proposed QCD and has not been revealed in previous literature. We find that the first-order QPT  is fundamentally tied to the competition between two ferromagnetic phases with long-range spin order in $x$- and $y$-axis, respectively. We also predict that a universal first-order QPT exists in an interacting spin system with competing ferromagnetic orders. 

We also numerically show that, at the first-order phase transition point of the detector, the sensitivity function $\chi$ diverges with $N^2$-scaling, where $N$ is the spin number. This scaling is much faster than previous first-order phase transitions~\cite{gammelmark2011phase,raghunandan2018high} and provides extraordinary high sensitivity for weak signal detection. To understand the microscopic mechanism of the QPTs, we display the fundamental changes in the ground-state wave function during the phase transitions using Husimi $Q$-functions. 

For natural atomic systems, the superradiant QPT in the Dicke model has been ruled out by the no-go theorem ~\cite{rzazewski1975phase,rzazewski1979nogo}. Here, we show explicitly how to overcome this no-go theorem. We show that the spin-spin (atom-atom) interaction can decrease the strong atom-field coupling required by the superradiant QPT significantly. This allows the super-radiant QPT to occur.

Our work overcomes challenges in the simulation of quantum phase transition dynamics with weak signal perturbation. The dynamical evolution of a system near the phase transition point necessarily involves excited states which can show significant deviations from the conventional ground-state to ground-state transition behavior. Via direct time-dynamic numerical evaluation, we overcome this issue. We show the linear scaling in both the maximum quantum gain and the corresponding signal-to-quantum noise ratio (SQNR), which reveal high figures of merit of our QCD. We also use the time-dependent $Q$-functions to show the macroscopic changes in the bosonic output mode and the long-range spin order during the dynamical critical amplification in our QCD. Usually, the enhanced decay of the Loschmidit echo around the phase transition point is utilized to measure QPTs~\cite{quan2006decay,Heyl2013Dynamical}. However, this enhanced echo decay exists in both first-order and second-order QPTs and only describes the deviation in the wave function from the initial state during the QPT. The enhanced quantum gain around the phase transition point addressed in this paper is a unique and universal characteristic of first-order QPTs, which captures the fundamental change in the macroscopic order of the detector.  

Our proposed device can be obtained by engineering a multi-qubit system to exhibit an artificial phase transition. We note that simulation of quantum phase transitions has become a major recent area of interest.  Second-order Ising-like QPTs have been demonstrated in experiments with trapped ions~\cite{zhang2017observation}, cold atoms~\cite{bernien2017probing} and circuit QED~\cite{Harris2018Phase}. As QPTs occur at zero temperature, our proposed QCD may have higher signal-to-noise ratio and lower dark counting rate than detectors utilizing thermodynamic phase transitions. 

We also note, that the detection of the axion, which is the prominent dark-matter candidate, is based on measuring single microwave photons generated from axion-photon conversion process ~\cite{Sikivie1983experimental,stern2015cavity,ADMX2018}. Currently, efficient detection of propagating microwave single-photon pulses remains challenging, due to the extremely low energy carried by the pulse~\cite{inomata2016single,kono2018quantum,Besse2018singleshot}. Our proposed quantum signal detector based on first-order quantum phase transitions might pave a new path for microwave photon counting and axion detection. 

This paper is organized as follows. We first provide a general introduction to the QCD paradigm in Sec.~\ref{sec:QCD}. Then, we introduce the Dicke-LMG model as an explicit detector model to demonstrate the QCD in Sec.~\ref{sec:Hamiltonian}. In Sec.~\ref{Sec:phase_diagram}, we establish two new magnetic order parameters (OPs) to characterize the quantum phases of the detector and present the complete phase diagram of the detector obtained from the mean field theory. By splitting the Dicke-LMG model into three sub-models, we study the first-order as well as second-order QPTs existing in the full model and the fundamental changes in the ground-state wave function during the phase transitions in Sec.~\ref{sec:QPT_QF}. In Sec.~\ref{sec:dynamical_amp}, we demonstrate the dynamical quantum critical amplification in the QCD by exploiting the giant sensitivity of the first-order QPT. In Sec.~\ref{sec:exp_realization}, we list some possible platforms to demonstrate our QCD. Finally, we collect our main conclusions in Sec.~\ref{sec:Conclusion}. The details of the mean-field theory are given in Appendix~\ref{sec:mean-field} and the numerical approach used in this paper is presented in Appendix~\ref{sec:numerical_simulation}.

\section{Quantum Critical Detector \label{sec:QCD}}
In our previous work~\cite{yang2018QCD}, we proposed a prototype QCD by exploiting the giant sensitivity in a first-order QPT. We note that originally, the critical point was defined specifically to denote the point where a continuous thermodynamic phase transition occurs.   Thus, the concept of quantum critical point and quantum criticality is traditionally used only for continuous quantum phase transitions~\cite{sachdev2007quantum}.  To avoid confusions in interpretation of ``critical"  for discontinuous QPTs in the detector context vs. the concept of ``critical" in the continuous QPT context, we denote the optimum bias point of the detector as the phase transition point hereafter.  It should however be noted that even at the discontinuous first-order QPT point, the free energy and entropy still change continuously with temperature. Our QCD is inspired by the routinely used classical critical detectors, such as SNSPD~\cite{gol2001picosecond,eisaman2011invited} and the bubble chamber~\cite{Glaser1952some}. The amplification of these classical detectors is based on the ultra-high sensitivity of the detector at the phase transition point of the thermodynamic (classical) phase transitions. Our QCD is the first quantum analog of the classical critical detectors. In this section, we explain the input signal, the amplification mechanism, and the macroscopic output signal of our QCD explicitly. We also explain why first-order QPTs are essential for quantum critical amplification.

To show the analogy between our proposed QCD with the conventional detectors, we deconstruct the SNSPD to explain the amplification scheme in classical critical detectors. The input of the SNSPD is a single-photon pulse---an extremely weak quantum signal. The core element of the SNSPD is a superconducting nanowire with typical width $100$ nm, thickness $8$ nm, and length $10\,{\rm \mu m}$~\cite{schuck2013waveguide,korzh2018demonstrating}. The current in the superconducting nanowire is biased very close to the critical current, thus even a single-photon pulse can break the superconductivity~\cite{jahani2019probabilistic}. The output signal is the voltage difference between the two ends of the superconducting nanowire. In the transduction (absorption) process, the incident single-photon pulse generates one resonantly excited electron. As the center frequency of the pulse is much larger than the energy gap of the superconductor, this highly excited electron will break hundreds of Cooper pairs via the strong electron-electron interaction. Then, the local temperature around the excited electron increases to form a hot-spot. This hot-spot diffuses and reduces the local density of the superconducting electrons within this small hot region. Finally, a phase transition from a superconductor to a normal metal in this small cross-section slab occurs and blocks the superconducting current in the nanowire. An observable output voltage pulse is generated to realize the amplification. Here, we see that the high sensitivity in the superconducting phase transition and the critical bias current play the key roles in weak quantum signal amplification.

Next, we show the analogy between a QCD and the SNSPD and explain the  macroscopic output signal of our proposed detector.

\subsection{Quantum critical amplification scheme}
The full measurement in our QCD is split into two main processes: transduction (absorption) and amplification.  In our proposed QCD, absorption of the incident weak signal leads to a small time-dependent variation in a relevant parameter of the detector system instead of the temperature. The amplification is realized by the QPT triggered by this parameter variation, in contrast to the thermodynamic phase transition triggered by the temperature change in an SNSPD. 

The input signal of our QCD can be an arbitrary quantum weak signal, such as a single-photon pulse. After absorption of this input signal, a small time-dependent variation in the detector parameter is generated. Similar to the classical critical detectors, we also need to bias our QCD very close to the phase transition point to guarantee that this small parameter variation can cross the phase boundary to trigger a QPT. As explained later, when a first-order QPT occurs, a large output change can be obtained to complete the amplification. We see that the amplification scheme in the QCD is the same as the classical critical detector, with only replacing the thermodynamic phase transition with a QPT.

In practice, we can engineer and select which parameter of the detector to be changed by the input signal. One example is the interaction strength within the detector, e.g. the spin-boson coupling or spin-spin coupling shown in this paper. It can also be a small effective magnetic field change coming from the magnetic dipole of the absorber, such as a nitrogen-vacancy center~\cite{yang2019Single}. The dynamics of the absorption of a quantum pulse can be theoretically incorporated into the dynamical quantum critical amplification~\cite{young2018limits,yang2018concept}. Without loss of generality, we model the transduction (absorption) process as a temporal change in the detector parameter, which is assumed to be proportional to the input signal absorption probability for simplicity. For a specific realization of the QCD, the transduction process can be numerically simulated via the quantum pulse scattering theory~\cite{yang2018concept,yang2019SPF}.

\subsection{Macroscopic output signal}
The output amplified signal of a QCD is the macroscopic change in one of the OPs. One simple example demonstrated in this paper is the superradiant OP, i.e., the macroscopic excitation in a single bosonic mode. Before the absorption of the incident weak signal, the detector is biased in the ground state of the phase, in which the bosonic mode is in the vacuum state. Finally, the bosonic mode evolves to a state with macroscopic excitations after the first-order QPT is triggered by the input signal. The macroscopic population in the bosonic mode can be read out via a classical device directly.

The detector model and the corresponding readout channel presented in this paper is only one explicit example to show the amplification mechanism of the QCD. In practice, we can engineer the QPTs, readout channel, and especially the interaction between the detector and the input signal to detect different kinds of particles, such as photons, charged particles, axions, etc. Another output signal of the prototype QCD in this paper could be the magnetic noise change, i.e., readout the in-plane magnetic fluctuations of spins with spin noise spectroscopy~\cite{zapasskii2013spin}. . 

\subsection{Essential role of the first-order quantum phase transition}
\begin{figure}
\includegraphics[width=8cm]{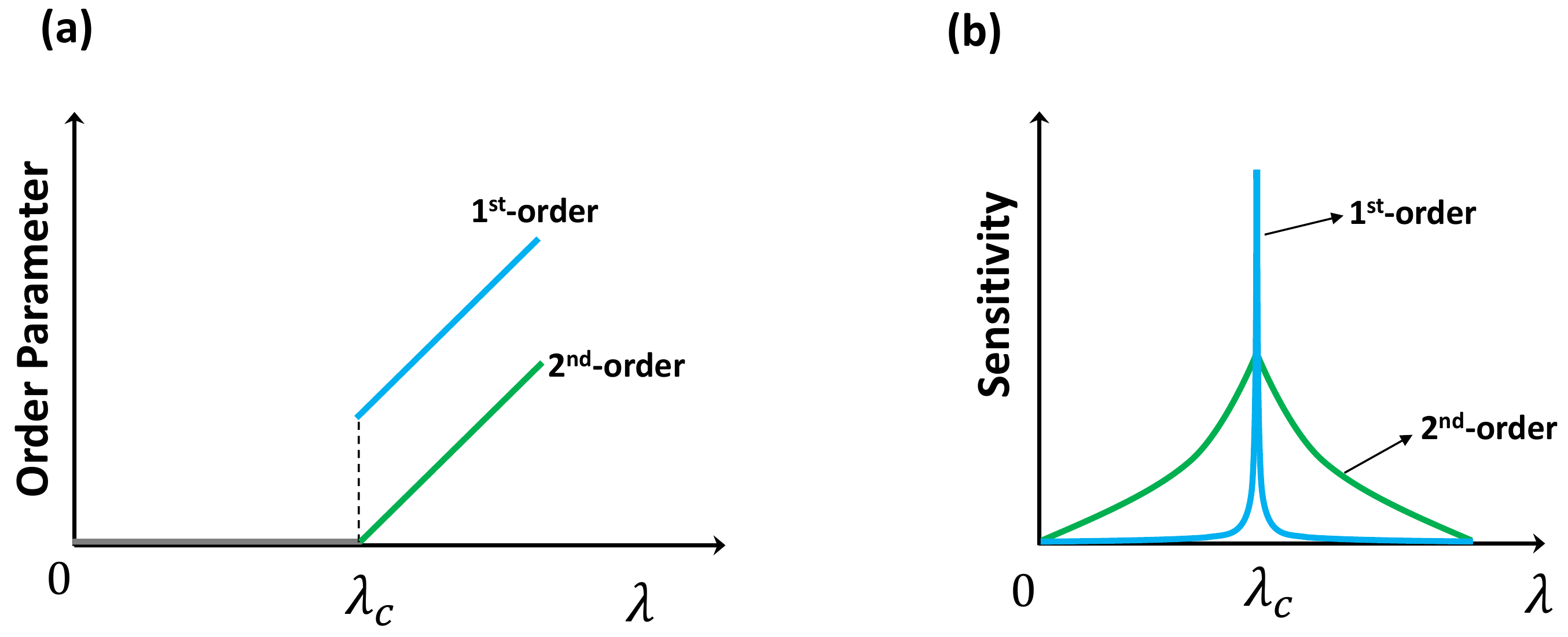}\caption{\label{fig:PhaseOrder} Schematics of the difference between first- and second-order quantum phase transitions with the order parameter (OP) in panel (a) and the sensitivity characterized by the first derivative of the order parameter ($d {\rm OP}/d\lambda$) in panel (b). Here, $\lambda_c$ denotes the phase transition points.}
\end{figure}

Now, we explain why first-order QPTs are essential for QCDs. A QPT describes an abrupt change in the ground state of a many-body system at zero temperature~\cite{sachdev2007quantum}. Most of the QPTs discovered in physical systems are of second-order, like the QPTs in Ising model~\cite{lieb1961two}, the Hubbard model~\cite{hubbard1963electron,Gu2004entanglement} and Bose-Hubbard model~\cite{Fisher1989boson}, the LMG model~\cite{lipkin1965validity,meshkov1965validity,glick1965validity}, the Dicke model~\cite{hepp1973superradiant,wang1973phase,hioe1973phase,hepp1973equilibrium}, etc. Here, we emphasize that first-order instead of second-order QPTs are required for quantum critical amplification. We compare the differences between first- and second-order QPTs schematically in Fig.~\ref{fig:PhaseOrder}. From panel (a), we see that the order parameter in second-order (or higher-order) QPTs changes continuously at the phase transition point $\lambda_c$. No macroscopic change in the OPs, which functions as the output signal of a QCD, exists during a second-order QPT. Even at the phase transition point, a large parameter variation is required to obtain an observable change in the system. This large parameter variation cannot be induced by a weak input signal, such as a single-photon pulse. Thus, a high quantum gain cannot be obtained using second-order QPTs, which limits their practical applicability for weak signal amplification. In contrast to the continuous phase transitions, the order parameter changes discontinuously at the phase transition point in a first-order QPT. Thus, even a very small parameter change at the phase transition can lead to a significantly large change in the values of the OP (the output signal). This elucidates the great potential of first-order QPTs for high-gain amplifiers. 

To characterize the sensitivity of a phase transition system, we defined a sensitivity function as the first derivative of the order parameter. From panel (b), we see that a non-analytical "kink" exists in the sensitivity function of both first- and second-order QPTs. However, the height of the sensitivity function at the first-order QPT point, which diverges in the thermodynamic limit, is much higher than that of a second-order QPT with a finite peak. Thus, first-order QPT can have much higher sensitivity and provides a natural platform for quantum metrology~\cite{giovannetti2004quantum,giovannetti2006quantum}, quantum amplification~\cite{caves1982quantum}, and new types of single-photon detectors~\cite{eisaman2011invited}. 

We also emphasize that the discontinuous jump and the diverging sensitivity in first-order QPTs result from a macroscopic order change in the ground states of two neighbouring quantum phases. Due to the vanishing energy gap at the phase transition point, the transition between the ground states of two quantum phases cannot be realized by a unitary adiabatic operation~\cite{dziarmaga2010dynamics}. By varying a parameter accross the phase transition point with time, one cannot induce a QPT from one ground state to the other in practice. Thus, a dynamical detection event may have totally different sensitivity scaling. Few first-order QPTs have already been found, like the Dicke-Ising model~\cite{lee2004first,gammelmark2011phase,zhang2014nonlocal}, the anti-ferromagnetic Ising chain~\cite{Ovchinnikov2003anti,Amin2009first,zhang2014quantum}, the LMG model in the zero-field limit~\cite{Vidal2004Entangle1}, the quantum Ising model with a four-spin exchange interaction~\cite{del2016nonequilium}, etc. Nevertheless, the dynamics of these QPTs around the phase transition point has not been revealed. The application of these first-order QPTs is, therefore, an open problem, as practical detection events and amplifications are fundamentally dynamical processes. 

In the following, we will present our prototype QCD in detail. First, we introduce the model Hamiltonian of our QCD. Via the mean-field theory, we derive the complete phase diagram of the detector. We also show the details of the QPTs within the detector with numerical simulations, specifically the fundamental change in the ground-state wave function. Finally, we display the dynamical change in the OPs and the wave function of the detector to show the quantum critical amplification process within the detector.       

\section{Model Hamiltonian for the Quantum Critical Detector \label{sec:Hamiltonian}}
\begin{figure}
\includegraphics[width=8cm]{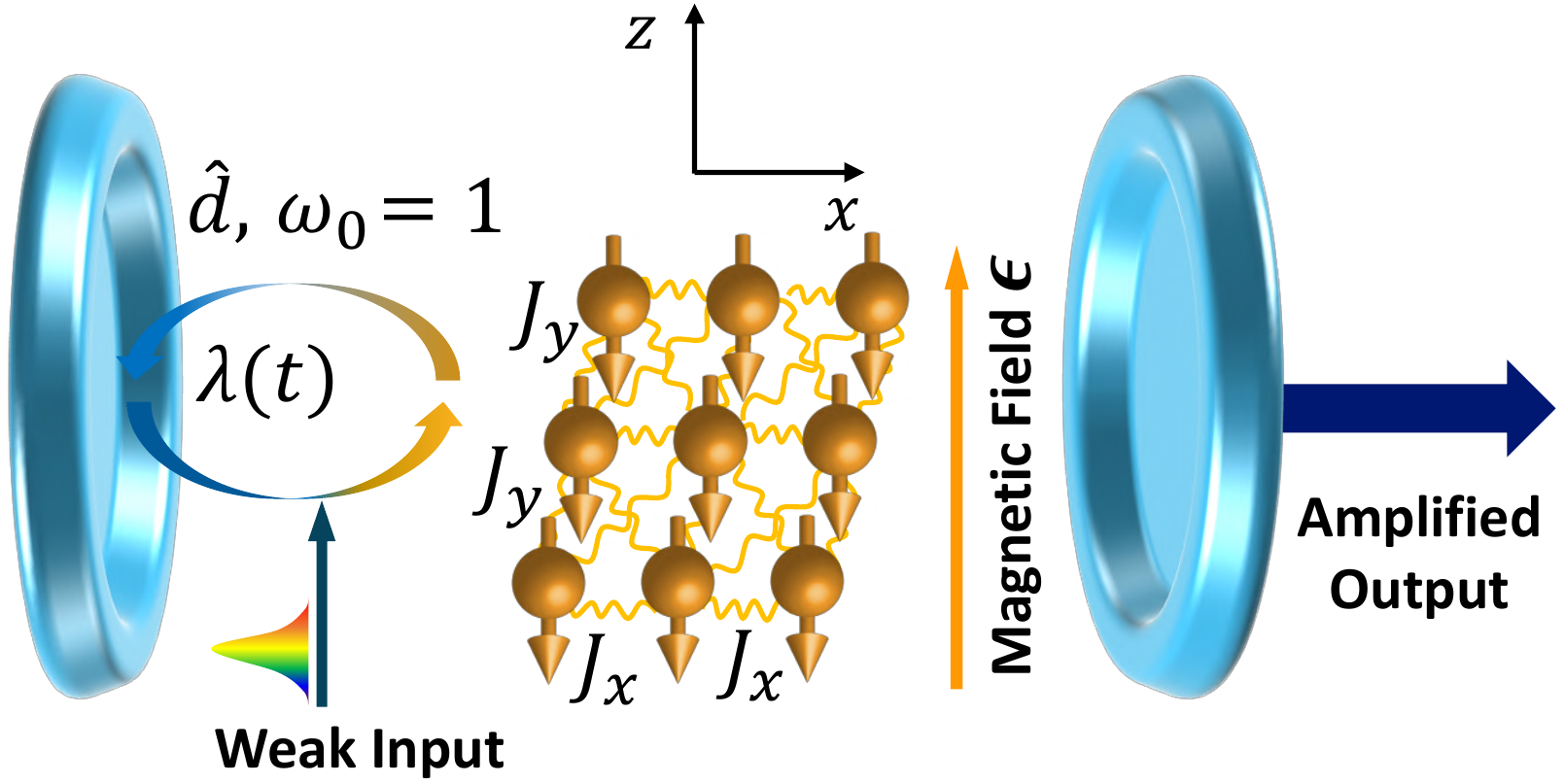}
\caption{\label{fig:schematic} Schematic of our proposed quantum critical detector (QCD). The bosonic mode (resonant cavity $\hat{d}$-mode) with frequency $\omega_{0}=1$ is the output mode. The spins are immersed in a homogeneous magnetic field along $z$-axis inducing an energy splitting $\epsilon$. The spin-boson coupling $\lambda$ is in $x$-direction and the all-to-all spin-spin coupling $J$ is in the $xy$-plane. The input weak signal leads to a small time-dependent variation in spin-boson coupling $\lambda(t)$ (or the spin-spin coupling) and triggers a first-order quantum phase transition if the system is optimally biased around the phase transition point. The energy pre-stored in the spins  is transferred to the bosonic mode and realizes the amplification in our QCD. }
\end{figure}

In our previous work~\cite{yang2018QCD}, we introduced a first-order QPT model, which is the underlying reason for amplification in the QCD. This model is composed of a bosonic mode and a spin ensemble with homogeneous long-range dipole-dipole interaction along only one direction ($y$-direction). Here, we extend this model to a more general case with three-dimensional homogeneous couplings (see Fig.~\ref{fig:schematic}),
\begin{align}
\!\!\!\!\!H & \!=\!\hat{d}^{\dagger}\hat{d}\!+\!\!\frac{\lambda}{\sqrt{N}}\!\!\sum_{j=1}^{N}\!\hat{\sigma}_{j}^{x}(\hat{d}\!+\!\hat{d}^{\dagger})\!+\!\frac{\epsilon}{2}\!\sum_{j=1}^{N}\!\hat{\sigma}_{j}^{z}\!-\!\!\sum_{j<k}^{N}\!\sum_{\alpha}\!\frac{\tilde{J}_{\alpha}}{N}\hat{\sigma}_{j}^{\alpha}\hat{\sigma}_{k}^{\alpha}\!.\!\!\!\label{eq:H_full}
\end{align}
Here, $\hat{d}(\hat{d}^{\dagger})$ denotes the output bosonic mode. Its frequency has been taken as the unit of the energy
$\omega_{0}=1$ and all the other parameters in the Hamiltonian have
been re-scaled by $\omega_{0}$. The operators $\hat{\sigma}_{j}^{\alpha}\ (\alpha=x,y,z)$
are the Pauli matrices of the $j$th spin. A magnetic field is applied
along the $z$-direction inducing an energy splitting $\epsilon$ between
spin states $\left|\uparrow\right\rangle _{j}$ and $\left|\downarrow\right\rangle _{j}$
. The coupling between the bosonic mode and the spin ensemble is along the
$x$-direction with homogeneous coupling strength $\lambda$. The last
term characterizes the all-to-all homogeneous dipolar coupling $\tilde{J}_{\alpha}$
between the spins, which is significantly different from the nearest-neighbor coupling in the traditional Ising models~\cite{sachdev2007quantum}. 

After defining the collective angular momentum
operators of the $N$ spins 
\begin{equation}
\hat{S}_{\alpha}=\frac{1}{2}\sum_{j=1}^{N}\hat{\sigma}_{j}^{\alpha},\ \alpha=x,y,z    
\end{equation}
we re-express our model Hamiltonian as,
\begin{equation}
\hat{H}=\hat{d}^{\dagger}\hat{d}\!+\!\frac{2\lambda}{\sqrt{N}}\hat{S}_{x}(\hat{d}\!+\!\hat{d}^{\dagger})\!+\!\epsilon\hat{S}_{z}\!-\frac{2}{N}(J_{x}\hat{S}_{x}^{2}+J_{y}\hat{S}_{y}^{2}).\!\label{eq:H_full1}
\end{equation}
Here, we have used the angular momentum conservation relation $\hat{S}_{x}^{2}+\hat{S}_{y}^{2}+\hat{S}_{z}^{2}=N(2N+1)/4$
and defined the new homogeneous dipolar coupling strength $J_{x}=\tilde{J}_{x}-\tilde{J_{z}}$
and $J_{y}=\tilde{J}_{y}-\tilde{J_{z}}$. Since the total angular momentum of the spins is conserved, we can perform the calculation in a subspace spanned by the Dicke states~\cite{dicke1954coherence}. Then, the $N$-spin ensemble is now equivalent to a single particle with spin-$N/2$. For simplicity, we only consider the case of ferromagnetic coupling and all the parameters in the Hamiltonian (\ref{eq:H_full1}) are positive real numbers. The rich physics of the antiferromagnetic coupling model (with negative dipolar coupling $J_x$ and/or $J_y$) will not be present in this paper. The first three terms compose the traditional Dicke model without the rotating-wave approximation~\cite{dicke1954coherence}
and the last three terms form the well known LMG model~\cite{lipkin1965validity,meshkov1965validity,glick1965validity}. Hence, we call this model the Dicke-LMG model. 

Our detector model (\ref{eq:H_full1}) is not a simple combination of the Dicke model and the LMG model. New quantum phases and new QPTs, especially a first-order QPT, emerge in this model. For second-order QPTs, substantial changes occur only in the ground-state wave function, which makes it extremely hard to detect the criticality~\cite{zhang2008detection,zhang2009direct}. Specifically, single-shot deterministic readout is required  for pulse signal detection instead of multiple repetitive probabilistic measurements. But, for first-order QPTs, discontinuous changes exists in directly measurable quantities at the phase transition point, which makes the quantum singularity significantly more detectable. Thus, first-order QPTs in a many-body system are of fundamental interest and possess much more application potential in quantum metrology and quantum detection. 

In the traditional Dicke model~\cite{dicke1954coherence}, a second-order superradiant QPT occurs when the spin-boson coupling exceeds the QPT point $\lambda_{c,{\rm II}}\equiv\sqrt{\epsilon}/2$~\cite{hepp1973superradiant,wang1973phase,hioe1973phase,hepp1973equilibrium,Strack2011Dicke}.
However, this superradiant QPT will be forbidden by the Thomas-Reiche-Kuhn (TRK) sum rule  in a natural-atom system~\cite{rzazewski1975phase,rzazewski1979nogo} or an artificial qubit system~\cite{nataf2010nogo}. However, this problem is easily overcome in our model with the help of
the dipolar interaction $J_{x}$ along $x$-axis. 

In the LMG model, second-order magnetic QPTs between the paramagnetic phase and ferromagnetic phase have been
found~\cite{lipkin1965validity,meshkov1965validity,glick1965validity}.
In 2004, Vidal \textit{et. al.} show the existence of a first-order QPT in LMG model in the zero-magnetic-field limit $\epsilon\rightarrow0$~\cite{Vidal2004Entangle1}. However, in this paper, we will show that the first-order QPT exists even for the finite magnetic field ($\epsilon\neq 0$) case. This first-order QPT results from the competition between two ferromagnetic phases and can be exploited for quantum amplification. In the previous literature, this first-order QPT has not been revealed because the traditional magnetic OP (the mean magnetization of the spins) being selected. In the following, we will show this magnetic OP cannot characterize the first-order QPT and we will introduce two new magnetic OPs to characterize the quantum phases in our model and to demonstrate the corresponding QPTs.

\section{Phase Diagram of Detector via Mean-Field Theory\label{Sec:phase_diagram}}
The key element of our QCD is the first-order-QPT-based quantum amplification. To utilize a QPT system as a novel amplification source, one first needs a clear phase diagram and the explicit boundaries between the quantum phases of a system. In this section, we introduce two magnetic OPs to characterize the first-order QPT of the interacting spin system. Then, we present the complete phase diagram of the Dicke-LMG model~(\ref{eq:H_full1}), which is obtained by employing a mean-field theory.

\subsection{Order Parameters}
In different phases, matter usually has different long-range orders. Thus, an OP is utilized to measure the degree of order in a phase transition system, such as the density change in solid-liquid-gas transitions. The OPs normally range between zero in one phase and nonzero in the other. 

To characterize the first-order magnetic QPTs, we now introduce two new magnetic OPs:
\begin{equation}
\zeta_{M,x}\equiv\frac{\langle\hat{S}_{x}^{2}\rangle_{0}}{N^{2}},
\end{equation}
and 
\begin{equation}
\zeta_{M,y}\equiv\frac{\langle\hat{S}_{y}^{2}\rangle_{0}}{N^{2}}.
\end{equation}
In the ground state of the paramagnetic phase, all the spins are
polarized along the negative $z$-axis, i.e., $|\downarrow\downarrow\downarrow\dots\downarrow\rangle_{z}$.
In this state, the mean values of these magnetic OPs are zero. When the system goes to the ferromagnetic phase, the magnetic OPs increase from zero to a finite value due to the macroscopic polarization of spins in the $xy$-plane. 

These two magnetic OPs actually characterize the spin fluctuations in the $xy$-plane:
\begin{equation}
\zeta_{M,x}=\frac{1}{N^2}(\Delta S_x)^2\equiv \frac{1}{N^2}(\langle\hat{S}_{x}^2\rangle_{0}-\langle\hat{S}_{x}\rangle_{0}^2),  
\end{equation} 
and
\begin{equation}
\zeta_{M,y}=\frac{1}{N^2}(\Delta S_y)^2\equiv\frac{1}{N^2}(\langle\hat{S}_{y}^2\rangle_{0}-\langle\hat{S}_{y}\rangle_{0}^2),  
\end{equation} 
Here, we have used the fact that the mean magnetizations in the $xy$ plane $M_x\equiv\langle\hat{S}_{x}\rangle_{0}/N$ and $M_y\equiv\langle\hat{S}_{y}\rangle_{0}/N$ are zero in the ferromagnetic phases. Because the two degenerate ground states~$|\rightarrow\rightarrow\rightarrow\cdots\rangle$
and $|\leftarrow\leftarrow\leftarrow\cdots\rangle$ have opposite polarizations~\cite{sachdev2007quantum,Dziarmaga2005Dynamics,Zurek2005Dynamics}. The change in the transverse magnetic fluctuations during a magnetic QPT can be probed experimentally through spin noise spectroscopy~\cite{zapasskii2013spin}. 

The superradiant phase can be well characterized by the superradiant
OP, which is defined as the rescaled excitation
number in the bosonic mode~\cite{hepp1973superradiant,wang1973phase}
\begin{equation}
\zeta_{S}=\frac{\langle\hat{d}^{\dagger}\hat{d}\rangle_{0}}{N}.\label{eq:sup-rad_order_para}
\end{equation}

According to the Ginzburg–Landau theory, we should choose the spontaneous magnetization in $xy$-plane (i.e., $\sqrt{\zeta_{M,x}}$ and $\sqrt{\zeta_{M,y}}$) as the OPs for magnetic phase transition~\cite{domb1996critical,pfeuty1970one} and the amplitude of the bosonic mode ($\sqrt{\zeta_S}$) as the OP for the superradiant phase transition. However, we do not use these traditional OPs for the following three reasons: (1) we find that the high-order correlation $\langle \hat{S}_x^2\hat{S}_y^2\rangle_0$ diverges at the first-order magnetic phase transition point, but $\langle \hat{S}_x\hat{S}_y\rangle_0$ shows no singularity~\cite{yang2019Single}; (2) $\zeta_{M,x}$ and $\zeta_{M,y}$
are equivalent to $\langle \hat{\sigma}_{i}^x\hat{\sigma}_{i+1}^x\rangle_0$, which is usually selected as the OP of the Ising spin chain~\cite{sachdev2007quantum}; (3) usually the particle number of the bosonic mode is measured in experiment instead of its amplitude.

Next, we calculate the mean values of these OPs in each quantum phase via a mean-field theory and then we show the full phase diagram of the Dicke-LMG model. 

\begin{table*}
\centering
\renewcommand{\arraystretch}{1.5}
\begin{tabular}{| >{\centering\arraybackslash}m{0.4in} | >{\centering\arraybackslash}m{1.6in} | >{\centering\arraybackslash}m{1.7in} | >{\centering\arraybackslash}m{1in} |
>{\centering\arraybackslash}m{0.9in} |
>{\centering\arraybackslash}m{1in} |}

\hline 
 & Stability Conditions & Ground States $|\sqrt{N}\alpha\rangle \otimes\left|\theta,\phi\right\rangle $ & $\zeta_{M,x}=\langle\hat{S}_{x}^{2}\rangle/N^{2}$ & $\zeta_{M,y}=\langle\hat{S}_{y}^{2}\rangle/N^{2}$ & $\zeta_{S}=\langle\hat{d}^{\dagger}\hat{d}\rangle/N$\\
\hline 
PN Phase & $\epsilon>4\lambda^{2}+2J_{x},\ \epsilon>2J_{y}$ & $\left|0\right\rangle \left|0,\phi_{0}\right\rangle $ & $0$ & $0$ & $0$\\
\hline 
FN Phase & $J_{y}>\epsilon/2,\ J_{y}>2\lambda^{2}+J_{x}$ & $\left|0\right\rangle \left|\theta_{0},\phi_{0}\right\rangle ,\ \phi_{0}=\frac{\pi}{2},\frac{3\pi}{2}$  & $0$ & $\frac{1}{4}(1\!-\!\frac{\epsilon^{2}}{4J_{y}^{2}})$ & $0$\\
\hline
FS Phase & $4\lambda^{2}+2J_{x}>\epsilon,\ 2\lambda^{2}+J_{x}>J_{y}$ & $|\!-\!\sqrt{N}\alpha_{0}e^{i\phi_{0}}\!\rangle\!\left|\theta_{0},\phi_{0}\right\rangle \!,\ \phi_{0}\!=\!0,\!\pi$  & $\frac{1}{4}[1\!-\!\frac{\epsilon^{2}}{(4\lambda^{2}+2J_{x})^{2}}\!]$ & $0$ & $\lambda^{2}[1\!-\!\frac{\epsilon^{2}}{(4\lambda^{2}+2J_{x})^{2}}]$\\
\hline 
\end{tabular}
\caption{\label{tab:OrderParameters}A summary of stability conditions, the
ground states, and the mean value of the order parameters in different
quantum phases. State $|\sqrt{N}\alpha\rangle \otimes\left|\theta,\phi\right\rangle $
denotes the tensor product of the bosonic coherent state and a coherent
spin state. Ferromagnetic phases have two degenerate states with
different azimuth angle $\phi_{0}$ and the ground state in this phase can be any superposition of these two degenerate ground states.}
\end{table*}

\subsection{Mean-Field Theory}
A many-body system with interactions is generally very difficult to solve exactly. The mean-field theory was invented to address this problem. In this method, the Hamiltonian is expanded in terms of the magnitude of fluctuations around the mean of the operators. The mean-field theory is the zeroth-order expansion of the Hamiltonian. The minimum of the mean-field value of the Hamiltonian gives the ground-state energy of the system. The first-order term will be of scale $1/N$, which can be neglected in the large-$N$ limit for the ground-state problem. Recently, Zhang et. al. presented an elegant mean-field theory to calculate the quantum phases in a many-body system~\cite{zhang2014quantum}. In this subsection, this method is exploited for the Dicke-LMG model in (\ref{eq:H_full1}) to obtain the ground state of the whole system and the phase boundary between different quantum phases. We will also verify this mean-field phase diagram with exact numerical simulation in the next section.

The spin-boson coupling in our Hamiltonian (\ref{eq:H_full1}) functions as a displacement of the bosonic mode. Therefore, in a mean-field ground state, the operator $\hat{d}$ can be replaced with a complex number
$\sqrt{N}\alpha$ by assuming the bosonic mode is a coherent state $|\sqrt{N}\alpha\rangle $~\cite{Emary2003chaos}. The ground-state of the LMG model in the thermodynamic limit $N\rightarrow\infty$ is proven to be a coherent spin state~\cite{castanos2006classical,Ribeiro2008Exact}. Thus, it is reasonable to replace the spin operators, $(\hat{\sigma}_{j}^{x},\hat{\sigma}_{j}^{y},\hat{\sigma}_{j}^{z})$, with a classical Bloch vector, $\vec{n}=(\sin\theta\cos\phi,\sin\theta\sin\phi,\cos\theta)$~\cite{botet1983large,dusuel2005continuous}.

To be consistent with the mean-field theory, we modified the definition
of the coherent spin state~\cite{radcliffe1971some,arecchi1972atomic} as,
\begin{equation}
\left|\theta,\phi\right\rangle =e^{i\theta(\hat{S}_{x}\sin\phi-\hat{S}_{y}\cos\phi)}\left|N/2,N/2\right\rangle .
\end{equation}
Thus, the angle $\theta$ is identical to the polar angle of a
spherical coordinate. Here, $\left|N/2,N/2\right\rangle $ is the
Dicke state with all spins in the up state $\left|\uparrow\right\rangle _{z}$.
In the following, we denote the ground state of the whole system as $|\sqrt{N}\alpha\rangle \otimes\left|\theta,\phi\right\rangle ,$
i.e., the direct product of a bosonic coherent state $|\sqrt{N}\alpha\rangle $
and a coherent spin state $\left|\theta,\phi\right\rangle$.

The scaled ground-state energy $E(\alpha,\theta,\phi)=\langle H\rangle_0/N$
is given by\begin{widetext}
\begin{equation}
E(\alpha,\alpha^*,\theta,\phi)=|\alpha|^{2}+\frac{\epsilon}{2}\cos\theta+\lambda(\alpha+\alpha^{*})\sin\theta\cos\phi-\frac{J_{x}}{2}\sin^{2}\theta\cos^{2}\phi-\frac{J_{y}}{2}\sin^{2}\theta\sin^{2}\phi,
\end{equation}
\end{widetext}
where a constant has been neglected. To minimize the ground-state energy, we need to set
the first derivatives of $E$ with respect to $\alpha$, $\alpha^*$, $\theta$, and $\phi$ to zero.
Since all the parameters in Hamiltonian (\ref{eq:H_full1}) are assumed
to be positive, it is easy to verify that, in the ground state, the amplitude
of the bosonic mode should be real, i.e., ${\rm Im}\alpha=0$. Then,
minimization conditions give the following equilibrium constraints,\begin{widetext}
\begin{align}
\frac{\partial}{\partial\alpha}E & =2\alpha+2\lambda\sin\theta\cos\phi=0,\label{eq:Derivative1}\\
\frac{\partial}{\partial\theta}E & =-\frac{\epsilon}{2}\sin\theta+2\lambda\alpha\cos\theta\cos\phi-(J_{x}\cos^{2}\phi+J_{y}\sin^{2}\phi)\sin\theta\cos\theta=0,\label{eq:Derivative2}\\
\frac{\partial}{\partial\phi}E & =-2\lambda\alpha\sin\theta\sin\phi+(J_{x}-J_{y})\sin^{2}\theta\sin\phi\cos\phi=0,\label{eq:Derivative3}
\end{align}
\end{widetext}where $\theta\in[0,\pi]$ and $\phi\in[0,2\pi)$. To guarantee that the minimums of $E(\alpha,\theta,\phi)$ are obtained, one needs to calculate the second derivatives.
The ground-state stability is determined by $3\times3$ Hessian matrix
\begin{align}
\mathcal{M} & =\left[\begin{array}{ccc}
\frac{\partial^{2}E}{\partial\alpha^{2}} & \frac{\partial^{2}E}{\partial\alpha\partial\theta} & \frac{\partial^{2}E}{\partial\alpha\partial\phi}\\
\frac{\partial^{2}E}{\partial\theta\partial\alpha} & \frac{\partial^{2}E}{\partial\theta^{2}} & \frac{\partial^{2}E}{\partial\theta\partial\phi}\\
\frac{\partial^{2}E}{\partial\phi\partial\alpha} & \frac{\partial^{2}E}{\partial\phi\partial\theta} & \frac{\partial^{2}E}{\partial\phi^{2}}
\end{array}\right].
\end{align}
The ground states are stable only if $\mathcal{M}$ is positive definite,
i.e., all eigenvalues of $\mathcal{M}$ are non-negative. 

We now construct the phase diagram with the help of the three OPs we defined,
\begin{align}
\zeta_{S}=\frac{\langle\hat{a}^{\dagger}\hat{a}\rangle_{0}}{N} & =\lambda^{2}\sin^{2}\theta\cos^{2}\phi,\\
\zeta_{M,x}=\frac{\langle\hat{S}_{x}^{2}\rangle_{0}}{N^{2}} & =\frac{1}{4}\sin^{2}\theta\cos^{2}\phi,\\
\zeta_{M,y}=\frac{\langle\hat{S}_{y}^{2}\rangle_{0}}{N^{2}} & =\frac{1}{4}\sin^{2}\theta\sin^{2}\phi.
\end{align}
The magnetization in $z$-direction can also be calculated easily
\begin{equation}
M_{z}=\frac{\langle\hat{S}_{z}\rangle_{0}}{N}=\frac{1}{2}\cos\theta.
\end{equation}
According to the values of the OPs, we find three quantum phases in the Dicke-LMG model: paramagnetic-normal (PN) phase, ferromagnetic-normal (FN) phase, and ferromagnetic-superradiant (FS) phase. In Table~\ref{tab:OrderParameters}, we list the stability conditions (phase boundaries), ground states, and the mean value of the three OPs on the ground state of each phase. More details can be found in Appendix~\ref{sec:mean-field}. We emphasize that the first-order magnetic QPT can be properly characterized only by the two magnetic OPs we introduced.

\begin{figure}
\includegraphics[width=8.5cm]{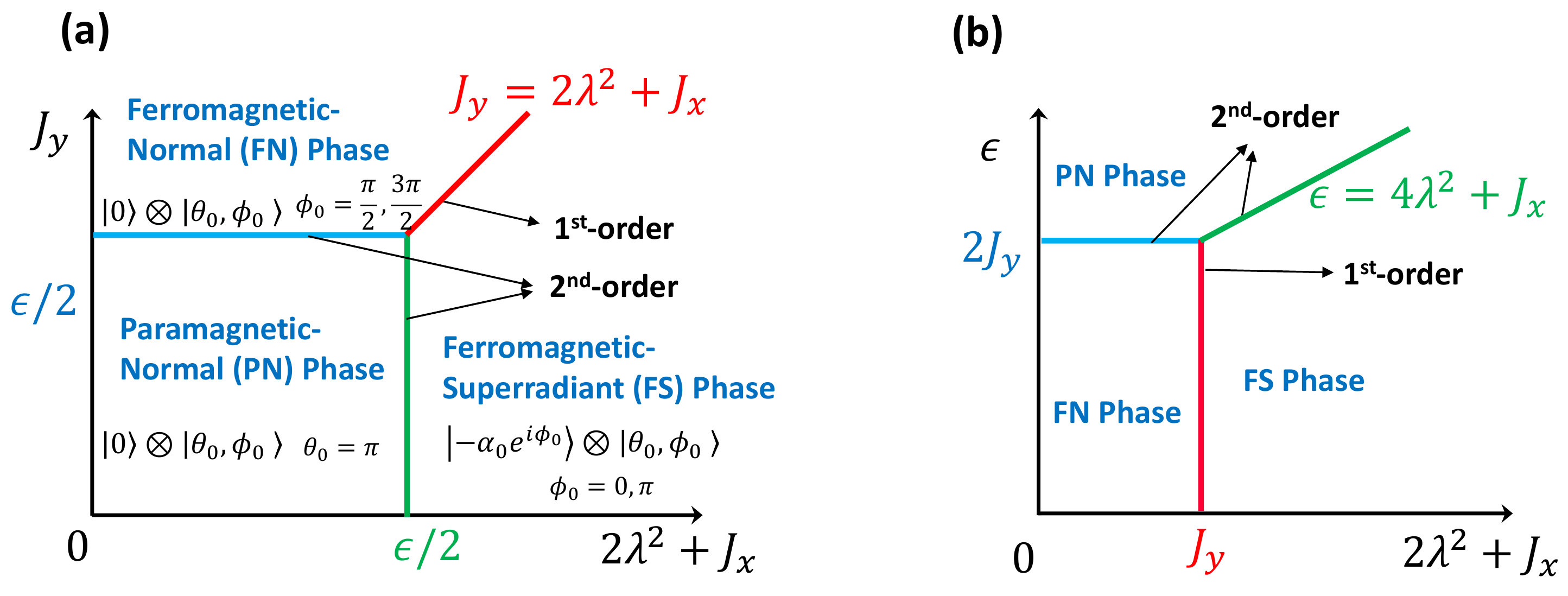}
\caption{\label{fig:phase_diagram} Phase diagram of the Dicke-LMG model and the ground state in each phase. In panel (a), we plot the phase diagram in the $(2\lambda^2+J_x,J_{y})$ plane. A second-order
phase transition from the PN phase to the FN phase occurs when increasing the spin-spin coupling $J_{y}$ to cross the blue line $J_{y}=\epsilon$/2. A second-order phase transition from the PN phase to the FS phase happens when increasing the spin-boson coupling $\lambda$ or spin-spin coupling $J_{x}$ to cross the green line $4\lambda^{2}+J_{x}=\epsilon$. In the strong-coupling regime with $J_{y}>\epsilon/2$ and $2\lambda^{2}+J_{x}>\epsilon/2$,
the QPT between the FN and the FS phases is of first-order, which of significant interest for quantum amplification. In panel (b) we
show the same phase diagram in the $(2\lambda^2+J_x,\epsilon)$ plane. This figure shows no first-order QPT happens if we only vary the magnetic field strength $\epsilon$.}
\end{figure}

To clearly reveal the relationship between these quantum phases, we display the phase diagram of the Dicke-LMG model in Fig.~\ref{fig:phase_diagram}. In panel (a), we plot the phase diagram in the $(2\lambda^{2}+J_{x},\ J_{y})$
plane and also show the ground state of each phase. The boundaries between the three quantum phases are depicted by the blue, green, and red lines.
There exists a unique triple-point $(2\lambda^{2}+J_{x}=\epsilon/2,J_{y}=\epsilon/2)$,
where the three lines intersect. A second
order QPT from the PN phase to the FN phase occurs when increasing the spin-spin coupling $J_{y}$ across the blue line $J_{y}=\epsilon/2$. The magnetic OP $\zeta_{M,y}$ changes from $0$ to a finite value continuously after the QPT. A second order QPT from the PN phase to the FS phase happens when increasing the spin-boson coupling $\lambda$ (or spin-spin coupling $J_{x}$) across the green line $2\lambda^{2}+J_{x}=\epsilon/2$. Both the superradiant OPs ($\zeta_{S}$) and the magnetic OP ($\zeta_{M,x}$) change from $0$ to a finite value continuously during this QPT. Particularly, the QPT from the FN phase to the FS phase, crossing the red line $2\lambda^2+J_x=J_y$, is of a first-order transition. This occurs in the strong-coupling region when both the $x$-direction coupling $2\lambda^{2}+J_{x}>\epsilon/2$ and the $y$-direction coupling $J_{y}>\epsilon/2$ exceed the second-order QPT points. The OP $\zeta_{M,y}$ suddenly drops from a finite value to zero and, at the same time, both $\zeta_{S}$ and $\zeta_{M,x}$ discontinuously jump from zero to finite values. This first-order QPT plays a crucial role in quantum amplification of our QCD.

We emphasize that the first-order QPT in our detector model fundamentally results from the competition between two ferromagnetic phases. To clarify this, we draw the same phase diagram in the $(\ 2\lambda^{2}+J_{x},\epsilon)$-plane in panel (b). As the $\epsilon$-axis is parallel to the first-order QPT boundary (the red line), the first-order QPT will not happen if we only vary the magnetic field strength $\epsilon$. We can also verify that the first-order phase transition represented in the previous Dicke-Ising model~\cite{lee2004first} has the same origin as the one we addressed here. \textbf{We predict that there is a universal first-order QPT in interacting spin systems from the nearest-neighbor short-range coupling to the all-to-all long-range coupling,}
\begin{equation}
\hat{H}=\frac{1}{2}\epsilon\sum_j\hat{\sigma}_j^{z}-\frac{1}{n}\sum_{\langle i<j \rangle}(J_{x}\hat{\sigma}_i^x\hat{\sigma}_j^x + J_{y}\hat{\sigma}_i^y\hat{\sigma}_j^y),  \end{equation}
where $\langle i<j\rangle$ run for the $n$ nearest neighbors and $n=1$ for Ising XY model and $n=(N-1)\approx N$ for the LMG model. The first-order QPT in this interacting spins model can be verified with the mean-field theory addressed above.

In the next section, we numerically demonstrate all the QPTs existing in our detector model and display the fundamental changes in the ground-state wave function during the QPTs with Husimi $Q$-functions.

\section{Quantum Phase Transitions and Ground-State Properties of the Detector \label{sec:QPT_QF}}
To reveal the microscopic mechanism of amplification in the first-order-QPT-based QCD, we first need to  understand the underlying physics of the QPTs in the model and especially the  fundamental change in the ground state wave function of the system during the QPTs. In the Dicke-LMG model, QPTs occurs at the boundaries of the phase diagram given in Fig.~\ref{fig:phase_diagram}. To deeply understand all the quantum phases and classify the order of the QPTs correctly, we split the full model in Eq.~(\ref{eq:H_full1}) into three sub-models: LMG model, Dicke-LMGx model, and Dicke-LMGy model. In the following subsections, we will study these sub-models separately. We follow a simple recipe, first present the numerical demonstration of the phase diagram, then show the details of the QPTs, and finally display the quasi-probability distribution of the ground states. 

\subsection{LMG Model}

In this subsection, we first consider the simplest sub-model---the LMG model with vanishing spin-boson coupling ($\lambda=0$)
in Hamiltonian (\ref{eq:H_full1}),
\begin{equation}
\hat{H}=\epsilon\hat{S}_{z}-\frac{2}{N}(J_{x}\hat{S}_{x}^{2}+J_{y}\hat{S}_{y}^{2}).\label{eq:H_LMG}
\end{equation}
The LMG model was first proposed
to describe the shape phase transition in nuclei~\cite{lipkin1965validity,meshkov1965validity,glick1965validity}. As an exactly solvable QPT model~\cite{pan1999analytical,links2003algebraic,ortiz2005exactly,Ribeiro2007thermo},
LMG model has attracted significant attentions in  ground-state entanglement~\cite{Vidal2004entangle,Vidal2004Entangle1}, spin
squeezing~\cite{ma2009fisher,ma2011quantum}, criticality detection~\cite{quan2007quantum,Kwok2008quantum}, heat-engine efficiency enhancement~\cite{ma2017quantum},
dynamics of the QPT~\cite{Morrison2008Dynamical},
ect. However, the first-order QPT in LMG model has not been fully
revealed due to the improperly selected magnetic OP as the net magnetization~$M_{z}$~\cite{Vidal2004entangle,Vidal2004Entangle1}.
Here, with the help of the magnetic OPs $\zeta_{M,x}$ and $\zeta_{M,y}$ we introduced in the previous section, we numerically show the first-order QPT as well as the well-understood
second-order ones. Utilizing the Husimi spin $Q$-function, we also
verify that the ground-state of the LMG model is indeed a coherent spin state. 

\begin{figure}
\includegraphics[width=8cm]{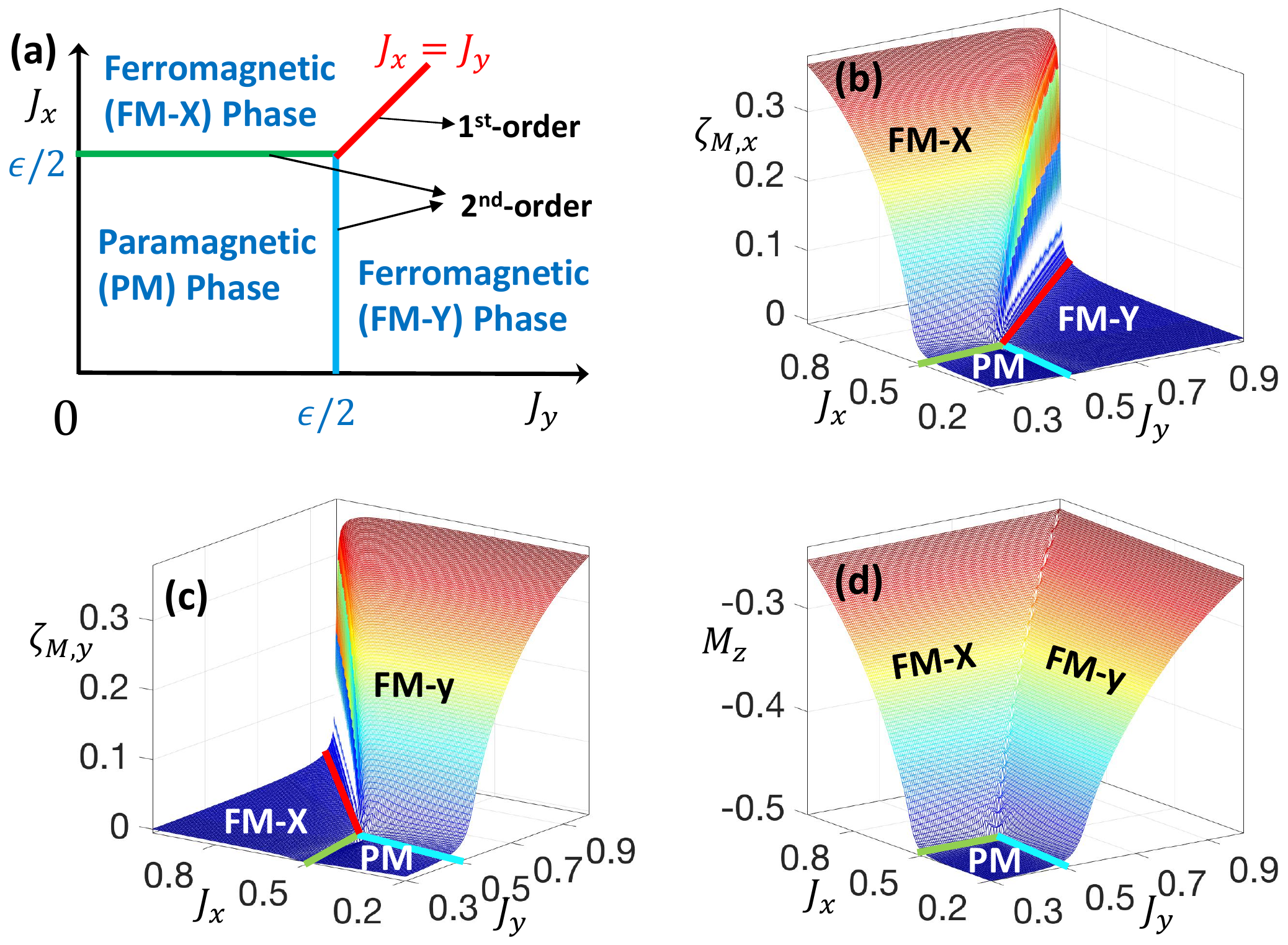}\caption{\label{fig:PD_LMG} The phase diagram of
the LMG model is shown in panel (a). The numerical demonstration of the order parameter $\zeta_{M,x}=\langle\hat{S}_{x}^{2}\rangle_{0}/N^{2}$ is
in panel (b) and the magnetic order parameter $\zeta_{M,x}=\langle\hat{S}_{x}^{2}\rangle_{0}/N^{2}$
is in panel (c). In panel (d), we show the traditional magnetic order parameter (the mean magnetization) $M_z$. Here, the energy splitting of the spins is set as $\epsilon=1$
and the spin number in this figure is $N=80$.}
\end{figure}

We now show the complete phase diagram of the LMG model, which has not been revealed in previous literature. The phase diagram of the LMG model extracted from the mean-field theory is displayed in Fig.~\ref{fig:PD_LMG} (a), which can also be obtained by setting $\lambda=0$ in Fig.~\ref{fig:phase_diagram}. The LMG model has one paramagnetic (PM) phase and two ferromagnetic phases (FM-X and FM-Y). The numerical demonstrations of the phase diagrams are shown in Fig.~\ref{fig:PD_LMG} (b) and (c). If both the spin-spin coupling in $x$-direction ($J_x<J_{xc,{\rm II}}$) and  $y$-direction ($J_y<J_{yc,{\rm II}}$) are below their corresponding phase transition coupling strengths ($J_{xc,{\rm II}}\equiv \epsilon/2$ and $J_{yc,{\rm II}}\equiv \epsilon/2$), both the two magnetic OPs ($\zeta_{M,x}$ and $\zeta_{M,y}$) are zero, thus the spins are in the PM phase. The well-known second-order QPT from the PM phase to the FM-X (FM-Y) phase occurs when $J_x$ ($J_y$) crosses the phase boundary denoted by the green (blue) line in panel (a). But in the strong  spin-spin coupling regime with $J_x>J_{xc,{\rm II}}$ and $J_y>J_{yc,{\rm II}}$, the sharp changes in the OPs at the boundary $J_x=J_y$ indicates a first-order QPT, which has not been revealed in previous literature. This first-order QPT can also be verified in experiments by measuring the change in the magnetic fluctuations along $x$ and $y$ directions. We also plot the net magnetization in $x$-direction in panel (d). With the two magnetic OPs we introduced, we can easily verify that the anisotropic transition in the Ising $XY$ chain~\cite{Bunder1999effect} is also a first-order QPT.

\begin{figure}
\includegraphics[width=8cm]{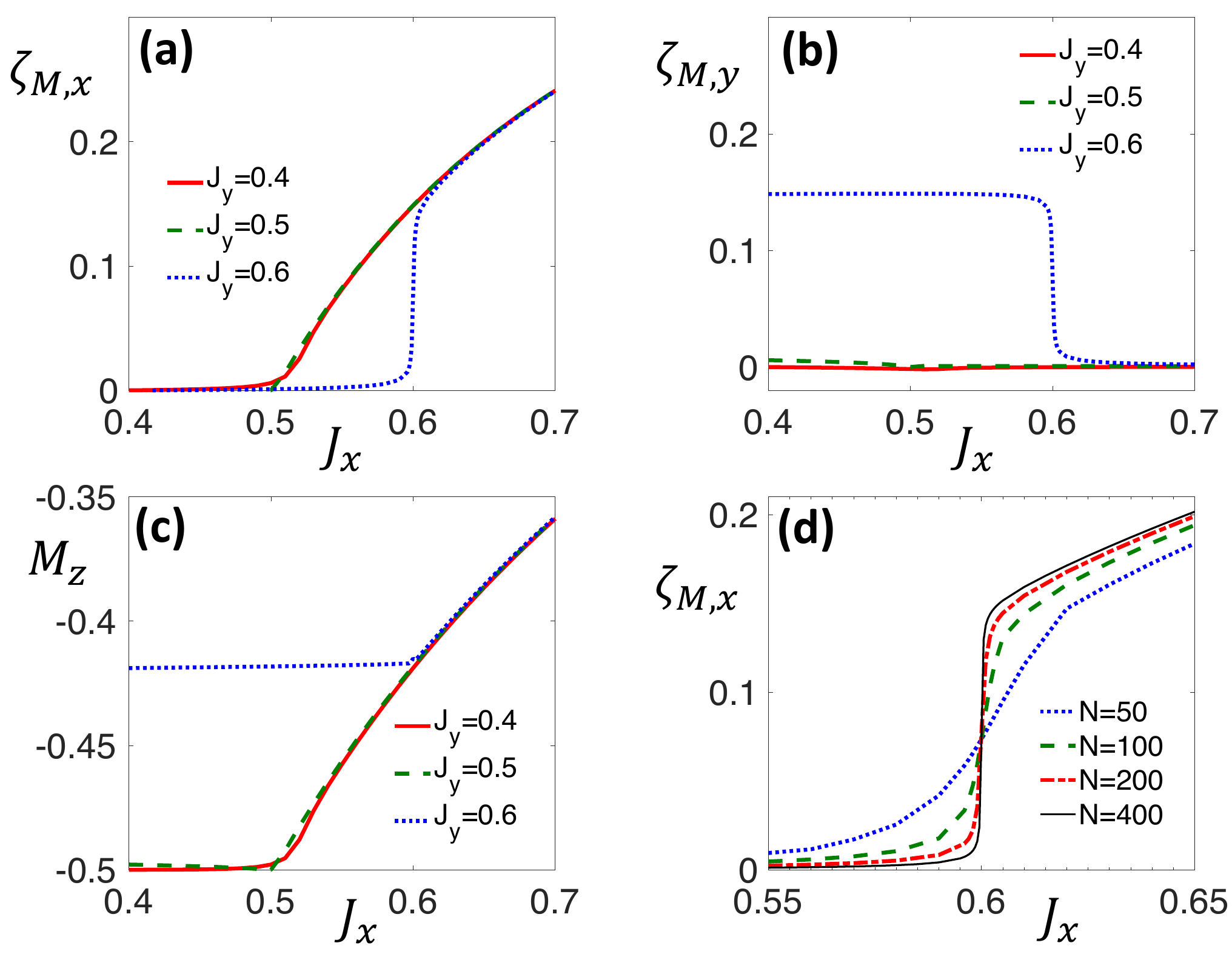}\caption{\label{fig:MQPT_LMG}Numerical demonstration of the second- and first-order
quantum phase transitions. In panel (a), (b), and (c), we show the
MOPs $\zeta_{M,x}$, $\zeta_{M,y}$, and $M_{z}$, respectively, as
functions of the spin-spin coupling $J_{x}$ for different $J_{y}$.
Here, the magnetic field strength is set to be $\epsilon=1$ and the
spin number is $N=200$. When $J_{y}$ is smaller than second order
QPT coupling $J_{yc,{\rm II}}=\epsilon/2=0.5$, the second-order
QPT happens at $J_{xc,{\rm II}}=\epsilon/2=0.5$ (see the red and green lines).
For $J_{y}>J_{c,{\rm II}}$, the first-order QPT occurs at $J_{xc,{\rm I}}=J_{y}$
(see the blue line). In panel (d), we plot $\zeta_{M,x}$ with $J_{y}=0.6>J_{yc,{\rm II}}$
for different spin number $N$. The diverging slope at the phase transition
point indicates the occurrence of a first-order instead of a second-order
QPT.}
\end{figure}

To reveal the details of the QPTs, we plot the magnetic OPs as functions of spin-spin coupling $J_{x}$ for different spin-spin coupling $J_{y}$ in Fig.~\ref{fig:MQPT_LMG}. From the red-solid and green-dashed lines in Fig.~\ref{fig:MQPT_LMG} (a), we can see that second-order QPTs occur at $J_{xc,{\rm II}}\equiv\epsilon/2$ when the spin-spin coupling along the $y$ axis is below the phase transition point $J_{y}\leq J_{yc,{\rm II}}\equiv\epsilon/2$. If we fix $J_x\leq J_{xc,{\rm II}}\equiv\epsilon/2$, but increase $J_y$ across the phase transition point $J_{yc,{\rm II}}$, similar second-order QPTs (data not shown) from the PM phase to the FM-Y phase also occur. In the strong spin-spin coupling case when  $J_{y}$ is larger than the second-order QPT strength $J_y>J_{yc,{\rm II}}$, a QPT occurs at a new phase transition point $J_{xc,{\rm I}}\equiv J_{y}$ [see the blue-dotted line in Fig.~\ref{fig:MQPT_LMG} (a)], which we verify as first-order later. In Fig.~\ref{fig:MQPT_LMG} (b), we show that only after a first-order QPT occurs, there exists a sudden drop in the magnetic OP $\zeta_{M,y}$ at the phase transition point $J_{xc,{\rm I}}$ (blue-dotted line), as the system transitions from the FM-Y phase to the FM-X phase. Thus, this first-order QPT results from the competition between the two ferromagnetic phases. In Fig.~\ref{fig:MQPT_LMG} (c), we show that the first-order QPT cannot be revealed by the traditional magnetic OP $M_{z}$, which changes continuously in both second- and first-order QPTs.

\begin{figure}
\includegraphics[width=8cm]{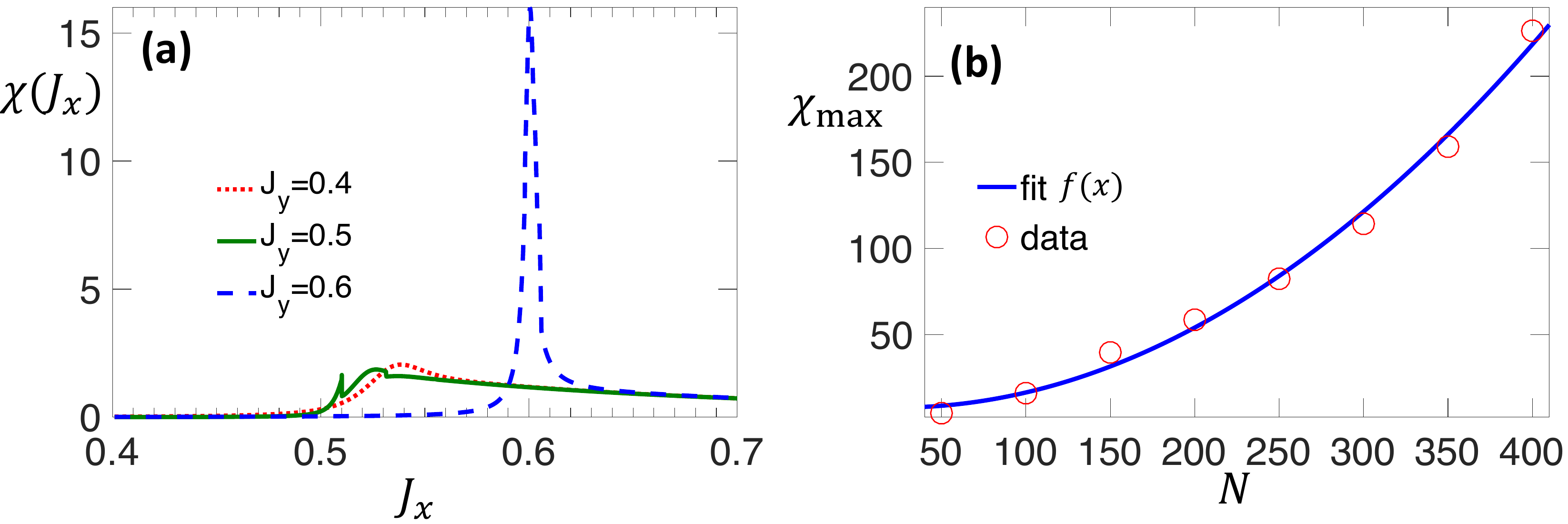}
\caption{\label{fig:chi_LMG}Numerical verification of the first-order QPT from FM-Y phase to FM-X phase. In panel (a), we plot the sensitivity $\chi(J_x)$ for different spin-spin coupling $J_y$. Only if $J_y>J_{yc,{\rm II}}=0.5$, large sensitivity at the phase transition point can be obtained (see the blue-dashed line). Here, the other parameters are $\epsilon =1$, spin number $N=100$, and the location of the phase transition points are marked by the black-dashed lines. In panel (b), we plot the maximum of the sensitivity $\chi_{\rm max}$ for different spin number $N$ with fixed $J_y=0.6$. According to the fit function $f(x)=0.0015x^2-0.069x+8.38$, $\chi_{\rm max}$ diverges with the spin number in $N^2$-scaling.}
\end{figure}

To verify the first-order magnetic QPT, we plot the magnetic OP $\zeta_{M,x}$ for different spin number $N$ but with fixed $J_{y}=0.6>J_{yc,{\rm II}}$ in Fig.~\ref{fig:MQPT_LMG} (d). The slope of $\zeta_{M,x}$ at the phase transition point $J_{xc,{\rm I}}=J_{y}$ increases with spin number $N$. Now, we define the sensitivity function of the LMG model as the first derivative of the magnetic OP $\zeta_{M,x}$, i.e. the sensitivity,
\begin{equation}
\chi(J_x)\equiv\frac{d\zeta_{M,x}}{dJ_x}=\frac{1}{N^2}\frac{d\langle\hat{S_x^2}\rangle_0}{dJ_x}.\label{eq:chi_J}
\end{equation}
From Fig.~\ref{fig:chi_LMG} (a), we see a very sharp peak located at the first-order QPT point  (the blue-dashed line), which is obtained with strong spin-spin coupling $J_y>J_{yc,{\rm II}}$. The peaks of the sensitivity function $\chi (J_x)$ for second-order QPTs (red and blue curves) are much lower than the first-order QPT. The position of the maximum of $\chi$ for second-order QPTs will approach the phase transition point $J_{xc,{\rm II}}$ asymptotically in the thermodynamic limit $N\rightarrow \infty$, but the height converges to a finite value. In Fig.~\ref{fig:chi_LMG} (b), we plot the maximum sensitivity $\chi_{\rm max}$ of the first-order QPT for different spin numbers. We see that $\chi_{\rm max}$ diverges with increasing spin number in $N^2$ scaling. Similar $N^2$ scaling will also be found later for the first-order QPT in the Dicke-LMGy model.

In the last part of this subsection, we shed light on the ground states of the three quantum phases. In 2005, Dusuel and Vidal studied the ground states of the LMG model in detail~\cite{dusuel2005continuous}. Via the Holstein-Primakoff transform, they found that in the thermodynamic limit $N\rightarrow\infty$, the ground state of the LMG model is exactly
a coherent spin state. Here, we numerically verify this result. Particularly, we show the fundamental changes of the ground-state wave functions during magnetic QPTs via the spin Husimi $Q$-function~\cite{Lee1984Qfunction}
\begin{equation}
Q(\theta,\phi)=\frac{2N+1}{4\pi}\left\langle \theta,\phi\right|\hat{\rho}_g\left|\theta,\phi\right\rangle,\label{eq:spinQF}
\end{equation}
where $\hat{\rho}_g$ is the ground-state density matrix of the spins. To vividly show the quasi-probability distribution of the ground state wavefunction, we transfer the spherical coordinates $(r,\theta,\phi)$
to the corresponding Cartesian coordinates $(x,y,z)$. Here, $r=Q(\theta,\phi)/{\rm max}[Q(\theta,\phi)]$
is the normalized $Q$-function, which is extensively used to characterize
the spin squeezing effect~\cite{Kitagawa1993squeezed}. 

\begin{figure}
\includegraphics[width=8cm]{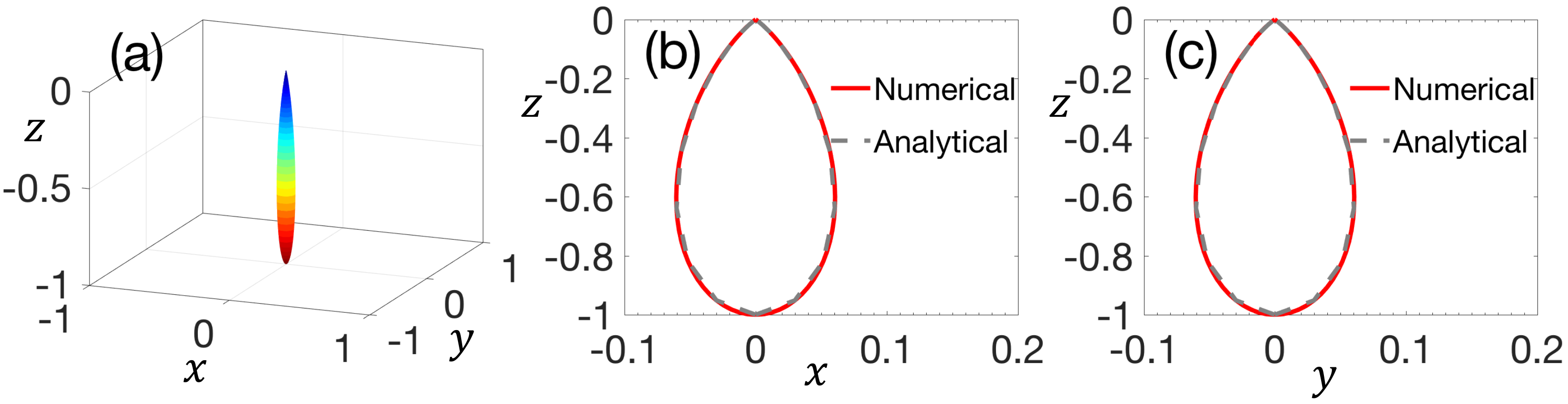}
\caption{\label{fig:QF_PM}(a) Husimi $Q$-function of the ground state $|\theta_{0}=\pi,\phi_{0}\rangle$
(i.e., the Dicke state $|N/2,-N/2\rangle$) of the spin in the paramagnetic
phase. (b) and (c) the cross sections of the $Q$-function in the
$xz$- and $yz$- planes. The red and gray lines are the numerical
and analytic results, respectively. Here, the parameters are taken
as $\epsilon=1$, $J_{x}=0$, $J_{y}=0$ and $N=200$.}
\end{figure}

The ground state of the PM phase is the Dicke state $|N/2,-N/2\rangle$ (i.e., the coherent spin state $|\theta_{0},\phi_{0}\rangle$ with
$\theta_{0}=\pi$ and totally undetermined $\phi_{0}$). In this state, all the spins are polarized along the negative $z$-axis. As shown in Fig.~\ref{fig:QF_PM} (a), the corresponding $Q$-function is a cigar-like structure lying along the negative $z$-axis. The cross sections of the $Q$-function in the $xz$ and $yz$ planes are displayed in Fig.~\ref{fig:QF_PM} (b) and (c), respectively. The red lines originate from the numerical ground states obtained by directly diagonalizing the LMG model and the gray lines are the analytic results of the Dicke state $\left|N/2,-N/2\right\rangle $. The numerical and analytic results exactly coincide with each other. From this quasi-probability distribution function, we can also see that the mean magnetization $M_z$ is a finite negative value, but the two order parameters $\zeta_{M,x}$ and $\zeta_{M,y}$ are negligibly small and approach zero in the limit $N\rightarrow\infty$.

\begin{figure}
\includegraphics[width=8cm]{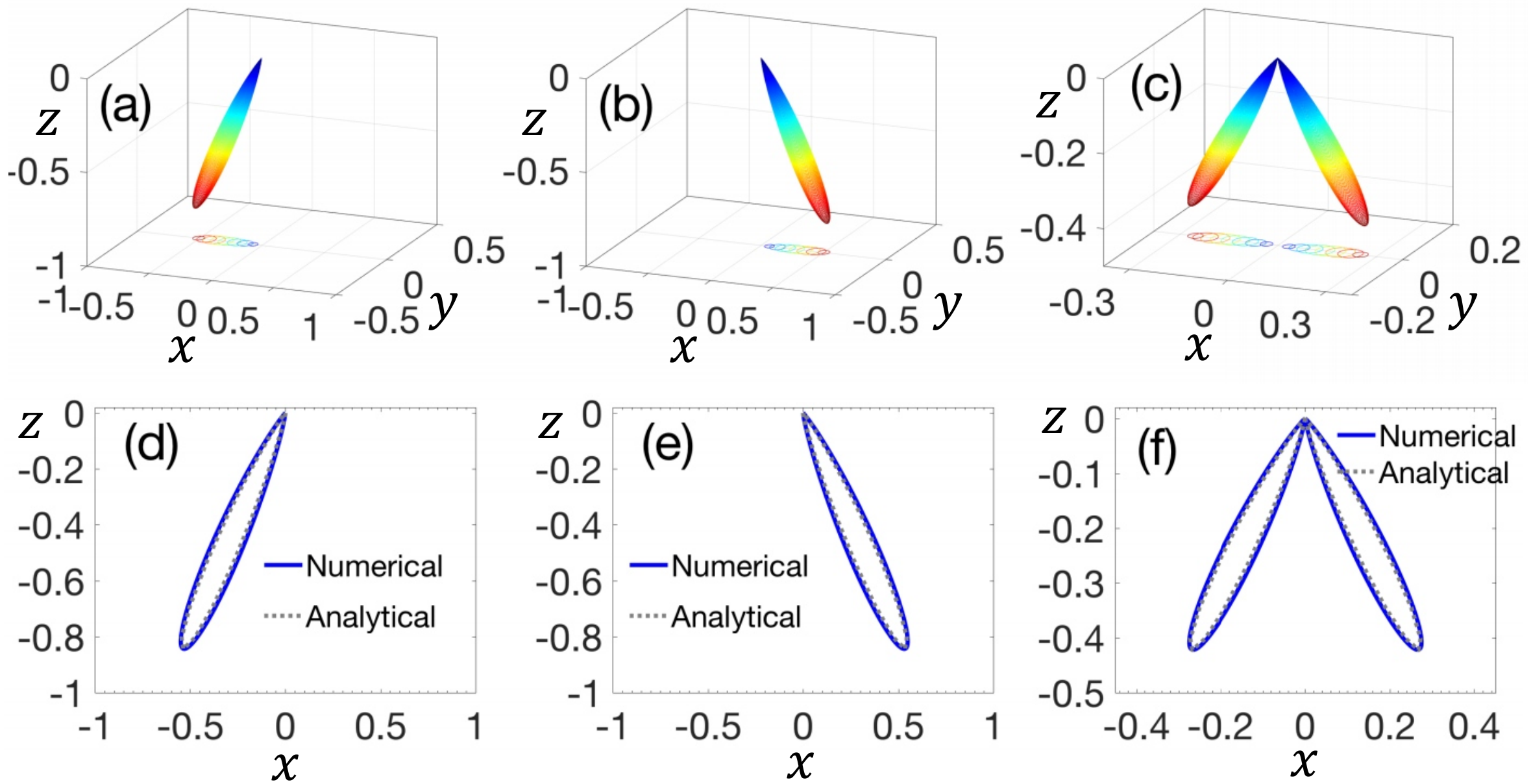}
\caption{\label{fig:QF_FMx} The Husimi $Q$-function of the ground states of the spin in the FM-X phase. Panels (a-c) are the $Q$ functions of the two degenerate states $|\theta_{0},\pi\rangle$, $|\theta_{0},0\rangle,$ and the superposition of these two states
$|\theta_{0},0\rangle+|\theta_{0},\pi\rangle$, respectively. The curves underneath are the contour projections of the corresponding $Q$-functions in $xy$-plane. The bottom row (d-f), displays cross sections of the $Q$-function in the $xz$-planes. The red and gray lines are the numerical and analytic results, respectively. Here, the parameters are taken as $\epsilon=1$, $J_{x}=0.6$, $J_{y}=0$ and $N=200$.}
\end{figure}

For the FM-X phase, there exist two degenerate ground states $|\theta_{0},0\rangle$
and $|\theta_{0},\pi\rangle$~\cite{dusuel2005continuous}, where
the polar angle $\theta_{0}$ is determined by the parameters of
the Hamiltonian as given in Eq.~(\ref{eq:theta_DLMGx}). The system can be in an arbitrary superposition of these two degenerate states. Thus, the ensemble mean of the magnetization along $x$-direction $\langle\hat{S}_{x}\rangle_{0}$ is zero, but the magnetic fluctuations characterized by the magnetic OP $\zeta_{M,x}$ is finite. The numerical simulation of the Husimi $Q$-functions of states $|\theta_{0},\pi\rangle$, $|\theta_{0},0\rangle$, and their symmetric quantum superposition $|G_{+}\rangle=(|\theta_{0},0\rangle+|\theta_{0},\pi\rangle)/\sqrt{2}$ are presented by the three columns in Fig.~\ref{fig:QF_FMx}, respectively. The cigar-like structures lie in the $xz$-plane, as the strong spin-spin coupling $J_x$ along the $x$-direction overwhelms the spin-spin coupling along the $y$-direction in the FM-X phase. From the $Q$-function, we can also see that $M_z$ is a finite negative value. The magnetic OP $\zeta_{M,y}$ is very small and will approach zero when $N\rightarrow\infty$. In the bottom row, we show the cross section of the $Q$-functions in the $xz$-plane. The blue-solid lines are obtained by the numerical ground states and the gray-dotted lines are the analytic results for the corresponding three coherent spin states. The perfect coincidence verifies that the ground state of the LMG model in FM-X phase is indeed a coherent spin state in the large $N$ limit.

\begin{figure}
\includegraphics[width=8cm]{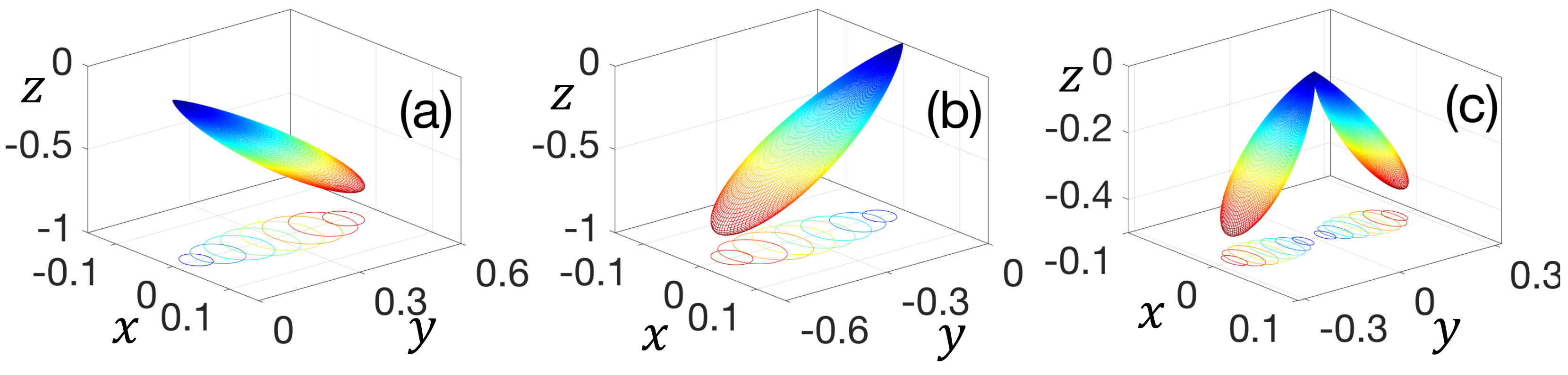}
\caption{\label{fig:QF_FMy} The Husimi $Q$-function of the ground states of the
spin in the FM-Y phase. The three columns correspond to the $Q$ functions of the two degenerate states $|\theta_{0},\pi/2\rangle$, $|\theta_{0},3\pi/2\rangle,$ and the superposition of these two states
$|\theta_{0},\pi/2\rangle+|\theta_{0},3\pi/2\rangle$, respectively. The curves underneath are the contour projections of the corresponding $Q$-functions in $xy$-plane. Here, the parameters are taken as $\epsilon=1$, $J_{x}=0$, $J_{y}=0.6$
and $N=200$.}
\end{figure}

Similar to the FM-X phase, the FM-Y phase also has two degenerate ground states $|\theta_{0},\pi/2\rangle$ and $|\theta_{0},3\pi/2\rangle$, where the polar angle $\theta_{0}$ is determined by the parameters of the Hamiltonian as given in Eq.~(\ref{eq:theta_LMGy}). The ensemble mean value of both $\langle\hat{S}_{x}\rangle_{0}$ and $\langle\hat{S}_{y}\rangle_{0}$ are zero. However, the magnetic fluctuations along $y$-axis characterized by the magnetic OP $\zeta_{M,y}$ is finite, which can be seen from the $Q$-function of states $|\theta_{0},\pi/2\rangle$, $|\theta_{0},3\pi/2\rangle$, and $|G_{+}\rangle=(|\theta_{0},\pi/2\rangle+|\theta_{0},3\pi/2\rangle)/\sqrt{2}$ in Fig.~\ref{fig:QF_FMy}. The cigar-like structures lie in the $yz$-plane, as the strong spin-spin coupling $J_y$ along the $y$-direction dominates in the FM-Y phase. In this case, $\zeta_{M,x}$ is very small and goes to zero for $N\rightarrow\infty$. 

In summary, we have revealed the first-order QPT in the traditional LMG model via the magnetic OPs $\zeta_{M,x}$ and $\zeta_{M,y}$. We predict that a similar first-order QPT should also exist in the XY Ising chain in a transverse field. The Husimi $Q$-functions show the fundamental difference between the ground state wave functions of the three quantum phases in the LMG model. We also numerically verified that the ground states of the LMG model are coherent spin states.

\subsection{Dicke-LMGx Model}

In this subsection, we mainly focus on the second-order superradiant QPT in the model,
\begin{equation}
\hat{H} = \hat{d}^{\dagger}\hat{d}+\frac{2\lambda}{\sqrt{N}}\hat{S}_{x}(\hat{d}+\hat{d}^{\dagger})+\epsilon\hat{S}_{z}-\frac{2}{N}J_{x}\hat{S}_{x}^{2}.\label{eq:H_DLMGx}
\end{equation}
This model can be obtained by setting the spin-spin coupling along $y$-direction to zero (i.e., $J_{y}=0$) in the Dicke-LMG model (\ref{eq:H_full1}). Thus, we call it the Dicke-LMGx model. In the original Dicke model~\cite{dicke1954coherence}, Dicke investigated the superradiance of an atomic ensemble composed of a large amount of indistinguishable two-level atoms as opposed to spins. Nevertheless, mathematically, indistinguishable two-level atoms are equivalent to spin-halves. Here, for the sake of consistency, we replace the atoms with spins.

The superradiant QPT in the Dicke model was first predicted by Hepp and Lieb in 1973~\cite{hepp1973superradiant}. Via analyzing the free energy per particle, they found there exists a thermodynamic phase transition as well as a QPT from the normal phase to the superradiant phase, in which macroscopic excitations exist in the ground state. Later, Wang and Hioe theoretically revisited this issue with a much simpler method~\cite{wang1973phase}. The QPT is easily shown with numerical simulation and has also been demonstrated in a recent experiment with a Bose-Einstein condensate (BEC)~\cite{baumann2010dicke}. However, the thermodynamic phase transition has not been demonstrated and the underlying mechanism is also missing. During our numerical simulation, we found that the phase transition temperature is strongly dependent on the order of two limits: the thermodynamic limit $N\rightarrow\infty$ and the Hilbert space cut-off corresponding to the bosonic mode $N_{\rm cutoff}\rightarrow\infty$. In the following, we only focus on the superradiant QPT.

\begin{figure}
\includegraphics[width=8cm]{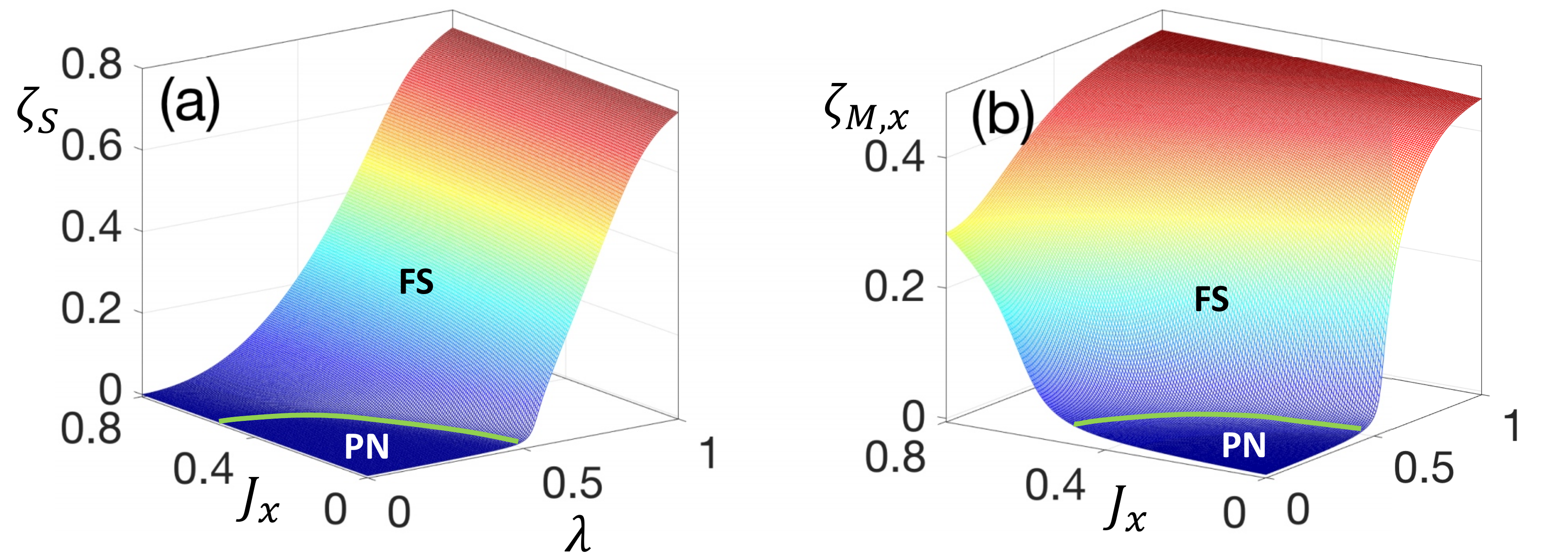}\caption{\label{fig:PD_DLMGx}Numerical demonstration of the phase diagram
of the Dicke-LMGx model with the superradiant order parameter $\zeta_{S}=\langle\hat{d}^{\dagger}\hat{d}\rangle_{0}/N$
in panel (a) and the magnetic order parameter $\zeta_{M,x}=\langle\hat{S}_{x}^{2}\rangle_{0}/N^{2}$
in panel (b). Here, the energy splitting of the spins is set as $\epsilon=1$,
the spin number in this figure is $N=40$ and the cut-off of the dimension
of the bosonic mode is also set as $40$.}
\end{figure}

There is a fundamental challenge to achieve the superradiant QPT in the Dicke model---the no-go theorem. The occurrence of the superradiant QPT is forbidden by the Thomas-Reiche-Kuhn (TRK) sum rule~\cite{rzazewski1975phase,rzazewski1979nogo}, as the atom-field (the spin-boson in our model) coupling cannot exceed the phase transition strength $\lambda_{c,{\rm II}}\equiv\sqrt{\epsilon}/2$ in a natural atomic system. There are many theoretical proposals to circumvent the no-go theorem. In 2007, Dimer \textit{et.
al.}, proposed to construct an effective Dicke Hamiltonian with a four-level atomic ensemble in an optical cavity~\cite{Dimer2007proposed}. Ciuti \textit{et. al.} proposed Cooper pair boxes capacitively coupled to a resonator~\cite{nataf2010nogo} or three-level systems~\cite{Baksic2013Superradiant}
to realize the superradiant QPT. Bastidas \textit{et. al.} suggested periodical modulating the spin-boson coupling to circumvent the no-go theorem~\cite{Bastidas2012non}. Recently, L\"u\textit{ et. al.}~\cite{lu2018single}, utilized the nonlinear coupling between a mechanical oscillator and an ancillary cavity mode to realize the superradiant QPT in the mechanical mode. In this subsection, we will show that this challenge can be easily overcome with the help of dipole-dipole interaction between spins (atoms), which has previously been disregarded. The spin-spin coupling lowers the phase transition spin-boson coupling strength required by the superradiant QPT.

The Dicke-LMGx model has two quantum phases: the PN phase and the FS phase. The phase diagram of the Dicke-LMGx model can be obtained by setting $J_{y}=0$ in Fig.~\ref{fig:phase_diagram}. We present the numerical simulation of the phase diagram with the superradiant OP $\zeta_{S}$ and the magnetic OP $\zeta_{M,x}$ in panels (a) and (b) in Fig.~\ref{fig:PD_DLMGx}, respectively. The other magnetic OP $\zeta_{M,y}$ is zero in both phases and has been ignored in this subsection. The phase boundary given by $4\lambda^{2}+2J_{x}=\epsilon$ is depicted by the green line which corresponds to the green line in Fig.~\ref{fig:phase_diagram} exactly. In the PN phase, both the superradiant OP $\zeta_{S}$ and the magnetic OP $\zeta_{M,x}$ are zero. In the FS phase, both $\zeta_{S}$ and $\zeta_{M,x}$
are finite except when $\lambda=0$. In panel (a), we see close to the $J_x$ axis, $\zeta_{S}$ marginally increases with $J_x$ after passing the boundary and its magnitude is negligible small even in the FS phase, as the magnetic  OP $\zeta_{S}$ is proportional to $\lambda^{2}$ [see Eq.~(\ref{eq:zeta_S})]. But $\zeta_{M,x}$ always increases to a relatively large value after the parameters cross the phase boundary as shown in panel (b).

\begin{figure}
\includegraphics[width=8cm]{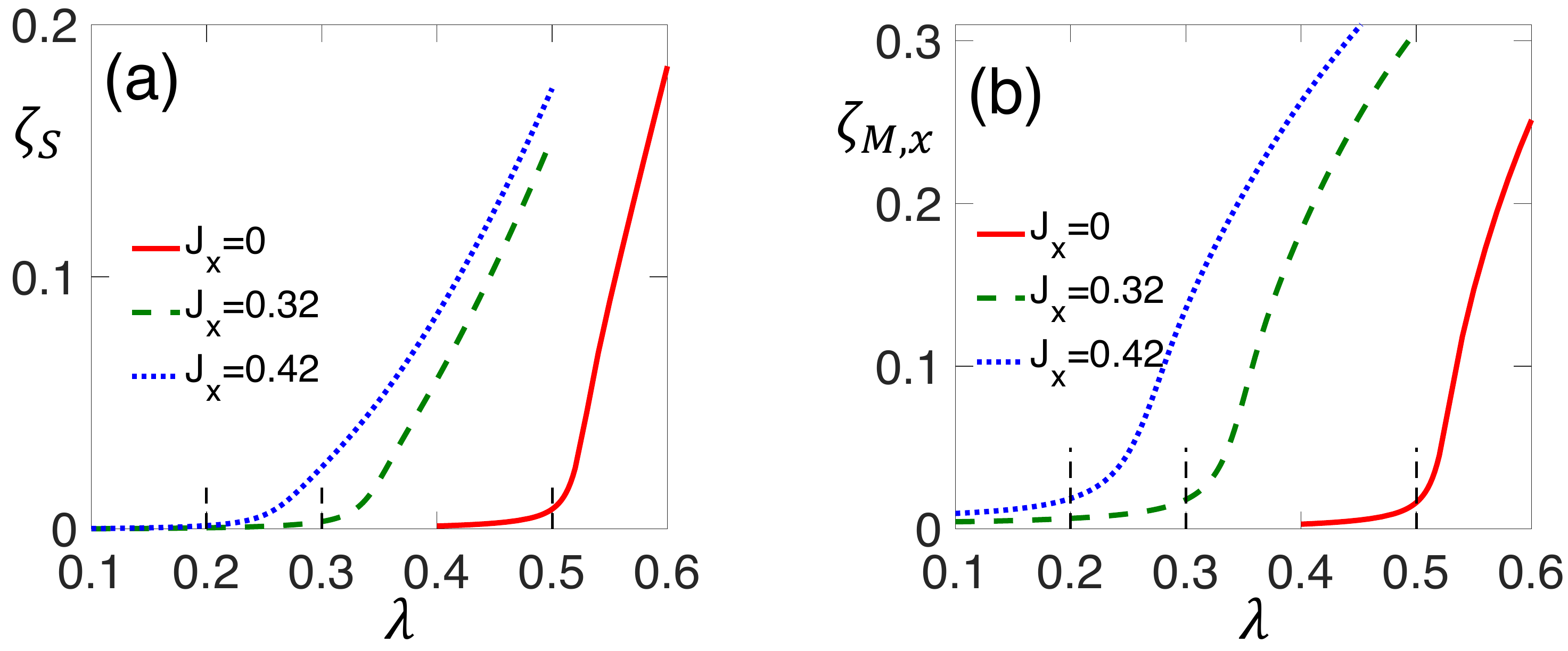}
\caption{\label{fig:QPT_DLMGx1} Numerical demonstration of the QPT from the paramagnetic-normal (PN) phase to the ferromagnetic-superradiant (FS) phase. In panels (a) and (b), we show the order parameters $\zeta_{S}$ and $\zeta_{M,y}$, respectively, for different spin-spin coupling $J_{x}$. The position of the phase transition spin-boson coupling $\lambda_{c,{\rm II}}\equiv\sqrt{\epsilon-2J_{x}}/2$ is marked by the black dashed lines. Here, the other parameters are set as $\epsilon=1$, $N=80$, and the bosonic mode cut-off is $80$.}
\end{figure}

\begin{figure}
\includegraphics[width=8cm]{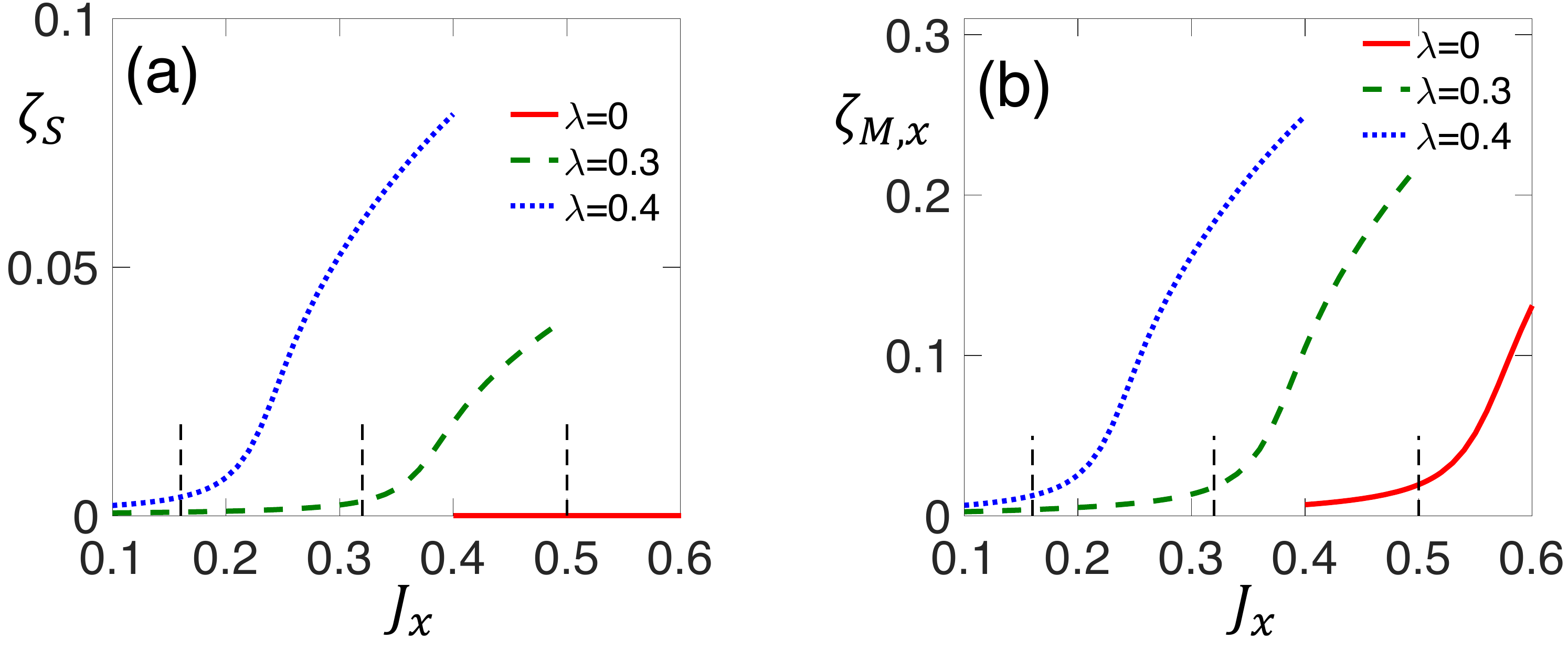}\caption{\label{fig:QPT_DLMGx2}Numerical demonstration of the QPT from the paramagnetic-normal phase to the ferromagnetic-superradiant (FS) phase. In panels (a) and (b), we show the order parameters $\zeta_{S}$ and $\zeta_{M,y}$, respectively, for different spin-boson coupling $\lambda$. The position of the phase transition spin-spin coupling $J_{xc,{\rm II}}=\epsilon/2-2\lambda^{2}$ is marked by the black dashed lines. Here, the other parameters are set as $\epsilon=1$, $N=80$, and the bosonic mode cut-off is $80$.}
\end{figure}

Now, we show how to overcome the no-go theorem in superradiant QPT. In Fig.~\ref{fig:QPT_DLMGx1}, we plot the OPs $\zeta_{S}$ and $\zeta_{M,x}$  in panels (a) and (b) as functions of spin-boson coupling $\lambda$ with fixed spin-spin coupling $J_{x}$. The position of the new phase transition spin-boson coupling $\lambda_{c,{\rm II}}\equiv\sqrt{\epsilon-2J_{x}}/2$ is marked by the vertical black dashed lines. For larger $J_{x}$ case, a relatively smaller spin-boson coupling $\lambda$ is required to trigger the QPT from the PN phase to the FS phase. With the help of the dipolar coupling along $x$-axis, a weak spin-boson coupling can still induce the superradiant QPT. Thus, the constraint from the no-go theorem in the Dicke model are easily overcome.  Although, to obtain a larger superradiant OP $\zeta_{S}$ (more excitations in the bosonic mode), strong spin-boson coupling is still needed. In Fig.~\ref{fig:QPT_DLMGx2}, we plot the OPs $\zeta_{S}$ and $\zeta_{M,x}$ as functions of spin-spin coupling $J_{x}$ with fixed spin-boson coupling $\lambda$ in panels (a) and (b), respectively. After the QPT, smaller $\zeta_{S}$ is obtained for smaller $\lambda$ but with the same $J_{x}$. Especially, in the case of $\lambda=0$, the superradiant OP $\zeta_{S}$ is still zero after the QPT when $J_{x}>J_{xc,{\rm II}}$ as shown by the red line in panel (a). In this case, the system is actually in the FN phase.

\begin{figure}
\includegraphics[width=8cm]{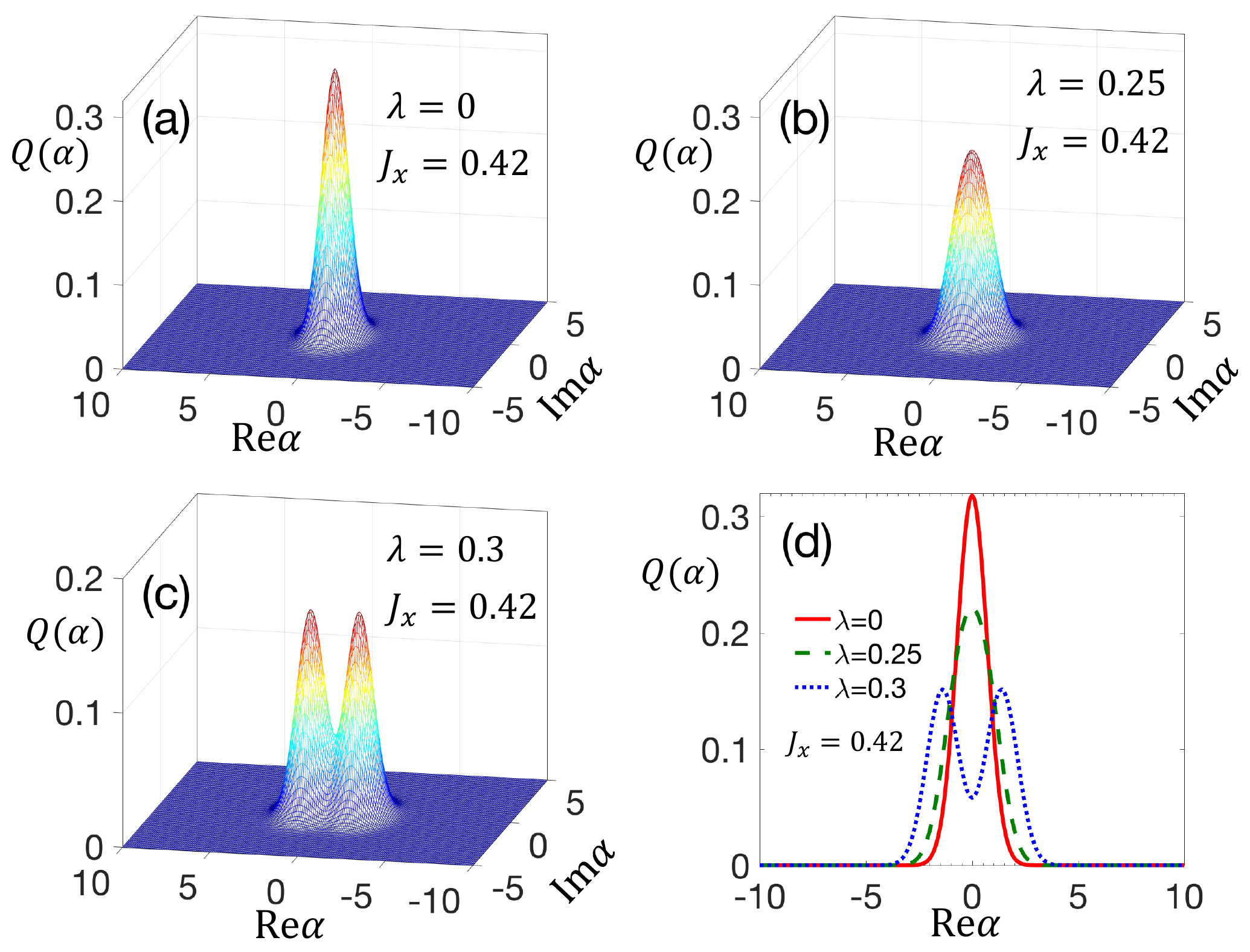}
\caption{\label{fig:BosonQF_DLMGx}The Husimi $Q(\alpha)$-function of the bosonic mode ground state for different parameters. Here, $\alpha=x+iy$
is a complex number, the magnetic field strength is set to be $\epsilon=1$,
the spin number is $N=80$, and the cut-off of the bosonic mode is
$80$. Panel (a) is for PN phase. Both panel (b) and (c) represent the FS
phase. The cross sections of the $Q$-function in the $xz$-plane
are shown in panel (d). In the FS phase, there are two degenerate states
$|\sqrt{N}\alpha_{0}\rangle\otimes|\theta_{0},\pi\rangle$ and $|-\sqrt{N}\alpha_{0}\rangle\otimes|\theta_{0},0\rangle$.
Here, we show the $Q$-function of the superposition state $|\sqrt{N}\alpha_{0}\rangle\otimes|\theta_{0},\pi\rangle+|-\sqrt{N}\alpha_{0}\rangle\otimes|\theta_{0},0\rangle$. }
\end{figure}

At the end of this subsection, let's shed light on the ground states of the system with the Husimi $Q$-functions. The $Q$-function of the spin is defined in Eq.~(\ref{eq:spinQF}) after tracing out the bosonic degrees of freedom, such that the $Q$-function of the bosonic mode is defined as
\begin{equation}
Q(\alpha)=\frac{1}{\pi}\rm{Tr}_{\rm spin}[\langle\alpha |\hat{\rho}_g|\alpha\rangle],
\end{equation}
where $\hat{\rho}_g$ is the ground-state density matrix of the entire system,  $|\alpha\rangle$ is the bosonic coherent state with amplitude $\alpha$, and $\rm{Tr}_{\rm spin}$ means tracing out the spin degrees of freedom. The ground state of the PN phase is $\left|0\right\rangle \otimes|\pi,\phi_{0}\rangle$, i.e.,
the bosonic mode is in the vacuum state and the spins are in the coherent spin state with $\theta_{0}=\pi$ and undetermined $\phi_{0}$. The FS phase has two degenerate ground states $|\sqrt{N}\alpha_{0}\rangle \otimes|\theta_{0},\pi\rangle$
and $|-\sqrt{N}\alpha_{0}\rangle \otimes|\theta_{0},0\rangle$,
where $\theta_{0}$ and $\alpha_{0}$ are determined by the parameters of the system as given in Eq.~(\ref{eq:theta_DLMGx}-\ref{eq:alpha_DLMGx}). In the FS phase, the system can be in an arbitrary superposition of these two degenerate states. Thus, the ensemble mean value of $\langle\hat{d}^{\dagger}\rangle_{0}$ (or
$\langle\hat{d}^{\dagger}\rangle_{0}$) and $\langle\hat{S}_{x}\rangle_{0}$ are zero. However, the mean of the coupling terms $\langle\hat{d}\hat{S}_{x}\rangle_{0}$
and $\langle\hat{d}^{\dagger}\hat{S}_{x}\rangle_{0}$ are finite negative values, which minimize the ground-state energy of the whole system. 

The ground state of the bosonic mode changes from the vacuum state to a coherent state after the superradiant QPT. The $Q$-function of the bosonic mode for different spin-boson couplings is displayed in Fig.~\ref{fig:BosonQF_DLMGx}. In panel (a), we show the $Q$-function of the bosonic mode in the ground state (the vacuum state) of the PN phase with $2J_{x}+4\lambda^{2}<\epsilon$. There is only one peak located at $\alpha=0$. For panels (b) and (c), the system is in the FS phase with $2J_{x}+4\lambda^{2}>\epsilon$ and the spin-boson coupling $\lambda$ leads to shifting of the peak along the real axis (as proven in Sec.~\ref{Sec:phase_diagram}). Here, we take the ground state of the FS phase as the symmetric quantum superposition state $|G_{+}\rangle=(|\sqrt{N}\alpha_{0}\rangle \otimes|\theta_{0},\pi\rangle+|-\sqrt{N}\alpha_{0}\rangle \otimes|\theta_{0},0\rangle)/\sqrt{2}$. In panel (b), the spin-coupling $\lambda$ is not large enough to split the $Q$-function into two separated peaks like in panel (c). This demonstrated more in panel (d), which shows the cross-section of the $Q$-function on the real-$\alpha$ plane. 

\begin{figure}
\includegraphics[width=8cm]{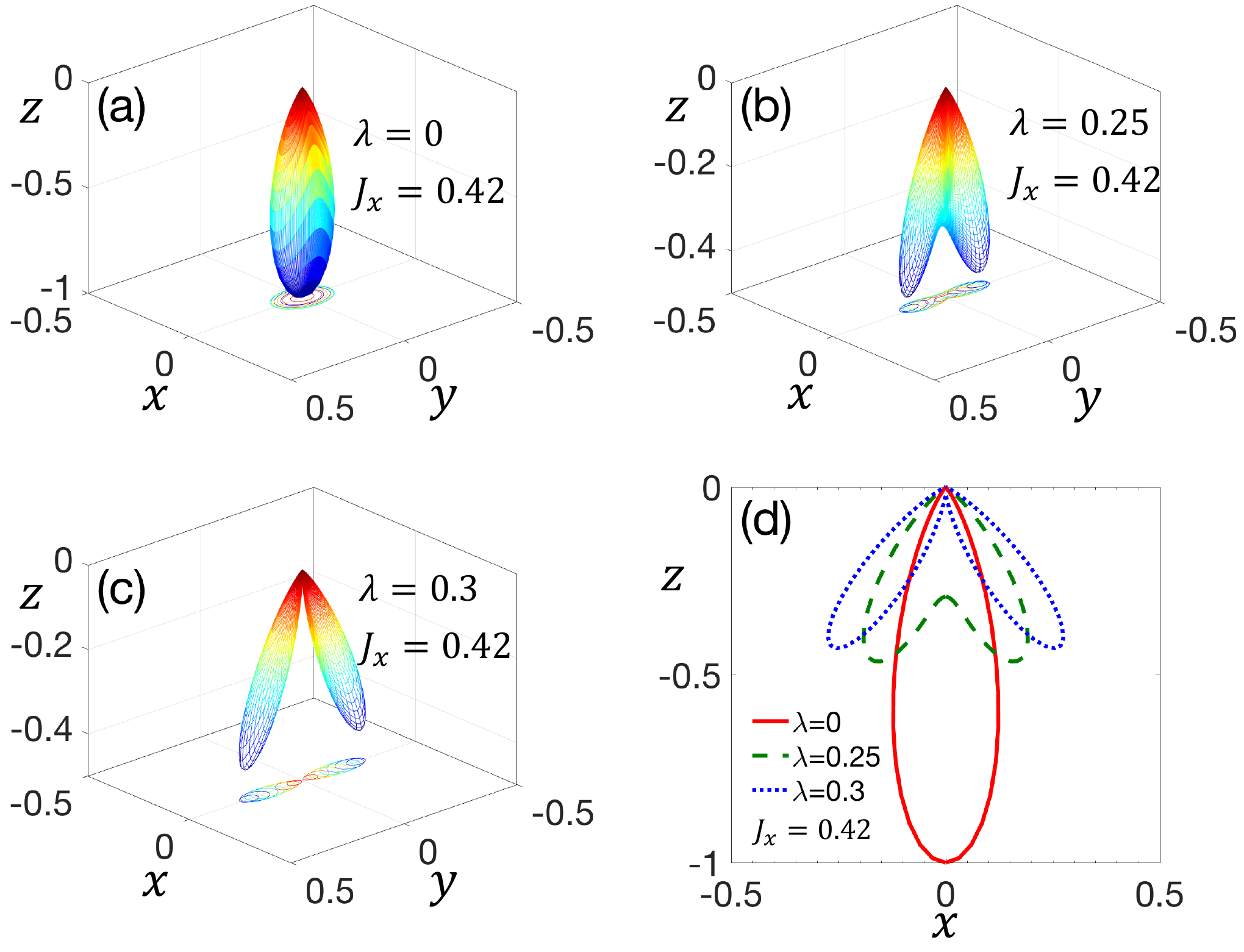}
\caption{\label{fig:SpinQF_DLMGx} The Husimi $Q(\theta,\phi)$-function of the ground states of the spins for different parameters. Here, the surface is obtained by transfer the spherical coordinates $(r=Q(\theta,\phi),\theta,\phi)$ to the corresponding Cartesian coordinates $(x,y,z)$. The magnetic field strength is set to be $\epsilon=1$, the spin number is $N=80$, and the cut-off of the bosonic mode is $80$. Panel (a) is for the PN phase. Both panel (b) and (c) are for the FS phase. The cross sections of the $Q$-function in the $xz$-plane are shown in panel (d). In the FS phase, there are two degenerate states $|\sqrt{N}\alpha_{0}\rangle\otimes|\theta_{0},\pi\rangle$
and $|-\sqrt{N}\alpha_{0}\rangle\otimes|\theta_{0},0\rangle$. Here,
we show the $Q$-function of the superposition state $|\sqrt{N}\alpha_{0}\rangle\otimes|\theta_{0},\pi\rangle+|-\sqrt{N}\alpha_{0}\rangle\otimes|\theta_{0},0\rangle$. }
\end{figure}

The $Q$-function of the spins is displayed in Fig.~\ref{fig:SpinQF_DLMGx} with the same parameters as in Fig.~\ref{fig:BosonQF_DLMGx}. For the PN phase as shown in panel (a), the normalized spin $Q$-function is a cigar-like structure lying along the negative $z$-axis. For the FS phase
as shown in panels (b) and (c), the interaction along $x$-axis splits the $Q$-function in the $xz$-plane. The $Q$-function of the ground state $|G_{+}\rangle$ has two branches on the negative-$x$ and positive-$x$ half planes corresponding to the two degenerate ground states, respectively. For larger $\lambda$, larger polar angles $\theta_{0}$ are obtained as shown in panel (d). 

In summary, we presented the details of the second-order QPT in the Dicke-LMGx model via the superradiant OP $\zeta_{S}$ and the magnetic OP $\zeta_{M,x}$. We showed that the superradiant QPT in this model is immune to the no-go theorem. The spin-spin dipolar interaction along $x$-axis relaxes the constraint from the TRK sum rule. Although, to obtain a larger superradiant OP $\zeta_{S}$, strong spin-boson coupling is still needed. 

\subsection{Dicke-LMGy Model} \label{sec:DLMGy}
In our previous work~\cite{yang2018QCD}, we have revealed the
first-order QPT in the Dicke-LMGy model,
\begin{equation}
\hat{H} = \hat{d}^{\dagger}\hat{d}+\frac{2\lambda}{\sqrt{N}}\hat{S}_{x}(\hat{d}+\hat{d}^{\dagger})+\epsilon\hat{S}_{z}-\frac{2}{N}J_{y}\hat{S}_{y}^{2},\label{eq:H_DLMGy}
\end{equation}
with $J_{x}=0$ in Hamiltonian (\ref{eq:H_full1}). In this subsection, we present more details about the ground states and QPTs in this model.

\begin{figure}
\includegraphics[width=8.5cm]{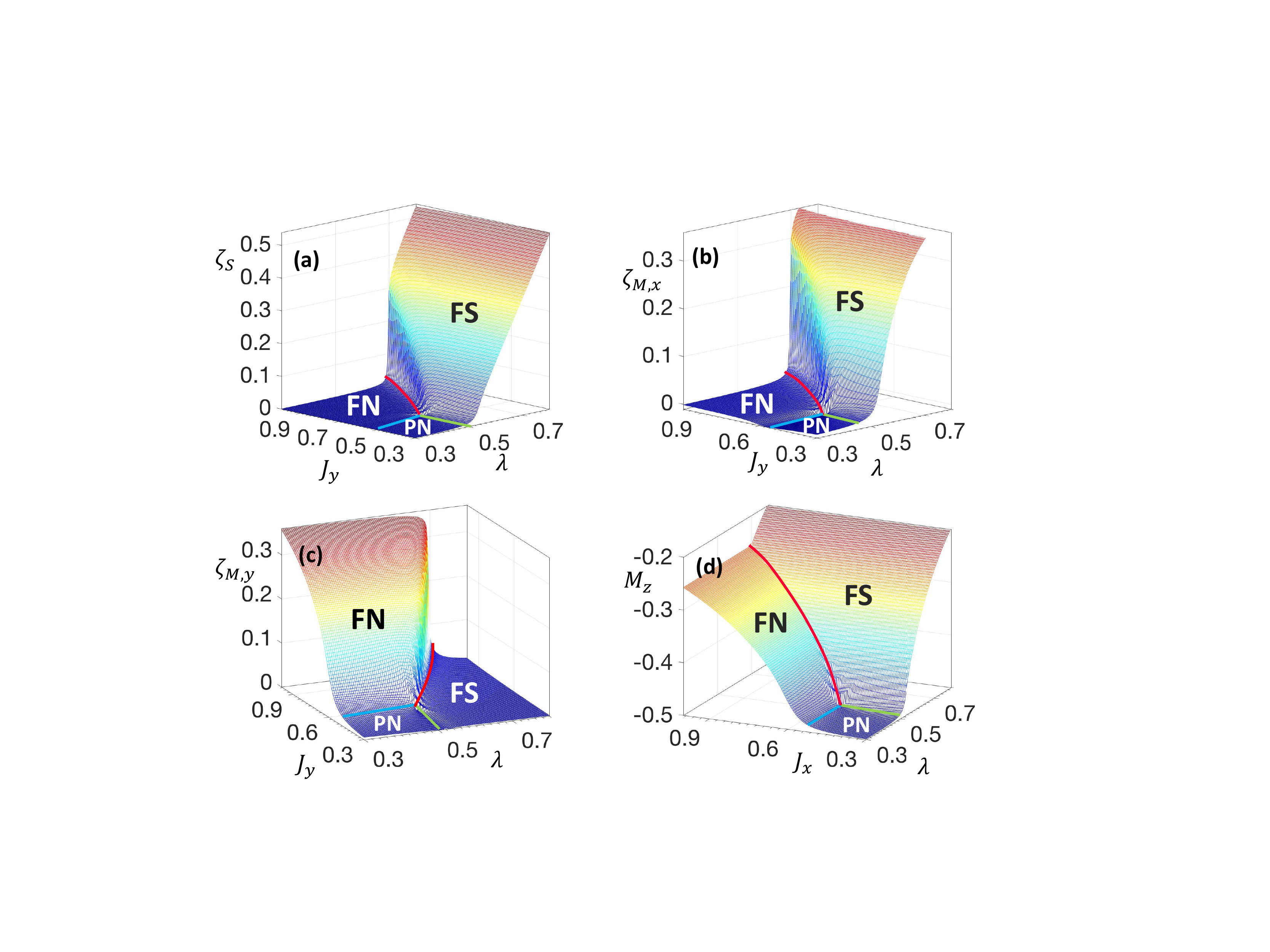}\caption{\label{fig:PD_DLMGy}Numerical demonstration of the phase diagram
of the Dicke-LMGx model. Panel (a-d) describe the order parameters $\zeta_{S}$, $\zeta_{M,x}$, $\zeta_{M,y}$, and $M_{z}$, respectively. Here, the energy splitting of the spins is set as $\epsilon=1$, the spin number in this figure is $N=40$ and the cut-off dimension of the bosonic mode is also set to $40$.}
\end{figure}

The Dicke-LMGy model has three quantum phases: PN phase, FN phase, and FS phase. The numerical demonstration of the phase diagram is given in Fig.~\ref{fig:PD_DLMGy}, which exactly coincides with the one obtained via the mean-field theory (see Fig.~\ref{fig:phase_diagram} with $J_{x}=0$). In panels (a) and (b), both the superradiant OP $\zeta_{S}$ and the magnetic OP $\zeta_{M,x}$ change continuously when the spin-boson
coupling $\lambda$ crosses the phase transition point  $\lambda_{c,{\rm II}}\equiv\sqrt{\epsilon}/2$ (the green line) if the spin-spin coupling $J_{y}$ is below the phase transition strength $J_{yc,{\rm II}}\equiv\epsilon/2$. This shows the second-order superradiant QPT from the PN phase to the FS phase similar to the one in the Dicke model~\cite{hepp1973superradiant,wang1973phase}. In panel (c), the magnetic OP $\zeta_{M,y}$ displays another second-order QPT from the PN phase to the FN phase when the spin-spin coupling $J_{y}$ crosses the phase transition point $J_{yc,{\rm II}}$ (the blue line) and the spin-boson coupling is below the phase transition strength $\lambda_{c,{\rm II}}$. This QPT coincides with the one found in the LMG model~\cite{lipkin1965validity,meshkov1965validity,glick1965validity}. However, discontinuous changes occur in all three OPs when the Hamiltonian parameters cross the red line $\lambda=\sqrt{J_y/2}$ in the strong-coupling region with $\lambda>\lambda_{c,{\rm II}}$ and $J_{y}>J_{yc,{\rm II}}$. This indicates a first-order QPT between the FN phase and the FS phase, which is of interest for quantum critical amplification. We finally plot the traditional magnetic OP $M_{z}$ in panel (d). Similar to the LMG model, $M_{z}$ cannot distinguish the FN and FS phases and it cannot characterize the first-order QPT either as no discontinuous change exists. 

\begin{figure}
\includegraphics[width=8cm]{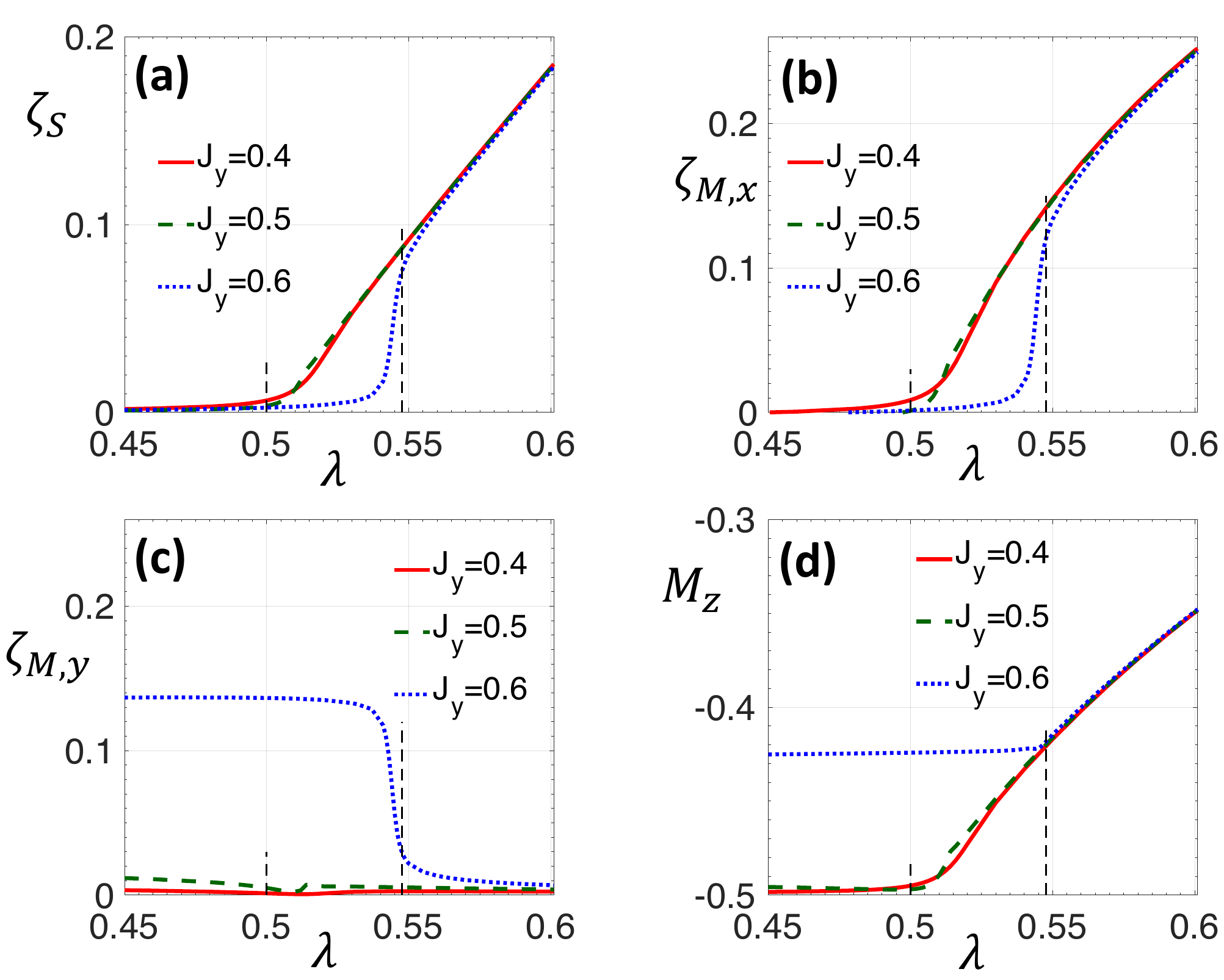}
\caption{\label{fig:QPT_DLMGy1}Numerical demonstration of the quantum phase transitions (QPTs) in the Dicke-LMGy model. In panel (a-d), we show the order parameters $\zeta_{S}$, $\zeta_{M,x}$, $\zeta_{M,y}$, and $M_{z}$, respectively for different spin-spin coupling $J_{x}$. Here, the other parameters are $\epsilon=1$, the spin number $N=80$, and the cut-off of the bosonic mode is $80$. The position of the phase transition spin-boson coupling $\lambda_{c,{\rm II}}=\sqrt{\epsilon}/2=0.5$ in the the second-order QPTs and the first-order phase transition coupling $\lambda_{c,{\rm I}}=\sqrt{J_{y}/2}$ in the  first-order QPTs are marked by the black dashed lines. }
\end{figure}

We show more details about the QPTs in the Dicke-LMGy model in Fig~\ref{fig:QPT_DLMGy1}. The superradiant OP $\zeta_{S}$ in panel (a) and the magnetic OP $\zeta_{M,x}$ in panel (b) behave similarly around the phase transition points. In the weak spin-spin coupling case $J_{y}\leq J_{yc,{\rm II}}$, second-order QPTs from the PN phase to the FS phase occurs at $\lambda_{c,{\rm II}}$ (see the red and green lines). Also from the red and green lines in panel (c), we see that the magnetic OP $\zeta_{M,y}$ is very small and changes marginally around the phase transition point $\lambda_{c,{\rm II}}$. This verifies that no magnetic order changes along the $y$-axis occurs in these second-order QPTs. From the blue lines in panels (a-c), we find that there exists a first-order QPT from the FN phase to the FS when the spin-spin coupling $J_{y}$ is greater than the second-order phase transition strength $J_{yc,{\rm II}}$. The OPs $\zeta_{S}$ and $\zeta_{M,x}$ jump from zero to a finite value at the phase transition point $\lambda_{c,{\rm I}}=\sqrt{J_{y}/2}$ and at the same time, $\zeta_{M,y}$ drops to zero. Thus, the first-order QPT results from the competition between the FS phase that arises from strong spin-boson coupling along the $x$-axis and the FN phase caused by the large spin-spin coupling along the $y$-axis. In panel (d), we show the traditional magnetic OP $M_z$. No discontinuous changes exist in $M_z$.

\begin{figure}
\includegraphics[width=8cm]{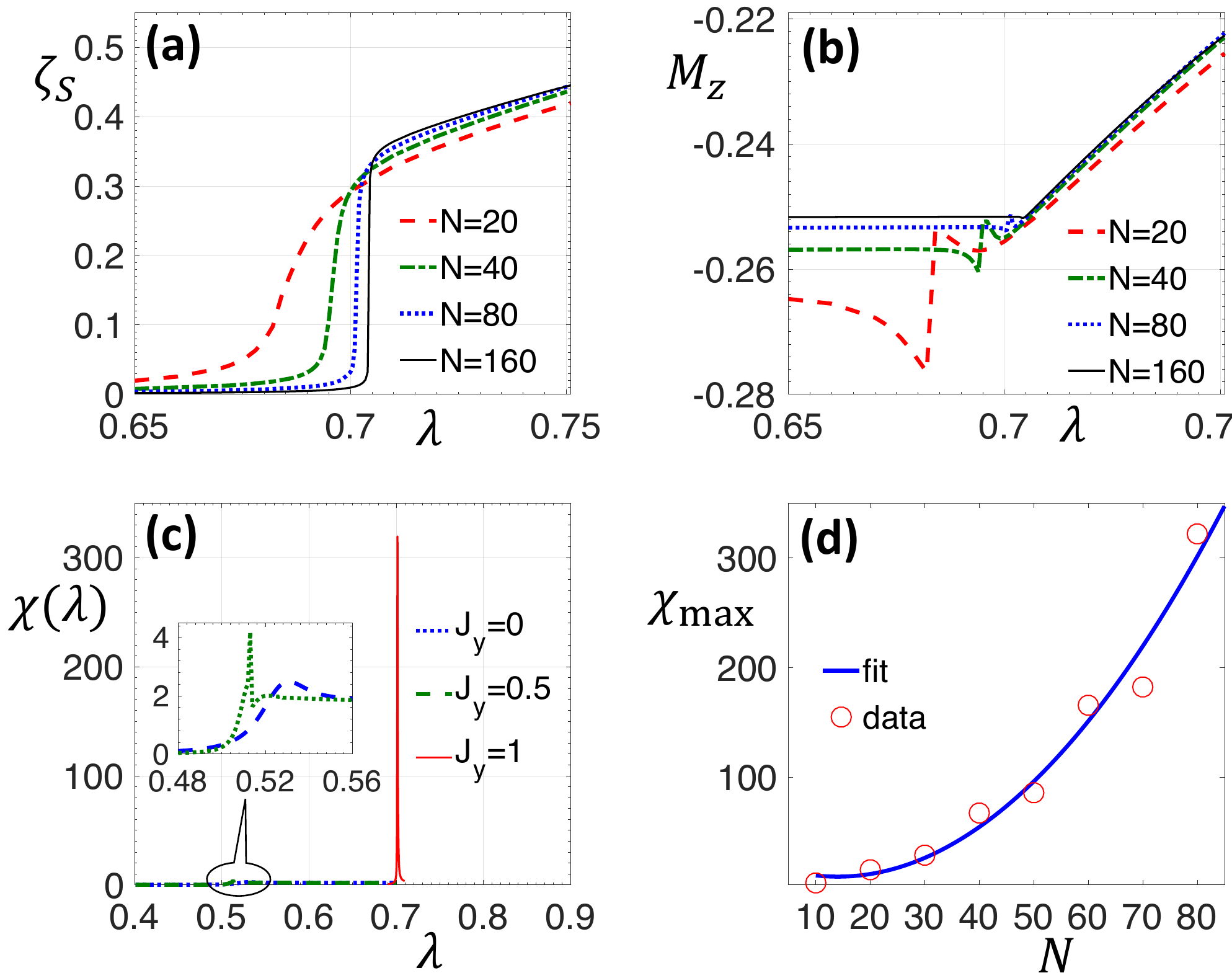}
\caption{\label{fig:QPT_DLMGy2}Verification of the first-order quantum phase transition (QPT) with fixed parameters $\epsilon=1$ and $J_{y}=1$. In panels (a) and (b), we plot $\zeta_{S}$ and $M_{z}$ as functions of $\lambda$, respectively, for different spin numbers. Here, the cut-off of the bosonic mode is set to the spin number $N$. In panel (c), we plot the sensitivity function defined in Eq.~(\ref{eq:chi_boson}) as a function of $\lambda$ with $N=80$. The blue and greed lines are for second-order QPTs with $J_y\leq J_{yc,{\rm II}}\equiv\epsilon/2=0.5$ and the red line is for a first-order QPT with $J_y>J_{yc,{\rm II}}$. The inset shows the details of the sensitivity function of second-order QPTs around the phase transition point $\lambda_{c,{\rm II}}\equiv\sqrt{\epsilon}/2=0.5$. In panel (d), we plot the maximum of $\chi$ in the first-order QPT with $J_{y}=1$. The red circles are the data from the numerical calculations and blue-solid line displays the corresponding fitting function $f(x)=0.067x^{2}-1.881x+22.62$. As we can see, the sensitivity diverges as $N^2$.}
\end{figure}

We verify that the FN-FS phase transition is indeed of first-order in Fig.~\ref{fig:QPT_DLMGy2}. Increasing the spin number $N$, the phase transition shows singular scaling behavior in panel (a). But in panel (b), we see that no such discontinuous change in $M_{z}$ at the phase transition point. Here, we utilize the first derivative of the superradiant OP $\zeta_S$ to characterize the sensitivity of the system at the phase transition point,
\begin{equation}
\chi(\lambda)=\frac{1}{N}\frac{d}{d\lambda}\langle\hat{d}^{\dagger}\hat{d}\rangle,\label{eq:chi_boson}
\end{equation}
where the factor 1/N is added for consistency with the magnetic
susceptibility. In panel (c), we plot the sensitivity as a function of spin-boson coupling $\lambda$ for different spin-spin coupling. Only the sensitivity function of the first-order QPT (the red line) has a sharp peak with diverging height at the phase transition point $\lambda_{c,{\rm I}}$. In panel (d), we plot the maximum of the sensitivity $\chi_{{\rm max}}$ (at the phase transition point) of the first-order QPT as a function of the spin number $N$. The sensitivity function $\chi_{{\rm max}}$ diverges with speed $\propto N^{2}$, which is different from the $\sqrt{N}$-scaling obtained in the previous first-order dissipative transition~\cite{raghunandan2018high} or the linear $N$ scaling in the first-order thermodynamic phase transition predicted by Imry~\cite{imry1980finite}.

\begin{figure}
\includegraphics[width=8cm]{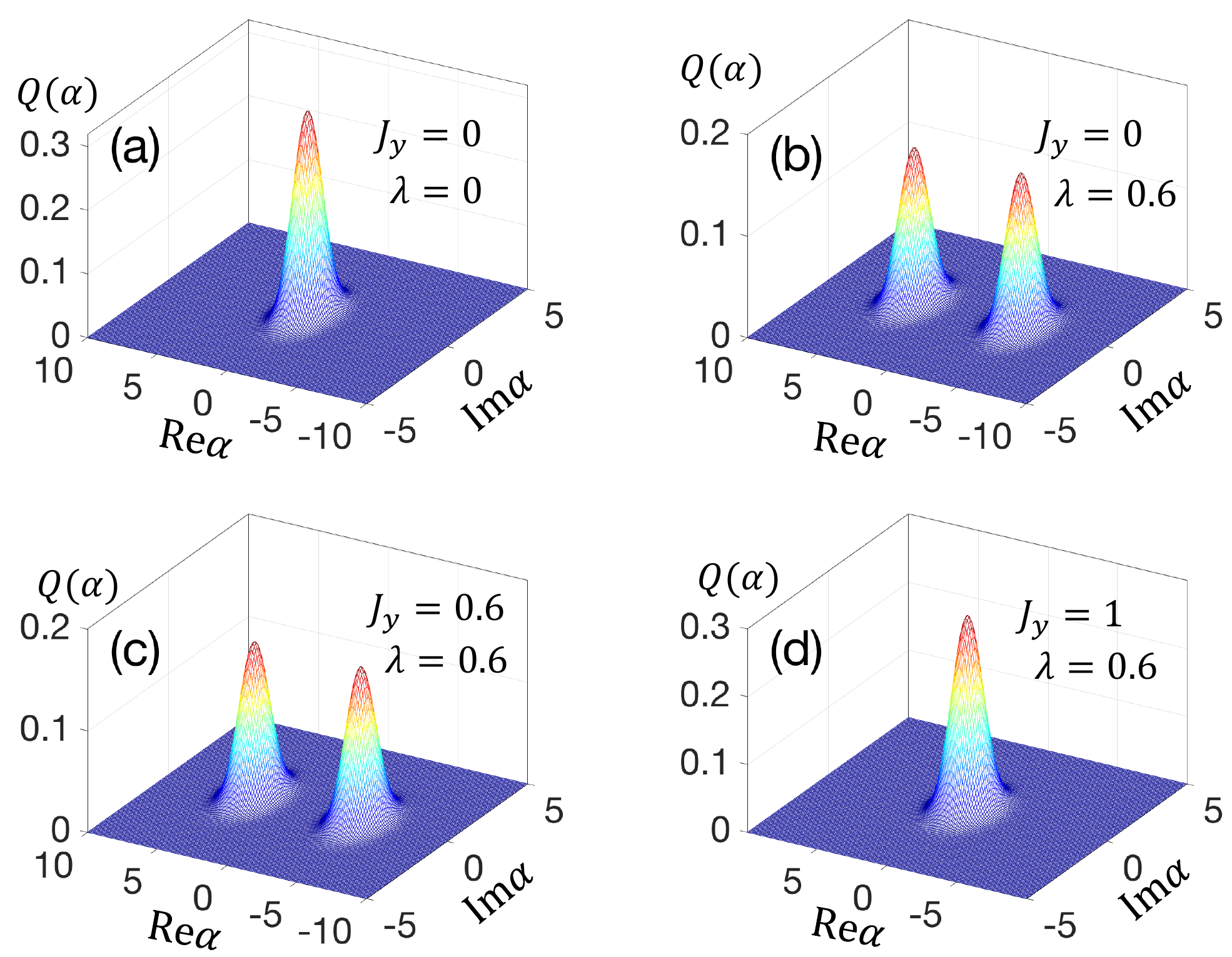}
\caption{\label{fig:QFBoson_DLMGy}The Husimi $Q$-function of the ground states of the bosonic mode in different phases. For panel (a), the system is in the PN phase with ground state $\left|0\right\rangle \otimes|\pi,\phi_{0}\rangle$.
Increasing spin-boson coupling $\lambda>\lambda_{c,{\rm II}}=0.5$, the system goes to the FS phase with two degenerate states $|\sqrt{N}\alpha_{0}\rangle\otimes|\theta_{0},\pi\rangle$
and $|-\sqrt{N}\alpha_{0}\rangle\otimes|\theta_{0},0\rangle$. The $Q$-function of the symmetric quantum superposition of these two FS ground states is displayed in (b). In panel (c), we increase the spin-spin coupling $J_{y}$, but not strong enough to cross the FS-FN boundary. Thus the $Q$-function is the same one as in panel (b). For $J_{y}>2\lambda^{2}$, the system goes to the FN phase with two degenerate ground states $|0\rangle\otimes|\theta_{0},\pi/2\rangle$ and $|0\rangle\otimes|\theta_{0},3\pi/2\rangle$.
The corresponding $Q$-function of the bosonic vacuum state is shown in (d).  Here, the other parameters are set as $\epsilon=1$, the spin number $N=80$, and the cut-off of the bosonic mode $80$.}
\end{figure}

\begin{figure}
\includegraphics[width=8.5cm]{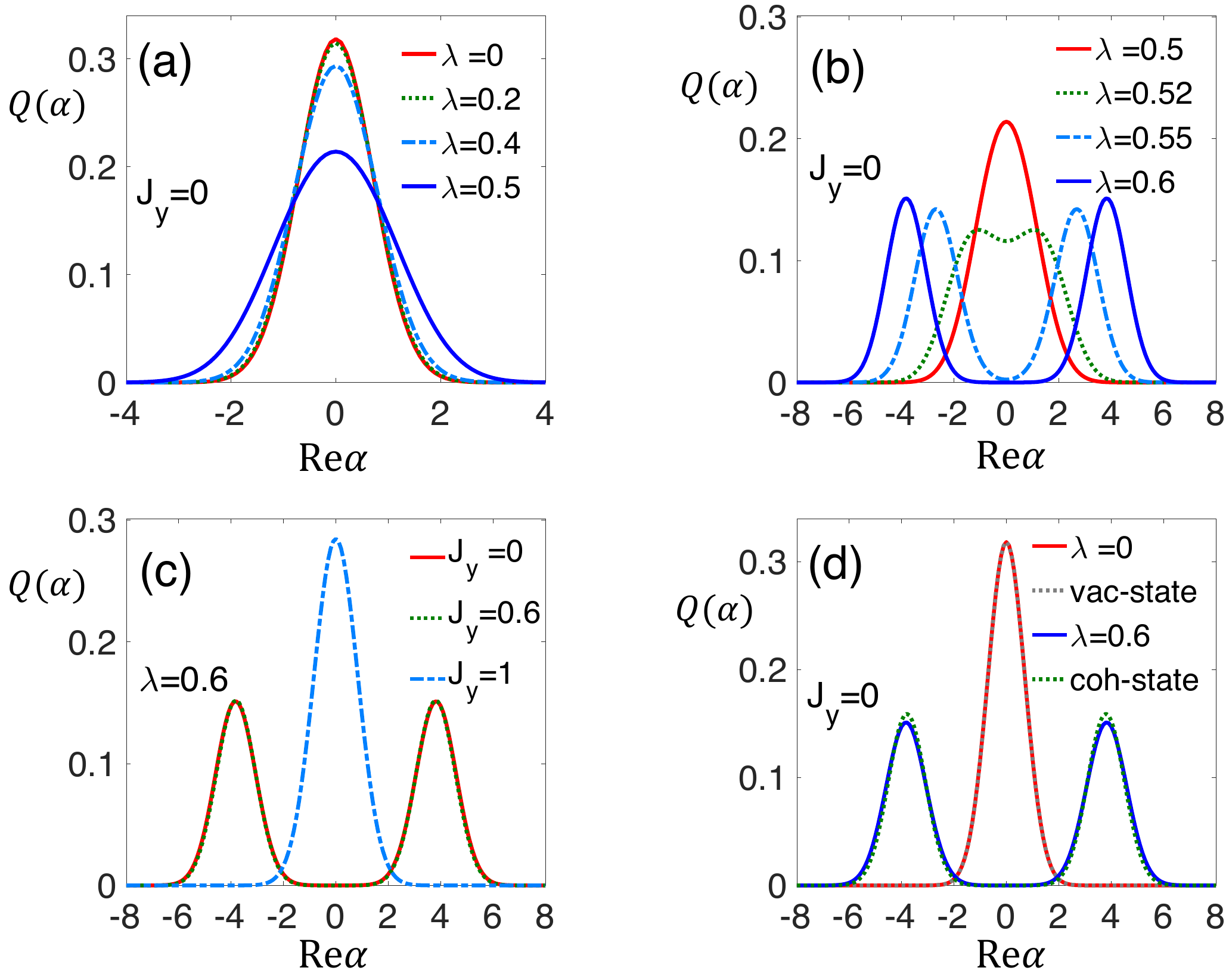}
\caption{\label{fig:QFBoson_DLMGy2D}In panels (a) and (b), we plot the cross section of the $Q$-functions for the ground states of the PN phase and the FS phase, respectively. In panel (c), we show the changes of the $Q$-function, when the system goes from the FS phase to the FN phase as the spin-spin coupling $J_y$ increases from zero (the red-solid line) to $1$ (the blue-dashed line). In panel (d), we compare the $Q$-functions of the ground states obtained by numerical diagonalization with the analytic results from the corresponding vacuum state and coherent state $|\alpha_0\rangle$ obtained by the mean-field theory. Here, the other parameters are set as $\epsilon=1$, the spin number $N=80$, and the cut-off of the bosonic mode $80$.}

\end{figure}

Let's examine the ground-state wave functions of the three
quantum phases using the Husimi $Q$-function. Similar to the Dicke-LMGx model, the ground state of the PN phase is $|0\rangle \otimes|\pi,\phi_{0}\rangle$ and the FS phase has two degenerate ground states $|\sqrt{N}\alpha_{0}\rangle \otimes|\theta_{0},\pi\rangle$
and $|-\sqrt{N}\alpha_{0}\rangle \otimes|\theta_{0},0\rangle$,
where $\theta_{0}$ and $\alpha_{0}$ are determined by the parameters of the system as given in Eq.~(\ref{eq:theta_DLMGx}-\ref{eq:alpha_DLMGx}). The FN phase also has two degenerate states $|0\rangle \otimes|\theta_{0},\pi/2\rangle $
and $\left|0\right\rangle \otimes|\theta_{0},3\pi/2\rangle $,
where $\theta_{0}$ is determined by Eq. (\ref{eq:theta_LMGy}). In the following, we always choose the ground states of the FS phase and FN phase as the symmetric quantum superposition of their corresponding two degenerate ground states. 

The $Q$-function of the bosonic mode $Q(\alpha)$ in different phases are displayed in Fig.~\ref{fig:QFBoson_DLMGy}. For panel (a), both the spin-boson coupling and the spin-spin coupling are below the second-order phase transition points, i.e., $\lambda<\lambda_{c,{\rm II}}\equiv \sqrt{\epsilon}/2$ and $J_{y}<J_{yc,{\rm II}}\equiv\epsilon/2$. Thus, the system is in the PN phase. The bosonic mode is in the vacuum state and the corresponding $Q$-function has only one peak located at $\alpha=0$. In panels (b), we increase the spin-boson coupling to exceed the second-order superradiant QPT point ($\lambda >\lambda_{c,{\rm II}}$), but fix the spin-spin coupling at zero. In this case, the system is in the FS phase. There are two peaks in $Q(\alpha)$ on the real axis corresponding to the two coherent states $|-\sqrt{N}\alpha_{0}\rangle $ and $|\sqrt{N}\alpha_{0}\rangle $, respectively. In panel (c), we increase the spin-spin coupling $J_{y}$, but not enough to suppress the FS phase, i.e., $J_y<2\lambda^2$. The system is still in the FS phase and the ground state does not change. Thus the $Q$-function is exactly the same as the one in panel (b). This can be more clearly seen in Fig.~\ref{fig:QFBoson_DLMGy2D} (c). In panel (d) the spin-spin coupling crosses the FS-FN phase boundary ($J_y>2\lambda^2$) and it is also greater than second-order magnetic QPT point $J_y>J_{yc,{\rm II}}$. The system therefore goes to the FN phase. For the ground state of the FN phase, the bosonic mode is in the vacuum state $\left|0\right\rangle $ and its $Q$-function is exactly the same as panel (a).

\begin{figure}
\includegraphics[width=8cm]{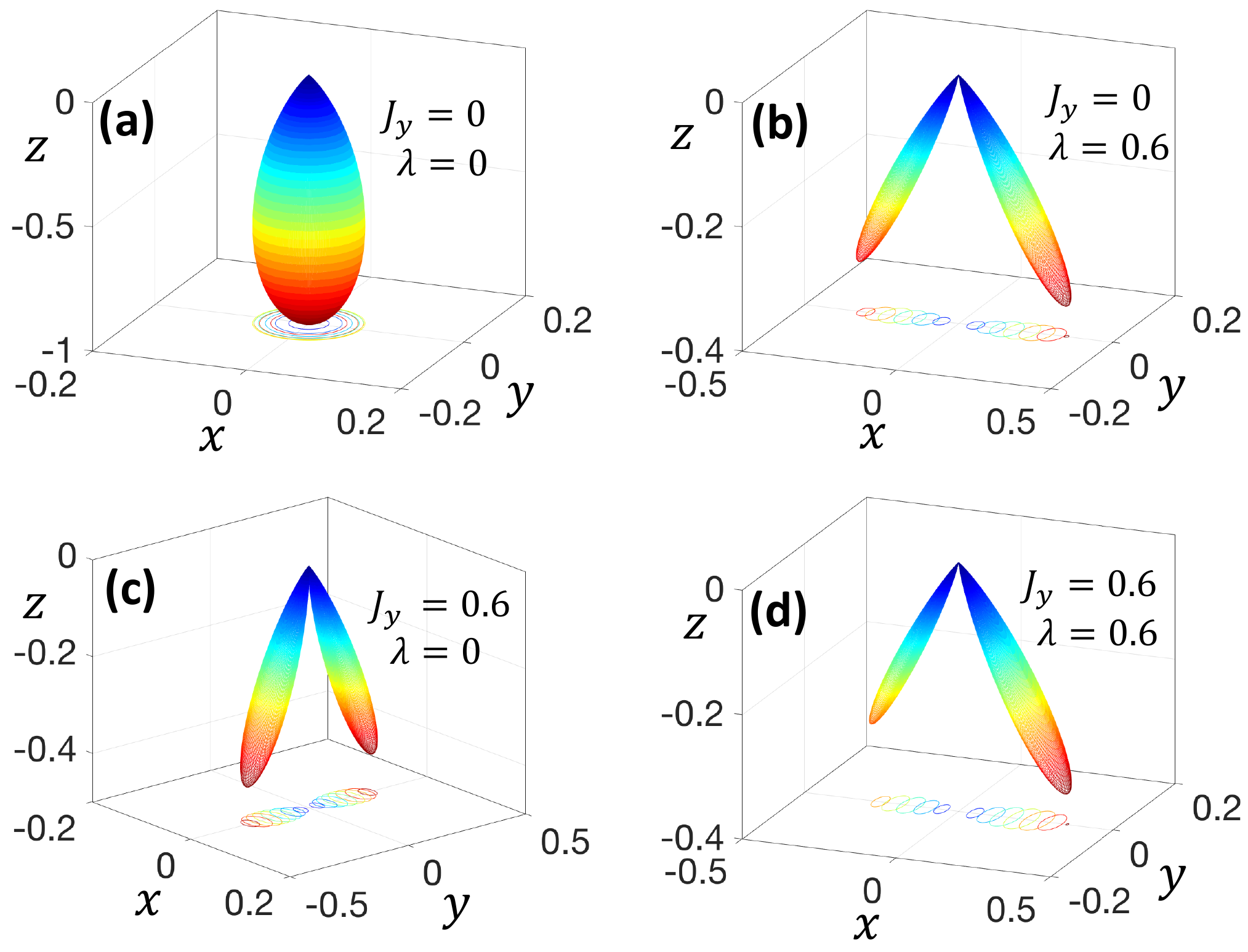}\caption{\label{fig:QFSpin_DLMGy}The Husimi $Q(\theta,\phi)$-function of the ground states of the spins for different phases. Here, the surface is obtained by transferring the spherical coordinates $(r=Q(\theta,\phi),\theta,\phi)$ to the corresponding Cartesian coordinates $(x,y,z)$. Panel (a-c) display
the $Q$-functions for the PN phase, FS phase, and FN phase, respectively. In panel (d), the strong spin-boson coupling $\lambda$ suppresses the FN phase and the system transitions back to the FS phase with the similar $Q$-function in panel (b). The other parameters are set as $\epsilon=1$, the spin number $N=80$, and the cut-off of the bosonic mode $80$.}
\end{figure}

To show how the bosonic $Q$-function changes with the parameters, we plot the cross section of $Q(\alpha)$ in the real-$\alpha$ plane in Fig~\ref{fig:QFBoson_DLMGy2D}. In panel (a), we see that the height of $Q(\alpha)$ in the PN phase decreases with the spin-boson coupling $\lambda$ accompanied by increasing width. This indicates that, for finite spin number $N$, the ground state of the PN phase will deviate from the vacuum state slightly and excitation in the bosonic mode increases with spin-boson coupling $\lambda$. However, the excitation number is not large enough to split $Q(\alpha)$ into two peaks. In panel (b), we see that when $\lambda$ passes the phase transition point $\lambda_{c,{\rm II}}$, the strong spin-boson coupling begins
to split $Q(\alpha)$ into two peaks and the separation between the two peaks increases with $\lambda$. In panel (c), the red line with $J_{y}=0$ exactly coincides with the blue line with $J_{y}=0.6$. This indicates that the spin-spin coupling $J_{y}$ will not affect the state of the bosonic mode before it passes the FS-FN boundary. Only after the QPT from the FS phase to the FN phase, $Q(\alpha)$ suddenly collapses to the vacuum state $Q$-function with a single peak at the origin. In panel (d), we verify the validity of the assumed ground states in the mean-field theory. The red line and gray lines are the $Q$-functions of the numerical ground state of the PN phase and the $Q$-function of the analytic mean-field vacuum state, respectively. The blue and the green lines are the $Q$-functions of the numerical ground state of FS phase and the $Q$-function of the mixed coherent state $(|-\sqrt{N}\alpha_{0}\rangle\langle-\sqrt{N}\alpha_{0}|+|\sqrt{N}\alpha_{0}\rangle\langle\sqrt{N}\alpha_{0}|)/2$ from the mean field theory. The numerical simulations coincide very well with the analytic results. This verifies the validity of the mean-field theory.

The $Q$-function of the spins in different phases are displayed in Fig.~\ref{fig:QFSpin_DLMGy}. In the PN phase, the ground state of the spins is $\left|\pi,\phi_{0}\right\rangle $ (i.e., the Dicke state $\left|N/2,-N/2\right\rangle $) with undetermined $\phi_{0}$. As shown in panel (a), the normalized spin $Q$-function is a cigar-like structure lying along the negative $z$-axis. In the FS phase, the strong spin-boson coupling along $x$-axis splits the $Q$-function into two cigars in the $xz$-plane as shown in panel (b). These two branches correspond to the two degenerate ground states with azimuth angles $\phi=0$ and $\phi=\pi$, respectively. In the FN phase as shown in panel (c), the strong spin-spin coupling along $y$-axis splits the $Q$-function into two cigars in the $yz$-plane, corresponding to the two degenerate states with azimuth angles $\phi=\pi/2$ and $\phi=3\pi/2$, respectively. In panel (d), we increase the spin-boson coupling $\lambda$ across the FN-FS phase boundary. The system goes back to the FS phase and the spin $Q$-function rotates back to the $xz$-plane as the one in panel (b).

In summary, we revealed the first-order as well as the second-order QPTs in the Dicke-LMGy model via the OPs $\zeta_{S}$, $\zeta_{M,x}$ and $\zeta_{M,y}$. Utilizing the diverging sensitivity $\chi(\lambda)$ at the phase transition point, we verified the QPT between the FN phase and the FS phase is indeed of first-order. This first-order QPT lays the foundation for the QCD demonstrated in the following. We also showed that the maximum of $\chi(\lambda)$ diverges with the spin number with speed $N^{2}$. We explicitly demonstrated the fundamental changes within the ground state wave functions in the QPTs via the bosonic and spin Husimi $Q$-functions.

In this section, by splitting the Dicke-LMG model (\ref{eq:H_full1}) into three sub-models, we studied the ground states of different quantum phases and their involved QPTs in detail. A full physical picture of the Dicke-LMG model can be constructed. In the following, we will show that the giant sensitivity existing in the first-order QPT can be utilized for quantum critical amplification. We also propose a class of biased detectors with an amplification scheme that exploits the quantum singularity in first-order QPTs.

\section{Dynamical Quantum Critical Amplification \label{sec:dynamical_amp}}
In Sec.~\ref{sec:QCD}, we explained the critical amplification scheme and the output signal of a general QCD. In this section, we take an explicit QCD described by the Dicke-LMGy model to demonstrate the dynamical quantum critical amplification process. We also show the dynamical change in the wave function of the detector to reveal microscopic variation during the amplification. 

In our QCD, the weak input signal to be measured functions as a control, similar to a single-photon detector~\cite{eisaman2011invited}. The transduction process of the detector is modeled as a weak input signal induced time-dependent variation of the spin-boson coupling strength $\lambda(t)=\lambda_{0}+\Delta\lambda\times P_{e}(t)$. Based on the phase diagram of the Dicke-LMGy model obtained in the previous section, we bias the spin-boson coupling $\lambda_0$ very close to the phase transition point $\lambda_{c,{\rm I}}=\sqrt{J_y/2}$. Thus, even a very small parameter variation $\Delta\lambda$ (amplitude) can trigger a first-order QPT from the FN phase to FS phase. The time-dependent variation of the spin-boson coupling $\Delta\lambda\times P(t)$ is assumed to proportional to the transduction probability~$P(t)$~\cite{yang2018concept}. The amplified output signal of the QCD is the macroscopic excitations in the bosonic mode. After the amplification, the bosonic mode evolves from the initial vacuum state to a coherent-like state with macroscopic excitations. 

We emphasize that the dynamical behavior of the first-order QPT system around the phase transition point determines whether the quantum singularity can be utilized as a resource for quantum amplification. The giant sensitivity of a first-order QPT shown in the previous section only exists in a transition between the ground states of two neighboring quantum phases, which cannot be connected via an adiabatic evolution~\cite{dziarmaga2010dynamics}. In our detector, the quantum critical amplification is realized by varying the spin-boson coupling across the phase boundary to trigger the first-order QPT. However, starting from the ground state of the FN phase, the detector can alternatively evolve to an arbitrary excited state instead of going to the ground state of the FS phase thereby completely degrading the critical amplification. In this section, to identify the existence of the quantum critical amplification during a dynamical process, we show the dynamics of the first-order QPT in our detector model with 80 spins via direct numerical time-evolution. We show that a linear scaling in the quantum gain and the SQNR of the QCD is obtained, instead of the $N^2$ sensitivity of the first-order QPT.

\begin{figure}
\includegraphics[width=8.5cm]{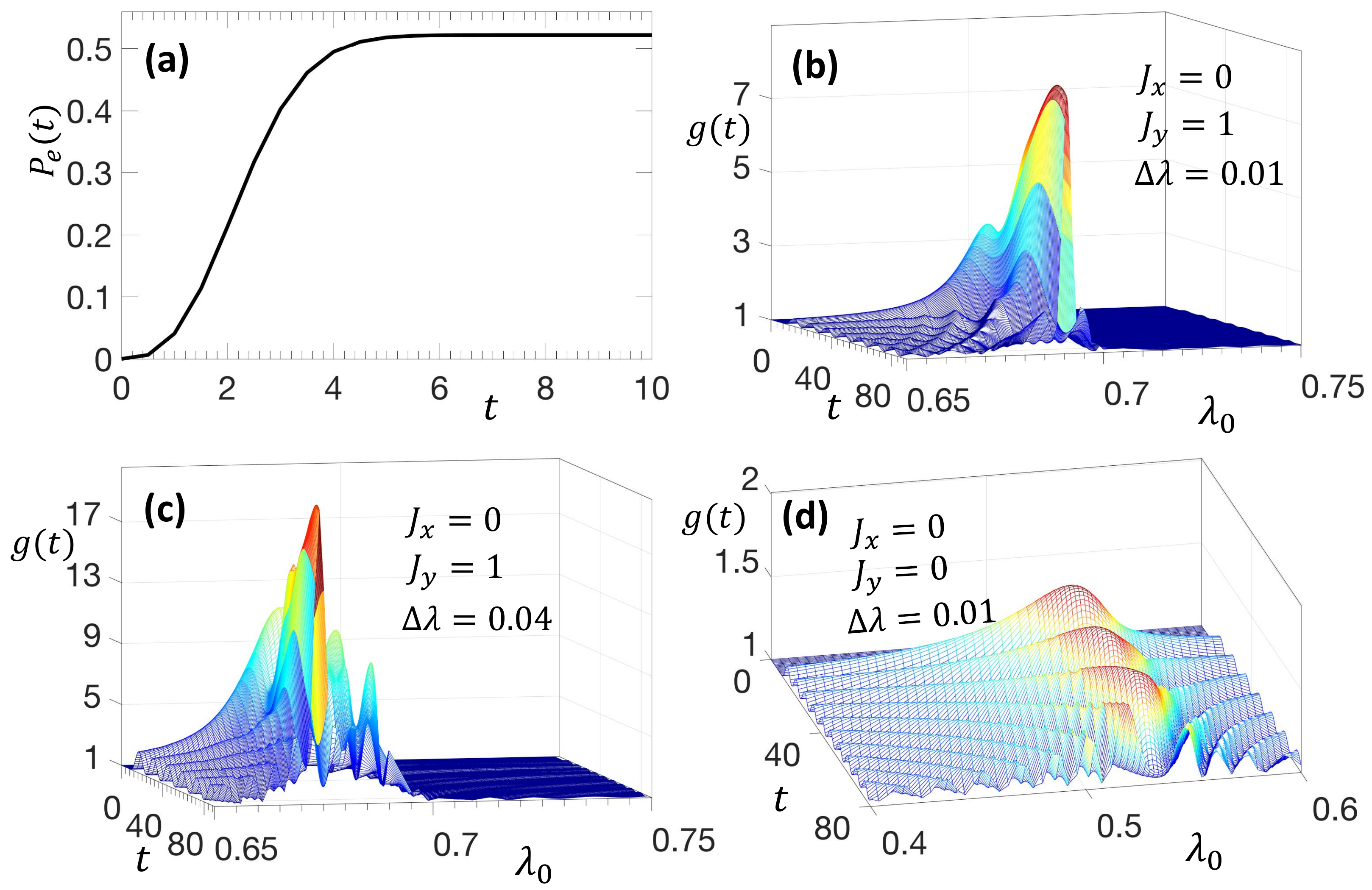}
\caption{\label{fig:dynamic_QPT}Demonstration of the first-order dynamical quantum phase transition via the time-dependent quantum
gain $g(t)=\langle\hat{d}^{\dagger}(t)\hat{d}(t)\rangle/\langle\hat{d}^{\dagger}(0)\hat{d}(0)\rangle$. In panel (a), we show the envelope $P_{e}(t)$ of the time-dependent parameter we used in this paper.
In panel (b), we show the dynamics of the quantum gain contributed from a first-order quantum phase transition (QPT). Here, the spin-spin coupling is fixed at $J_{y}=1>J_{yc,{\rm II}}$ and the spin-boson coupling varies with time $\lambda=\lambda_{0}+\Delta\lambda\times P_{e}(t)$. The amplitude of the small change in the parameter $\lambda$ is set to be $\Delta\lambda=0.01$ and the time-dependent envelope $P_{e}(t)$ is given in panel (a). In Panel (c), we increase the amplitude $\Delta\lambda$ to $0.04$. In panel (d), we show the time-dependent gain yielded by a second-order QPT with $J_y<J_{yc, {\rm II}}$. In this figure, the other parameters are taken as $\epsilon=1$, $J_{x}=0$, both the spin number $N$ and the bosonic mode cutoff the bosonic mode are $40$, and the time is in units of $1/\omega_{0}$.}
\end{figure}

\subsection{Amplification via First-Order Quantum Phase Transition \label{sec:dynamical_amp_A}}
We now introduce two quantities to characterize the performance of a QCD. The first one is the quantum gain (i.e., the amplification factor)
\begin{equation}
g(t)=\frac{\langle\psi(t)|\hat{d}^{\dagger}\hat{d}|\psi(t)\rangle}{\langle\psi(0)|\hat{d}^{\dagger}\hat{d}|\psi(0)\rangle},
\end{equation}
where $\psi(0)$ is the ground state of initial Hamiltonian $H(0)$ with bias $\lambda=\lambda_{0}$ and $\psi(t)$ is the wave function of the detector at time $t$. As shown later, the quantum gain can be used as a unique characteristic of first-order QPTs. To characterize
the quantum noise in our QCD, we define the signal-to-quantum noise ratio (SQNR) as~\cite{yuen1976states}
\begin{equation}
{\rm SQNR}=\langle\hat{d}^{\dagger}(t)\hat{d}(t)\rangle^{2}/\langle[\Delta\hat{d}^{\dagger}(t)\hat{d}(t)]^{2}\rangle,
\end{equation}
where $\langle[\Delta\hat{d}^{\dagger}(t)\hat{d}(t)]^{2}\rangle=\langle[\hat{d}^{\dagger}(t)\hat{d}(t)]^{2}\rangle-\langle\hat{d}^{\dagger}(t)\hat{d}(t)\rangle^{2}$
is the variance of the bosonic excitation number operator. 

The full dynamics of the whole system is governed by the time-dependent Hamiltonian $H(t)$, which is formed by replacing the constant spin-boson coupling $\lambda$ with the time-dependent one $\lambda(t)$ in Eq.~(\ref{eq:H_DLMGy}). The exact procedure to perform the numerical time evolution is given in Appendix.~\ref{sec:numerical_simulation}. Alternatively, we can also use the time-dependent input signal to control the spin-spin coupling [$J_x(t)$ or $J_y(t)$] to realize the quantum amplification with fixed spin-boson coupling (data not shown).

The dynamical amplification in the QCD is demonstrated by the time-dependent gain as a function of the bias spin-boson coupling
$\lambda_{0}$ and time in Fig.~\ref{fig:dynamic_QPT}. From the time-dependent gain in Fig.~\ref{fig:dynamic_QPT} (b), we find that the efficient amplification triggered by a very small parameter change can only obtained if the system is biased very close to the phase transition point. In the simulation, the spin-spin coupling is taken as $J_y=1>J_{c,{\rm II}}$ to prepare for a first-order QPT. and the amplitude of the small change of the spin-boson coupling is set to be $\Delta\lambda=0.01$. In Fig.~\ref{fig:dynamic_QPT} (c), we increase the variation amplitude $\Delta\lambda$ to $0.04$. We see that the location of the large-gain peak shifts and its height increases. In Fig.~\ref{fig:dynamic_QPT} (d), we show that efficient amplification can not be obtained from second-order QPTs with spin-spin coupling $J_y=0$.

\begin{figure}
\centering
\includegraphics[width=8.5cm]{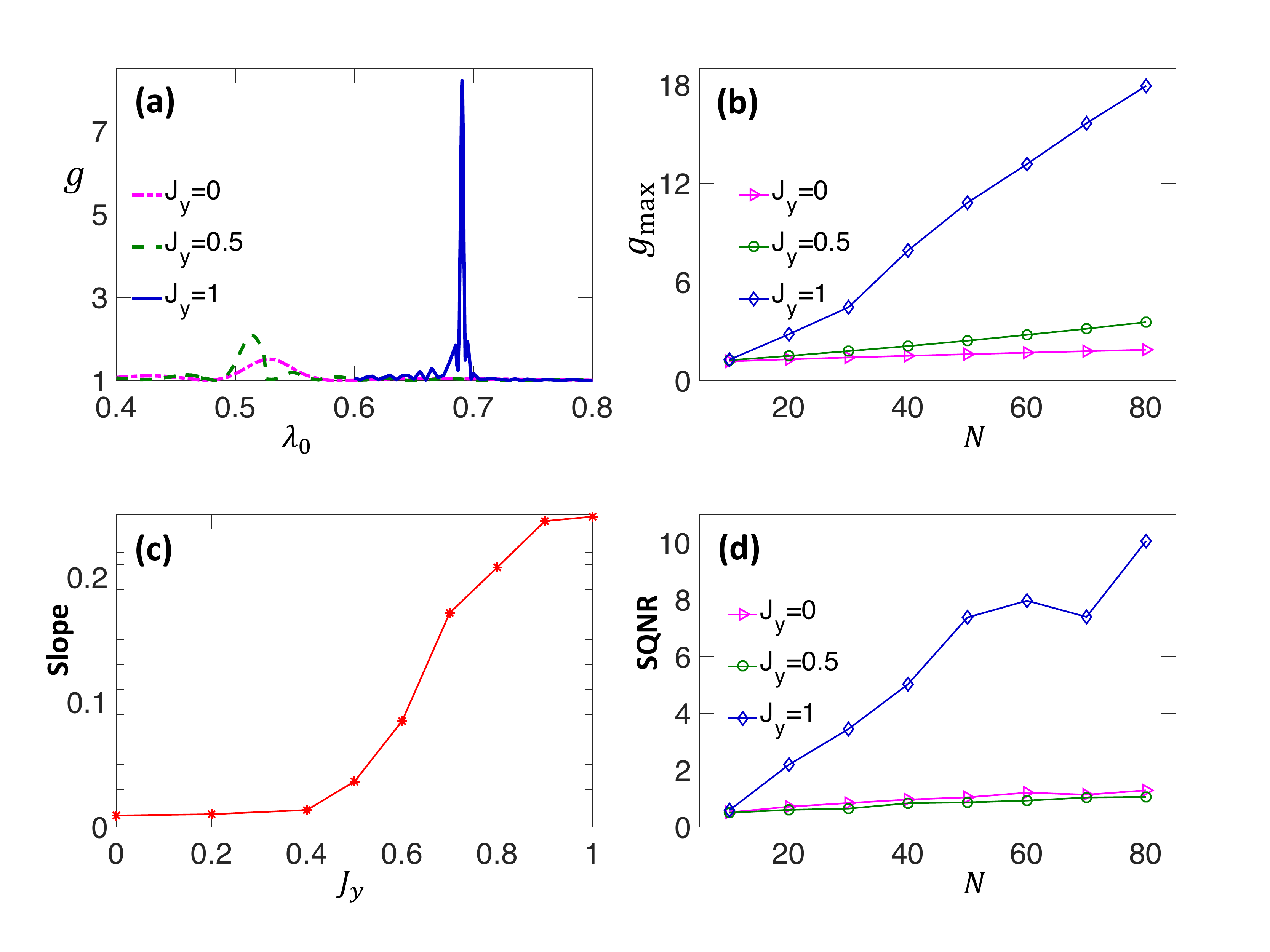}
\caption{\label{fig:Scaling}Contrast of the quantum gains $g$ from second-order and first-order quantum phase transitions (QPTs). In panel (a), we show the quantum  gain as a function of the bias spin-boson coupling $\lambda_0$ at the time when $g(t)$ reaches its first maximum. The pink and green lines with small spin-spin coupling $J_y\leq J_{yc,{\rm II}}$ are from second-order QPTs and the blue line with strong spin-spin coupling ($J_y>J_{yc,{\rm II}}$) represents the gain from the first-order QPT. Panel (b) shows the linear scaling of the maximum quantum gain with the spin number $N$ for different spin-spin coupling $J_{y}$. The corresponding signal-to-quantum noise ratios are shown in panel
(d). In panel (c), we show there is a phase-transition-like behavior
in the slope of the maximum gain when $J_{y}$ crosses the phase transition
point $J_{yc,{\rm II}}=0.5$.}
\end{figure}

We explicitly demonstrate the necessity of the first-order QPTs for dynamical quantum critical amplification in contrast with second-order QPTs. In Fig.~\ref{fig:Scaling} (a), we plot the gain as a function of the bias spin-boson coupling for different values of the spin-spin coupling $J_y$. We see that, for second-order QPTs with weak spin-spin coupling $J_y\leq J_{yc,{\rm II}}=\epsilon/2$ (the pink and green curves), there is only a low and flat peak in the quantum gain located around the second-order phase transition spin-boson coupling $\lambda_{c,{\rm II}}=\sqrt{\epsilon}/2$. But, for the first-order QPT with strong spin-spin coupling $J_y>J_{yc,{\rm II}}$ (the blue curve), a very sharp peak exists in the quantum gain around the first-order phase transition point $\lambda_{c,{\rm I}}=\sqrt{J/2}>\lambda_{c,{\rm II}}$. Thus, the first-order QPT is essential for amplification in our QCD. Here, we also see that the dynamics of the detector is highly sensitive to the initial bias $\lambda_0$. Similar to the enhanced decay of the Loschmidt echo by the criticality in a second-order QPT~\cite{quan2006decay}, the enhanced quantum gain in our QCD is a universal and unique characteristic of quantum singularity in a first-order QPT. 

To show the high figures of merit of our QCD, we present the scaling of the quantum amplification with the spin number $N$ in Fig.~\ref{fig:Scaling} (b-d). In panel (b), we see that the maximum gain is linearly proportional to $N$. We contrast the amplification resulting from the first-order QPT (the blue diamond line) and second-order phase transitions (the pink triangle line and the green circle line). The latter is negligibly small when compared with the first-order QPT. In panel (c), we derive the slope of the quantum gain~$dg_{{\rm max}}/dN$ as a function of the spin-spin coupling $J_y$. There exists a ``phase transition'' phenomenon in the slope at the same phase transition point $J_{yc,{\rm II}}$ of the transition from second-order to first-order QPTs. The corresponding SQNR for the three lines are displayed in panel (d). The amplification based on first-order QPT has much higher SQNR than that of second-order QPTs. Similar to the quantum gain (the rescaled excitation number in the output bosonic mode), the SQNR also increases linearly with the spin number. The SQNR of the final output state is consistent with the SQNR of a coherent-like state.

\begin{figure}
\includegraphics[width=8cm]{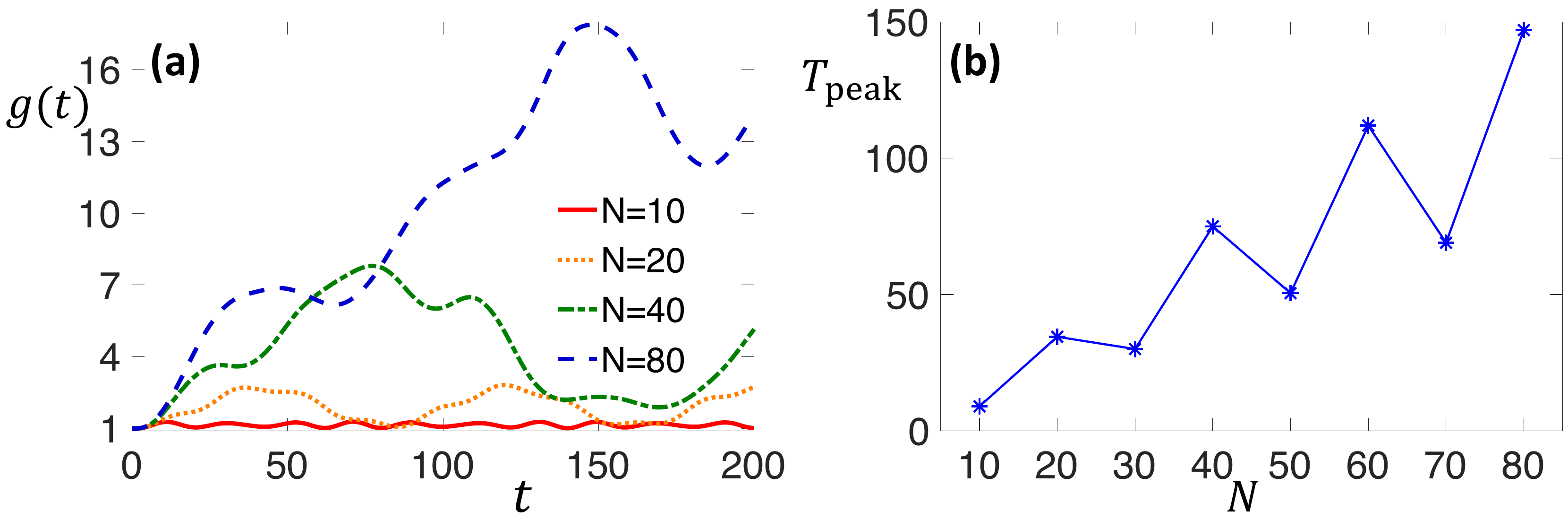}
\caption{\label{fig:dynamic_QPT_2D}The long-time dynamical behavior of the quantum gain.  In panel (a), we plot the quantum
gain $g(t)$ as a function of time with optimized bias spin-boson
coupling $\lambda_{0}$. Different lines denote different spin number
$N$. In panel (b), we show the time $T_{{\rm peak}}$ to reach the
first maximum of $g(t)$ for different values of the spin number $N$. The other
parameters in this figure are taken as $\epsilon=1$, $J_{x}=0$,
and $J_{y}=1$. }
\end{figure}

Let's analyze the long-time dynamical behavior of the detector. In Fig.~\ref{fig:dynamic_QPT_2D}(a), we plot the time-dependent quantum gain $g(t)$ at the optimized bias spin-boson coupling $\lambda_0$, which gives the maximum quantum gain. Here, different lines denote different spin numbers. We find that the quantum gain oscillates periodically as the whole system is closed. Starting from the FN phase, the detector swings to the FS phase and back. Furthermore, the time-evolution period is also dependent on the spin number $N$. In Fig.~\ref{fig:dynamic_QPT_2D}(b), we plot the time of the quantum gain $g(t)$ takes to reach its first maximum $T_{{\rm peak}}$ as a function of spin number. We see that $T_{{\rm peak}}$ is quasi-linearly proportional to $N$.

\subsection{Time Dynamics of The Wave Function}
For second-order QPTs, the substantial change only exists in the ground-state waves at the phase boundary. Thus, the Loschmidt echo, which is defined as the projection of the wave function on the initial state [see Eq(\ref{eq:LE})], is naturally selected to characterize the critical dynamics of the system. However, for first-order QPTs, the $Q$ function turns out be a more powerful tool to reveal the intrinsic changes within the detector during the dynamical critical amplification.  The numerical approach we introduced, which allows us to calculate the time-dependent Husimi $Q$-functions of both the bosonic mode and the spin ensemble,  holds a fundamental advantage for dynamical amplification and noise calculations.

\begin{figure}
\includegraphics[width=8.5cm]{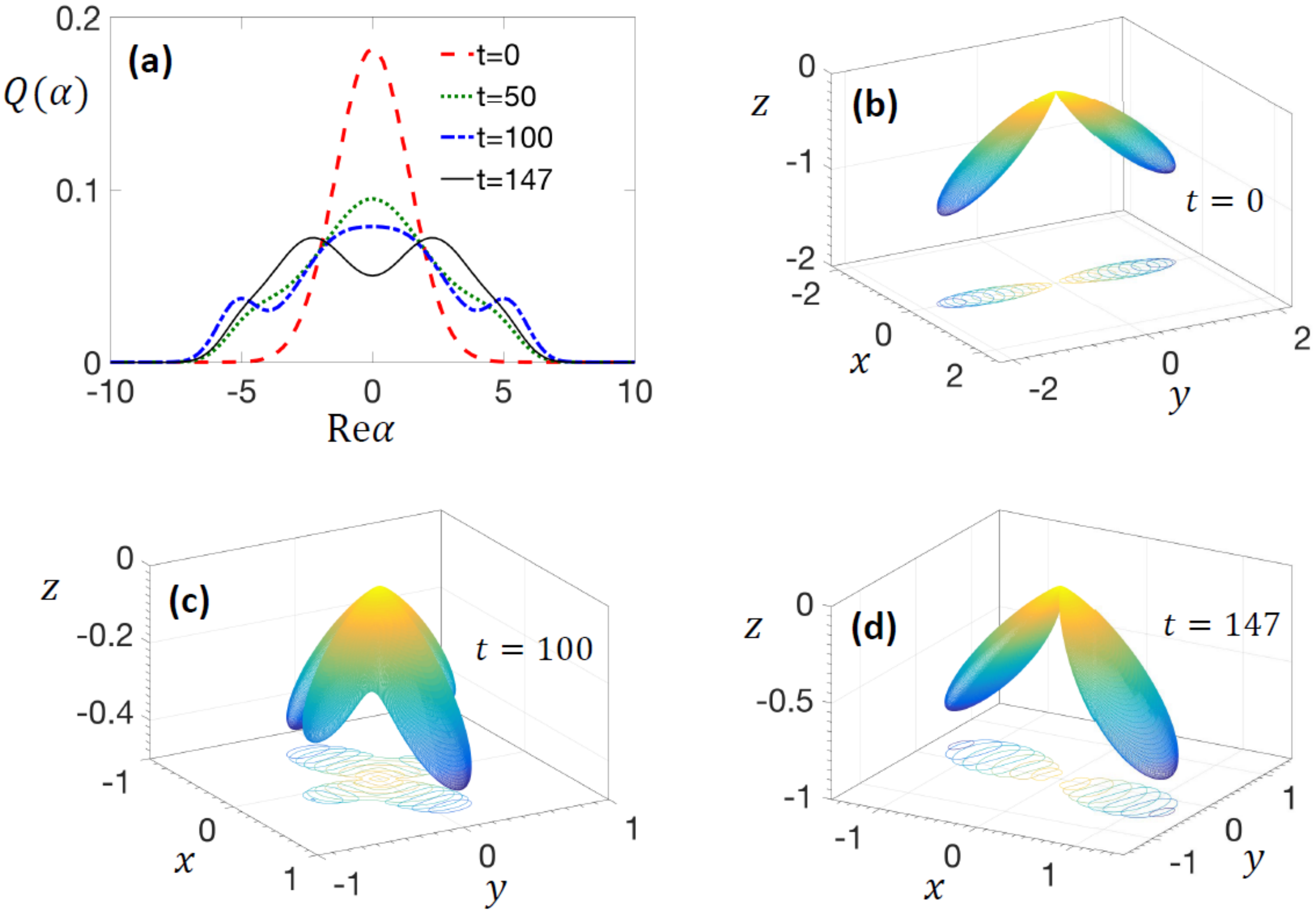}\caption{\label{fig:QF_Dynamics} Dynamics of the bosonic and spin $Q$-functions during the dynamical quantum phase transition. (a) The cross section of the bosonic $Q$-function in the real plane at different time. (b-d) The spin $Q$-function at different time. The parameters in this figure are taken as: the energy splitting of the spin $\epsilon=1$, the spin-spin coupling in $x$-direction $J_{x}=0$, the spin-spin coupling in $y$-direction $J_{y}=1$, spin number $N=80$, the cutoff the bosonic mode $80$, and the variation amplitude $\Delta \lambda=0.01$. The spin-boson coupling $\lambda_{0}$ is biased at the optimal spin-boson coupling in Fig.~\ref{fig:dynamic_QPT} (b).}
\end{figure}

We first look at the $Q$-function of the bosonic mode, which reveals the microscopic change in the quasi-distribution of the output mode.  In Fig.~\ref{fig:QF_Dynamics}  (a), we only show the cross section of the $Q$-function in the real plane at different time $t$, as the essential variation only occurs on the real axis. Initially, the system is in the FN phase and the bosonic mode is in an extremely low-excitation state. The corresponding $Q$-fucntion has a single peak at the origin. Then, the height of this peak begins to decease (green-dotted line). Finally, the central peak splits into two separated peaks (black line), which means the bosonic mode has been excited to a coherent-like state with large excitation numbers. The dynamics of the bosonic $Q$-function shows the transition from the normal phase to the superradiant phase. The macroscopic excitation in the bosonic mode functions as the amplified output signal of the QCD, which can be read out directly with a classical device. 

The dynamical transition of the spins from the FM-Y phase to the FM-X is displayed by the spin $Q$-function in Fig.~\ref{fig:QF_Dynamics} (b-c). Here, to show the dynamical change in the distribution function, we do not normalize the spin $Q$-function. Initially, the spins is in the FM-Y phase, thus the spins are polarized in the $yz$-plane as illustrated in panel (b). The two cigar-like structures correspond to the two degenerate ground states of the FM-Y phase. As shown in panel (c), two new branches in the $xz$-plane appear and swell with time. At the same time, the two branches in the $yz$-plane shrink and finally disappear [see panel (d)]. During the transition from the FM-Y phase to the FM-X pahse, the spin-noise in the $y$ axis decreases, but the the spin-noise in the $x$-direction increases. The dynamical change in the spin-fluctuations can also be probed experimentally through spin noise spectroscopy~\cite{zapasskii2013spin}.

\begin{figure}
\includegraphics[width=8.5cm]{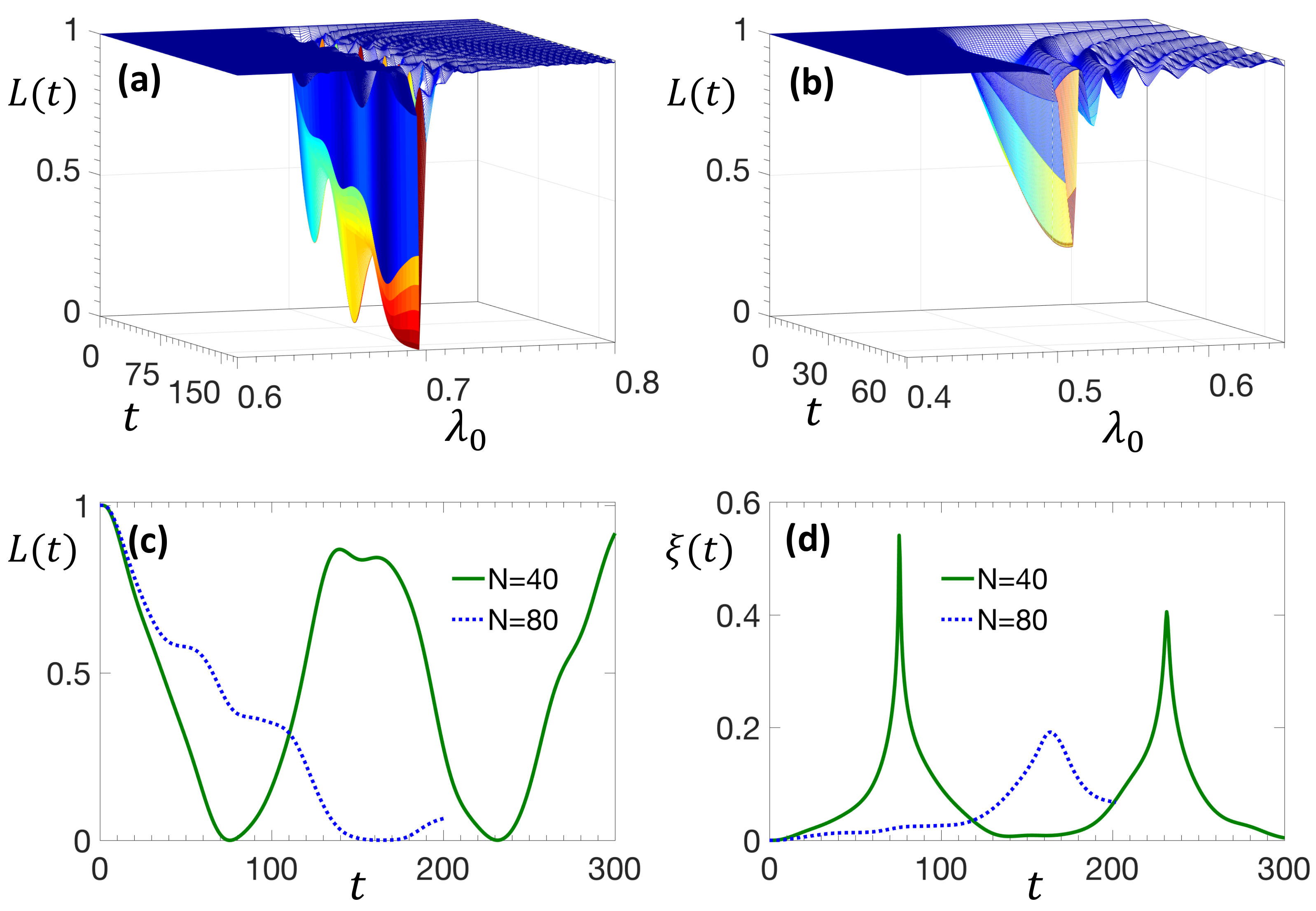}
\caption{\label{fig:LE_RF_dynamics} Dynamics of the Loschmidt echo $L(t)$ and the rate function $\xi (t)$. In panels (a) and (b) we show the enhanced decay of the LE during the first-order and second-order dynamical quantum phase transitions, respectively. In panels (c) and (d), we show the dynamics of the Loschmidt echo and the corresponding rate function of the first-order QPT. Here, the spin-boson coupling is biased at the optimal point, which leads to the maximum quantum gain. The other parameters in this figure are taken as: the energy splitting of the spin $\epsilon=1$, the spin-spin coupling in $x$-direction $J_{x}=0$, and the variation amplitude of the spin-boson coupling $\Delta \lambda=0.01$.}
\end{figure}

One of central concepts in wave function dynamics proposed in the previous literature is the Loschmidt echo,
\begin{equation}
L(t)\equiv |\langle\psi(0)|\psi(t)\rangle|^{2}, \label{eq:LE}
\end{equation}
where $|\psi(t)\rangle$ is the wave function of the system at time $t$. The Loschmidt echo describes the deviation of the wave function from the initial state. It was first introduced to characterize the decoherence of a quantum system as a correspondent of the classical chaotic system~\cite{Jalabert2001environ,Karkuszewski2002quantum,Cucchietti2003decoh}. In 2006, Quan \textit{et al.} proposed measuring the Loschmidt echo to show the quantum criticality of second-order QPT in a transverse Ising chain~\cite{quan2006decay}. Here, we can also use the Loschmidt echo to show the quantum singularity during a first-order QPT. From Fig.~\ref{fig:LE_RF_dynamics} (a), we see  that the decay of the Loschmidt echo is greatly enhanced around the first-order phase transition point $\lambda_{c,{\rm I}}$. This enhancement also exists in the second-order QPT in our detector model as shown in \ref{fig:LE_RF_dynamics} (b). Thus, the enhanced decay in the Loschmidt echo is a universal characteristic of a QPT but not unique for first-order QPTs. In contrast, the enhanced quantum gain around the first-order phase transition point shown in the previous subsection is the unique and universal characteristic of first-order QPTs.  

Based on the Loschmidt echo, one can define a rate function
\begin{equation}
\xi(t)\equiv-\frac{1}{N}\log L(t),
\end{equation}
to characterize the exponentially decay scaling of $L(t)$. In 2013, Heyl \textit{et. al.} found there exist non-analytic kinks in the rate function $\xi(t)$ and denoted these kinks as a universal behavior and a standard sign of a "quantum phase transition" occurring in the time domain---the dynamical QPT~\cite{Heyl2013Dynamical,heyl2018dynamical}. Direct observation of these kinks has also been realized in a trapped-ion quantum simulator~\cite{Jurcevic2017direct}. In Fig.~\ref{fig:LE_RF_dynamics}(c) and (d), we show the decay of the  Loschmidt echo and the corresponding rate function respectively. The green and blue curves describe the case with spin number $N=40$ and $N=80$ respectively. From panel (c), we see that the LE collapses and revives periodically. The revival period increases with spin number. In panel (d), we observe the non-analytical kinks in the rate function $\xi (t)$ during the first-order QPT. The kinking time exactly coincides with the time that the Loschmidt echo $L(t)$ reaches its minimum. Thus, the kinks of the rate function results from the dips (theoretically should be the zero-value points) of the Loschmidt echo $L(t)$. We emphasize that the wave function resulting in these non-analytical kinks is not the ground-state of the system in another phase. Converse to the $Q$-function, we also see that the Loschmidt echo cannot reveal the intrinsic change within the detector during the dynamical critical amplification.

\section{Experimental Implementation\label{sec:exp_realization}}

In the last decade, numerous quantum simulators have been demonstrated as powerful tools for modeling strongly correlated many-body systems. One of the most important applications of these simulators is to demonstrate the QPTs and the corresponding QPTs existing in quantum matter. Especially, the well-known Ising-type QPTs have been successfully simulated with Rydberg atoms~\cite{labuhn2016tunable,bernien2017probing}, trapped ions~\cite{monz2001quibts,islam2013emergence,zhang2017observation}, and superconducting qubits~\cite{song2017Super,Harris2018Phase}. In this section, we will propose possible experimental platforms for the first-order QPT in the Dicke-LMG model and the potential implementation of our QCD.

The Dicke superradiant phase transition was first demonstrated experimentally by Baumann \textit{et al.} in a system formed by a Bose-Einstein condensate coupled to an optical cavity~\cite{baumann2010dicke}. The main challenge to observe the superradiant QPT in experiment arises from the strong light-atom interaction (i.e., the spin-boson coupling in our model) required by the phase transition condition. The phase transition light-atom interaction to realize the superradiant phase is of the same order as the atomic transition frequency, which is extremely difficult to achieve. For natural atoms, this strong light-atom coupling is even forbidden by the TRK sum rule~\cite{rzazewski1975phase,rzazewski1979nogo}. One of the methods to circumvent this challenge is to construct an effective two-level-atom ensemble with significantly reduced transition frequency. This method was first proposed by Dimer \textit{et al.}~\cite{Dimer2007proposed}, in which the two working levels are composed of the two ground states of a four-level atom. The transition frequency and the atom-field coupling of the effective two-level atoms are determined by light-induced frequency shifts and Raman transition rates, respectively. In the experiment~\cite{baumann2010dicke}, a similar ideal has been applied to overcome the strong-coupling challenge. The effective two-level system is constructed with two orbital states with zero and finite momentum $k$, respectively. The energy splitting between the two effective states $\epsilon=2\omega_r$ is given by the recoil frequency $\omega_r=k^2/2m$ ($m$ is the mass of the atom). The coupling $\lambda$ between the effective two-level-atom ensemble and the output cavity mode is controlled by an extra pump laser field. When the field-atom coupling $\lambda$ exceeds the phase transition coupling, both the macroscopic excitation in the output mode and the long-range-order pattern in the atoms were observed.

The second-order QPT in the LMG model has only been demonstrated in a recent experiment with Dysprosium atoms (spin number $N=16$)~\cite{makhalov2019probing}. Here, the biggest challenge lies in achieving the long-range homogeneous coupling between the spins. The short-range or medium-range coupling between spins (or atoms) has been utilized to demonstrate the QPTs in an Ising-type model, such as the van der Waals interaction between Rydberg atoms~\cite{bernien2017probing,browaeys2016experimental}, Raman transition induced spin-spin interaction in trapped-ion system~\cite{zhang2017observation,molmer1999multi}, SQUID coupler mediated coupling between superconducting qubits~\cite{Harris2018Phase,van2005mediated}, etc. But these couplings decrease with the distance $r$ between the spins with the power law~$r^{-\alpha}$ ($\alpha\sim 1-6$). In the last two decades, several theoretical schemes have been proposed to realize the long-range homogeneous coupling in the LMG model. In the first scheme, the long-range homogeneous coupling is accomplished by eliminating the auxiliary bosonic mode in a driven Dicke model~\cite{Morrison2008collective}, which can be implemented in a cavity quantum electrodynamics (QED) setup~\cite{Morrison2008Dynamical}, BEC system~\cite{chen2009interaction}, and circuit-QED system~\cite{larson2010circuit}. Similar method has been utilized in the experiment in Ref~\cite{makhalov2019probing}. In another scheme~\cite{Unanyan2003decoh,molmer1999multi}, the necessary coupling between trapped ions is mediated by two pump lasers in the M\o lmer-S\o rensen configuration~\cite{sorenson1999quantum}. In the Lamb-Dicke regime, the coupling strength given by the resonant second-order process is insensitive to the vibration states. Recently, an important breakthrough in experiment has been attained. The long-range spin exchange interaction mediated by an optical cavity has been achieved with $10^5$ laser-cooled $^{87} \rm{Sr}$ atoms~\cite{Norcia2018cavity}.

Due to the required strong long-range spin-spin coupling between a large number of spins ($N>30$ spins are required to observe the first-order QPT), it is quite challenging to experimentally demonstrate a QCD based on the first-order QPT in the Dicke-LMG model. However, we note that a QCD based on the first-order QPT in the Ising XY model (with short-range interaction) predicted in Sec.~\ref{Sec:phase_diagram} can be demonstrated in the near future. Recently, the XY-type coupling has been realized with superconducting qubits~\cite{Dalmonte2015realizing}
and Rydberg atoms~\cite{lulu2017optical}. The significant advance in integration of superconducting qubit~\cite{houck2012chip} and trapping and fast assembling cold atom~\cite{endres2016atom} enables engineering of scalable quantum simulators. These quantum simulators form solid platforms for implementation of QCDs.

\section{Summary \label{sec:Conclusion}}
We present a new paradigm in weak-signal detection---the QCD, which implements an amplification scheme by exploiting the singularity of first-order QPTs. After transduction (absorption), the input weak signal induces a minor change in the detector parameter. The detector is pre-biased very close to the phase transition point of a first-order QPT. Thus, this small variation can trigger a QPT resulting in a macroscopic change in the order parameter of the detector. This macroscopic change functions as the output signal of the detector. 

We use a specific detector model---the Dicke-LMG model---to explain the working mechanism of a QCD explicitly. We introduce two magnetic order parameters to characterize the first-order QPT between two ferromagnetic phases. Specifically, we predict a universal first-order QPT in the interacting-spin system resulting from the competition of two different long-range spin orders. We show that the no-go theorem, which rules out the existence of the superradiant QPT, can be circumvented with spin-spin (atom-atom) interaction. We introduce the spin $Q$-function to display the macroscopic spin order and to show the fundamental change in the ground-state wave function during QPTs. We contrast the behaviors of the first-order and second-order QPTs in the detector. We show that first-order QPTs are essential for QCDs.

We define the quantum gain and SQNR to show the figures of merit of a QCD. We also utilize the time-dependent $Q$ function to show the intrinsic change within the detector during the dynamical quantum critical amplification. We found the decay of the Loschmidt echo will be enhanced at the phase transition points of both first- and second-order QPTs. However, only the enhanced quantum gain around the phase transition point can be utilized as the unique and universal measurable characteristic of a first-order QPT. We also show the non-analytical kinks in the rate function (exponent of the Loschmidt echo), which occurs at the time when the quantum gain of the detector reaches its maximum. 

In recent experiments, second-order QPTs, especially the Ising-type QPTs, have been successfully demonstrated with trapped ions~\cite{zhang2017observation}, cold atoms~\cite{bernien2017probing} and circuit QED~\cite{Harris2018Phase}. The recent advances in laser trapping and assembling of cold atoms~\cite{endres2016atom} shows that a quantum simulator with $>100$ spins can be built in the very near future. These systems form excellent platforms for the physical realization of our proposed QCD, especially the ones based on the first-order QPT in the Ising XY model.

\section*{acknowledgment}
The authors thank H.-T. Quan for for fruitful discussions. This work is supported by the DARPA DETECT ARO award (W911NF-18-1-0074). 

\appendix

\section{Quantum phases via Mean-Field Theorem\label{sec:mean-field}}

In this appendix, we provide the details to obtain the the phase
diagram via the mean field theory. The ground state of the whole system is given by the solutions of the equilibrium equations. For convenience, we first re-express
the equilibrium equations ~(\ref{eq:Derivative1}-\ref{eq:Derivative3})
as,
\begin{align}
\alpha & =-\lambda\sin\theta\cos\phi,\\
0 & \!=\!\left\{ \!\epsilon\!+\!2[(2\lambda^{2}\!+\!J_{x})\cos^{2}\phi\!+\!J_{y}\sin^{2}\phi]\cos\theta\!\right\} \!\sin\theta\!,\\
0 & =[2\lambda^{2}+J_{x}-J_{y}]\sin^{2}\theta\sin\phi\cos\phi.
\end{align}
As shown in the following, there exist three types of solutions corresponding
to three different quantum phases. 

The ground-state stability is determined by the $3\times3$ Hessian matrix
\begin{align}
\mathcal{M} & =\left[\begin{array}{ccc}
\frac{\partial^{2}E}{\partial\alpha^{2}} & \frac{\partial^{2}E}{\partial\alpha\partial\theta} & \frac{\partial^{2}E}{\partial\alpha\partial\phi}\\
\frac{\partial^{2}E}{\partial\theta\partial\alpha} & \frac{\partial^{2}E}{\partial\theta^{2}} & \frac{\partial^{2}E}{\partial\theta\partial\phi}\\
\frac{\partial^{2}E}{\partial\phi\partial\alpha} & \frac{\partial^{2}E}{\partial\phi\partial\theta} & \frac{\partial^{2}E}{\partial\phi^{2}}
\end{array}\right],
\end{align}
where the elements are given by
\[
\frac{\partial^{2}E}{\partial\alpha^{2}}=2,
\]
\[
\frac{\partial^{2}E}{\partial\theta^{2}}\!\!=\!\!-\frac{\epsilon}{2}\!\cos\theta-2\lambda\alpha\!\sin\theta\!\cos\phi-\!(\!J_{x}\!\cos^{2}\!\!\phi\!+\!J_{y}\!\sin^{2}\!\!\phi\!)\!\cos\!2\theta\!,
\]
\[
\frac{\partial^{2}E}{\partial\phi^{2}}=-2\lambda\alpha\sin\theta\cos\phi+(J_{x}-J_{y})\sin^{2}\theta\cos2\phi,
\]
\[
\frac{\partial^{2}E}{\partial\alpha\partial\theta}=\frac{\partial^{2}E}{\partial\theta\partial\alpha}=2\lambda\cos\theta\cos\phi,
\]
\[
\frac{\partial^{2}E}{\partial\alpha\partial\phi}=\frac{\partial^{2}E}{\partial\phi\partial\alpha}-2\lambda\sin\theta\sin\phi,
\]
\[
\frac{\partial^{2}E}{\partial\theta\partial\phi}\!\!=\!\!\frac{\partial^{2}E}{\partial\phi\partial\theta}\!\!=\!-2\lambda\alpha\cos\theta\sin\phi+\!\frac{1}{2}(\!J_{x}\!-\!J_{y}\!)\sin2\theta\sin2\phi.
\]
The boundaries between these quantum phases are finally determined by requirement of the positive definiteness of $\mathcal{M}$.

\textbf{(i) Paramagnetic-Normal phase}\textemdash{} The simplest solution
of the equilibrium equations is given by,
\begin{align}
\sin\theta_{0} & =0,\\
\cos\theta_{0} & =-1,\\
\alpha_{0} & =0.
\end{align}
The azimuthal angle $\phi_{0}$ is fully undetermined. Utilizing
the solution, the Hessian matrix is simplified with nonzero elements,
\begin{align*}
\mathcal{M}_{11} & =2,\\
\mathcal{M}_{22} & =\frac{\epsilon}{2}-J_{x}\cos^{2}\phi_{0}-J_{y}\sin^{2}\phi_{0},\\
\mathcal{M}_{12} & =\mathcal{M}_{21}=-2\lambda\cos\phi_{0}.
\end{align*}
The positive definiteness of $\mathcal{M}$ requires
\begin{equation}
\epsilon\geq(4\lambda^{2}+2J_{x})\cos^{2}\phi_{0}+2J_{y}\sin^{2}\phi_{0}.
\end{equation}
As $\phi_{0}$ is fully undetermined, thus the stability conditions
for this solution require $\epsilon>(4\lambda^{2}+2J_{x})$ and $\epsilon>2J_{y}$. 

For this solution, the OPs are given by,

\begin{align}
\frac{\langle\hat{a}^{\dagger}\hat{a}\rangle_{0}}{N} & =0,\ \frac{\langle\hat{S}_{x}^{2}\rangle_{0}}{N^{2}}=0,\ \frac{\langle\hat{S}_{y}^{2}\rangle_{0}}{N^{2}}=0,
\end{align}
and $M_{z}=-1/2$. In this case, all the spins are polarized along
the negative $z$-axis and there are no macroscopic excitations in
the bosonic mode. Thus, this solution gives the PN phase.

\textbf{(ii) Ferromagnetic-Normal phase}\textemdash{} The second solution
is
\begin{align}
\cos\theta_{0} & =-\frac{\epsilon}{2J_{y}},\label{eq:theta_LMGy}\\
\cos\phi_{0} & =0,\\
\alpha_{0} & =0,
\end{align}
requiring $2|J_{y}|\geq\epsilon$. However, the negative branch $J_{y}<0$
will be ruled out by the stablity condition later. Utilizing the solution,
the Hessian matrix is simplified with nonzero elements,
\begin{align*}
\mathcal{M}_{11} & =2,\\
\mathcal{M}_{22} & =-\frac{\epsilon}{2}\cos\theta_{0}-J_{y}\cos2\theta_{0},\\
\mathcal{M}_{33} & =(J_{y}-J_{x})\sin^{2}\theta_{0},\\
\mathcal{M}_{13} & =\mathcal{M}_{31}=-2\lambda\sin\theta_{0}\sin\phi_{0}.
\end{align*}
 It is found that $\mathcal{M}$ is positive definite only if $2J_{y}\geq\epsilon$
and $J_{y}\geq2\lambda^{2}+J_{x}$. This solution has two degenerate
branches with $\phi_{0}=\pi/2$ and $\phi_{0}=3\pi/2$ corresponding
to two degenerate ferromagnetic states with spins polarized along
positive and negative $y$-axis, respectively. This can be clearly
shown by the spin $Q$-function in Sec.~\ref{sec:QPT_QF}. 

The mean values of the order parameters are obtained
\begin{align}
\frac{\langle\hat{a}^{\dagger}\hat{a}\rangle_{0}}{N} & =0,\ \frac{\langle\hat{S}_{x}^{2}\rangle_{0}}{N^{2}}=0,\ \frac{\langle\hat{S}_{y}^{2}\rangle_{0}}{N}=\frac{1}{4}\left(1-\frac{\epsilon^{2}}{4J_{y}^{2}}\right),
\end{align}
and $M_{z}=-\epsilon/4J_{y}$. In this case, parts of spins are polarized
along the $y$-axis forming many small ferromagnetic domains, but
still there are no macroscopic excitations in the bosonic mode. Thus,
this solution gives the FN phase. 

It deserves to be pointed out that, the system can be in any superposition
of these two degenerate ferromagnetic states when the system transitions to the FN phase after a QPT. As a result, the ensemble mean value of the $y$ component of the total angular momentum in the ground state is
always zero $\langle\hat{S}_{y}\rangle_{0}=0$.

\textbf{(iii) Ferromagnetic-Superradiant phase}\textemdash{} The third
solution is given by
\begin{align}
\cos\theta_{0} & =-\frac{\epsilon}{4\lambda^{2}+2J_{x}},\label{eq:theta_DLMGx}\\
\sin\phi_{0} & =0,\\
\alpha & =-\lambda\sin\theta_{0}\cos\phi_{0},\label{eq:alpha_DLMGx}
\end{align}
 requiring $|4\lambda^{2}+2J_{x}|\geq\epsilon$. Similarly, the negative
branch is ruled out by the stability condition. Utilizing the solution,
the Hessian matrix is simplified with the nonzero elements,
\begin{align*}
\mathcal{M}_{11} & =2,\\
\mathcal{M}_{22} & =-\frac{\epsilon}{2}\cos\theta_{0}+2\lambda^{2}\sin^{2}\theta_{0}-J_{x}\cos2\theta_{0},\\
\mathcal{M}_{33} & =(2\lambda^{2}+J_{x}-J_{y})\sin^{2}\theta_{0},\\
\mathcal{M}_{12} & =\mathcal{M}_{21}=2\lambda\cos\theta_{0}\cos\phi_{0}.
\end{align*}
To guarantee the positive definiteness of $\mathcal{M}$, one requires
$4\lambda^{2}+2J_{x}\geq\epsilon$ and $2\lambda^{2}+J_{x}\geq J_{y}$.
Similarly, this solution has two degenerate branches with $\phi_{0}=0$
and $\phi_{0}=\pi$ corresponding to two degenerate ferromagnetic
states polarized along positive and negative $x$-axis, respectively.
In the case of $\phi_{0}=0$, $\langle\hat{S}_{x}\rangle_{0}$ is
positive but $\langle\hat{d}^{\dagger}+\hat{d}\rangle_{0}$ is negative.
While for $\phi=\pi$, $\langle\hat{S}_{x}\rangle_{0}$ is negative
but $\langle\hat{d}^{\dagger}+\hat{d}\rangle_{0}$ is positive. As
a result, the ensemble mean value of both $\langle\hat{S}_{x}\rangle_{0}$
and $\langle\hat{d}^{\dagger}+\hat{d}\rangle_{0}$ are zero, but the
mean of their product is a finite negative value $\langle\hat{S}_{x}(\hat{d}^{\dagger}+\hat{d})\rangle_{0}<0$
and this makes the total energy reach its mimimum. 

The mean values of the order parameters are obtained
\begin{align}
\frac{\langle\hat{a}^{\dagger}\hat{a}\rangle_{0}}{N} & =\lambda^{2}\left[1-\frac{\epsilon^{2}}{(4\lambda^{2}+2J_{x})^{2}}\right]>0,\label{eq:zeta_S}\\
\frac{\langle\hat{S}_{x}^{2}\rangle_{0}}{N} & =\frac{1}{4}\left[1-\frac{\epsilon^{2}}{(4\lambda^{2}+2J_{x})^{2}}\right],\\
\frac{\langle\hat{S}_{y}^{2}\rangle_{0}}{N} & =0,
\end{align}
and $M_{z}=-\epsilon/2(4\lambda^{2}+2J_{x})$. In this case, part
of spins are polarized along the $x$-axis forming many small ferromagnetic
domains. At the same time, there are macroscopic excitations in the
bosonic mode. Thus, this solution gives the FS phase.

\section{Numerical Simulation \label{sec:numerical_simulation}}

In the traditional multi-spin system, the dimension of the Hilbert space
$\propto2^{N}$ diverges exponentially with the spin number $N$,
which makes it extremely hard to directly simulate for the $N>20$ case.
In our tractable model, this challenge can be circumvented by performing
the calculation in the Dicke-state subspace with dimension $N+1$
. If we take the cutoff of the bosonic mode as $N_{b}$, the Hamlitonian
of the whole system can be expanded with the $N_{b}\times N$ states
$|n\rangle\otimes|N/2,m\rangle$. Here, $\left|n\right\rangle $ ($n=0,1,2,\dots,N_{b}-1$)
is the Fock state of the bosonic mode and $\left|N/2,m\right\rangle $
($m=-N/2,-N/2+1,\dots,N/2-1,N/2$) is a Dicke state~\cite{dicke1954coherence}.

\subsection{Eigen-State Spectrum and Ground-State Properities}

Under the basis \{$n\rangle\otimes|N/2,m\rangle$\}, any operator
can be expressed as a matrix. Especially, we can diagonalize the matrix
of the Hamiltonian $\hat{H}$ to get its eigen states and the eigen-energy
spectrum. Each eigen state is a vector with dimension $N_{b}\times N$
and the mean value of an arbitrary operator $\hat{O}$ can be easily
calculated with the eigen states obtained from the numerical diagonalization.

The $Q$-functions of the bosonic mode and the spins can also be obtained by constructing the bosonic coherent states and the coherent spin states in this basis.

By constructing the thermal equilibrium state density matrix,
\begin{equation}
\hat{\rho}(T)=\frac{1}{Z}e^{-\beta\hat{H}}=\frac{1}{Z}\hat{M}e^{-\beta\hat{D}}\hat{M}^{\dagger},
\end{equation}
we can also study the thermoequilibrium property of the system at
finite termperature. Here, $\beta=1/k_{B}T$ is the inverse temperature
and $Z={\rm Tr}\hat{\rho}={\rm Tr}\exp(-\beta\hat{D})$ is the partition
function. The unitary matrix $\hat{M}$ and the diagonal matrix $D$
are obtained from the eigendecomposition of the Hamiltonian
\begin{equation}
\hat{H}\hat{M}=\hat{D}\hat{M}.
\end{equation}

\subsection{Dynamics in the Hilbert Space}

In the dynamical amplification part, we need to study the dynamics
of the system under a time-dependent Hamiltonian. Analytically, we
can utilize the time-ordered expansion
\begin{eqnarray}
|\psi(t)\rangle & = &  \sum_{k=0}^{\infty}\frac{1}{k!}\int_{0}^{t}\cdots\int_{0}^{t}dt_{1}\cdots dt_{k}\nonumber \\
 &   & \times\mathcal{T}\left[\hat{H}(t_{k})\cdots\hat{H}(t_{2})\hat{H}(t_{1})\right]|\psi(0)\rangle,
\end{eqnarray}
to investigate the short-time behavior of the system. However, for a QPT
system, due to the vanishing energy gap, most interesting effects come
from the adiabatic long-time behavor. The numerical approach introduced
in the following holds a significant advantage for dynamical amplification
and noise calculations in the long-time limit.

By splitting the dynamical process into many small time intervals
($\Delta t_{1},\Delta t_{2},\dots,\Delta t_{k},\dots$), the time-evolution
can be approximated by
\begin{equation}
|\psi(t)\rangle=\cdots e^{-i\hat{H}(t_{k})\Delta t_{k}}\cdots e^{-i\hat{H}(t_{2})\Delta t_{2}}e^{-i\hat{H}(t_{1})\Delta t_{1}}|\psi(0)\rangle.\label{eq:TimeEvolution}
\end{equation}
The mean value of an arbitrary operator and its uncertainty can be
calcualted via the wave function $|\psi(t)\rangle$.

\subsection{Dynamics in The Louville Space}

The numerical method introduced in the previous subsection only works
for the time-evolution of pure states. In this subsection, we generalize
this approach to the mixed-state case. 

\subsubsection{Single-body system}

In a given basis spanned by $\{\left|n\right\rangle \}$ with dimension
$N$, the density matrix of an arbitrary state can always be expanded
as
\begin{equation}
\hat{\rho}=\sum_{mn}\rho_{mn}\left|m\right\rangle \left\langle n\right|,
\end{equation}
where $\rho_{mn}=\left\langle m\right|\hat{\rho}\left|n\right\rangle $
is the element of the density matrix $\hat{\rho}$. For a close system,
the dynamics of the density matrix is governed by the Louville-von
Neumann equation
\begin{equation}
\frac{d}{dt}\hat{\rho}(t)=-i[\hat{H}(t),\hat{\rho}(t)]\equiv\hat{\hat{\mathcal{L}}}(t)\rho(t),
\end{equation}
where $\hat{\hat{\mathcal{L}}}(t)$ is a superoperator ($N^{2}\times N^{2}$
matrix) in the Louville space. This equation has an equivalent form as the Schr\"odinger equation. Thus, we can use the same method in
Eq.~(\ref{eq:TimeEvolution}) to operate the time-evolution. For
a open system, its dynamics is governed by a master equation with
extra fluctuation-dissipation terms $\hat{\hat{\mathcal{L}}}_{B}$
coming from the bath. The numerical approach can be generalized to
quantum open system straightforwardly. The key step for numerical
simulation of the time-evolution in the Louville space is to construct
the matrix-form of the superoperator and the vector-form of the density
matrix. In the following, we will show how to do this exactly.

\textcolor{black}{To perform the time-evolution in the Louville space,
we can reshape the density matrix $\rho$ to a vector in the basis}
\begin{equation}
\left|\rho\right\rangle =\rho_{mn}\left|m,n\right\rangle ,
\end{equation}
with
\begin{equation}
\left|m,n\right\rangle \equiv\left|m\right\rangle \otimes\left\langle n\right|.
\end{equation}
In this basis, we create a new space with dimension $N^{2}$ and basis
$\left|m,n\right\rangle $, which is the tensor product of a ket vector
and a bra vector. In MATLAB, if you make the operation $\rho(:)$,
the density matrix $\rho$ will be automatically reshaped to a vector
in this space.\textbf{ }In this basis, we have the following Hilbert-Louville
correspondences:
\begin{align}
\hat{H}(t)\hat{\rho}(t) & \leftrightarrow\hat{\hat{H}}_{L}(t)\left|\rho(t)\right\rangle ,\\
\hat{\rho}(t)\hat{H}(t) & \leftrightarrow\hat{\hat{H}}_{R}(t)\left|\rho(t)\right\rangle ,
\end{align}
where the $N^{2}\times N^{2}$ superoperators are given by
\begin{align}
\hat{\hat{H}}_{L}(t) & =\hat{H}(t)\otimes\hat{I}_{R}\\
\hat{\hat{H}}_{R}(t) & =\hat{I}_{L}\otimes\hat{H}^{T}(t),
\end{align}
and $\hat{I}_{R}=\hat{I}_{L}$ is the identity matrix with dimension
$N$. Now, we give the following correspondences to construct superoperators,
\begin{align}
\hat{O}_{1}\hat{\rho} & \leftrightarrow\hat{O}_{1}\otimes\hat{I}_{R}\left|\rho\right\rangle \\
\hat{\rho}\hat{O}_{1} & \leftrightarrow\hat{I}_{L}\otimes\hat{O}_{1}^{T}\left|\rho\right\rangle \\
\hat{O}_{1}\hat{\rho}\hat{O}_{2} & \leftrightarrow\hat{O}_{1}\otimes\hat{O}_{2}^{T}\left|\rho\right\rangle 
\end{align}
We emphysize that the order of these two operators $\hat{O}_{1}$
and $\hat{O}_{2}$ depends on the reshaping of the density matrix $\rho$
to a vector $\left|\rho\right\rangle $.\textbf{ }One can also construct
a space with basis
\begin{equation}
\left|m,n\right\rangle =\left\langle m\right|\otimes\left|n\right\rangle ,
\end{equation}
by the operation $\rho'(:)$ in MATLAB. Then, we have the correspondence
in a inverse order: 
\begin{align*}
\hat{O}_{1}\hat{\rho} & \leftrightarrow\hat{I}_{R}\otimes\hat{O}_{1}^{T}\left|\rho\right\rangle \\
\hat{\rho}\hat{O}_{1} & \leftrightarrow\hat{O}_{1}\otimes\hat{I}_{R}\left|\rho\right\rangle \\
\hat{O}_{1}\hat{\rho}\hat{O}_{2} & \leftrightarrow\hat{O}_{2}\otimes\hat{O}_{1}^{T}\left|\rho\right\rangle 
\end{align*}
The mean value of any Hermitian operator in Hilbert space can be calculated
as the inner product of two vectors in Louville space
\begin{equation}
\langle\hat{O}\rangle={\rm Tr}[\hat{\rho}\hat{O}]={\rm Tr}[\hat{O}\hat{\rho}]=\left\langle O\right|\rho\rangle=\left\langle \rho\right|O\rangle.
\end{equation}

\subsubsection{Multi-body system}

If we have two system $A$ and $B$ with basis $\{\left|n_{a}\right\rangle \otimes\left|n_{b}\right\rangle \}$,
the density matrix of the combined system can be expanded as
\begin{equation}
\rho=\sum_{m_{a}n_{a}}\sum_{m_{b}n_{b}}\rho_{m_{a}m_{b};n_{a}n_{b}}\left|m_{a}m_{b}\right\rangle \left\langle n_{a}n_{b}\right|.
\end{equation}
In this case, there are two main ways to reshape the density matrix
to a vector:

\textbf{(i)} First do the tensor product of different systems $A$
and $B$, and then take the tensor product of bra and ket vectors:
\begin{equation}
\left|\rho\right\rangle =\sum_{m_{a}n_{a}}\sum_{m_{b}n_{b}}\rho_{m_{a}m_{b};n_{a}n_{b}}\left|m_{a},m_{b};n_{a},n_{b}\right\rangle ,
\end{equation}
with
\begin{equation}
\left|m_{a},m_{b};n_{a},n_{b}\right\rangle =\left(\left|m_{a}\right\rangle \otimes\left|m_{b}\right\rangle \right)\otimes\left(\left\langle n_{a}\right|\otimes\left\langle n_{b}\right|\right).
\end{equation}
In this case, the Hilbert-Louville correspondence is given by,
\begin{align}
\hat{A}\hat{B}\hat{\rho} & \leftrightarrow\left(\hat{A}\otimes\hat{B}\right)\otimes\left(\hat{I}_{A}\otimes\hat{I}_{B}\right)\left|\rho\right\rangle \\
\hat{\rho}\hat{A}\hat{B} & \leftrightarrow\left(\hat{I}_{A}\otimes\hat{I}_{B}\right)\otimes\left(\hat{A}\otimes\hat{B}\right)^{T}\left|\rho\right\rangle\\
\hat{A}\hat{\rho}\hat{B} & \leftrightarrow\left(\hat{A}\otimes\hat{I}_{B}\right)\otimes\left(\hat{I}_{A}\otimes\hat{B}\right)^{T}\left|\rho\right\rangle .
\end{align}

\textbf{(ii)} First do the tensor product of bra and ket vectors,
and then do the tensor product of different systems $A$ and $B$:
\begin{equation}
\left|\rho\right\rangle =\sum_{m_{a}n_{a}}\sum_{m_{b}n_{b}}\rho_{m_{a}n_{a};m_{b}n_{b}}\left|m_{a},n_{a};m_{b},n_{b}\right\rangle ,
\end{equation}
with
\begin{equation}
\left|m_{a},n_{a};m_{b},n_{b}\right\rangle =\left(\left|m_{a}\right\rangle \otimes\left\langle n_{a}\right|\right)\otimes\left(\left|m_{b}\right\rangle \otimes\left\langle n_{b}\right|\right).
\end{equation}
In this case, we have the Hilbert-Louville correspondence
\begin{align}
\hat{A}\hat{B}\hat{\rho} & \leftrightarrow\left(\hat{A}\otimes\hat{I}_{A}\right)\otimes\left(\hat{B}\otimes\hat{I}_{B}\right)\left|\rho\right\rangle \\
\hat{\rho}\hat{A}\hat{B} & \leftrightarrow\left(\hat{I}_{A}\otimes\hat{A}^{T}\right)\otimes\left(\hat{I}_{B}\otimes\hat{B}^{T}\right)\left|\rho\right\rangle \\
\hat{A}\hat{\rho}\hat{B} & \leftrightarrow\left(\hat{A}\otimes\hat{I}_{A}\right)\otimes\left(\hat{I}_{B}\otimes\hat{B}^{T}\right)\left|\rho\right\rangle .
\end{align}
For multi-body system, the rules are the same as two-body system.

\bibliography{main}

\end{document}